\documentclass[english]{elsarticle}
\usepackage[a4paper]{geometry}
\usepackage[T1]{fontenc}
\usepackage[latin9]{inputenc}
\usepackage{amsthm}
\usepackage{amsmath}
\usepackage{amsfonts}
\usepackage{mathabx}
\usepackage{amssymb}
\usepackage{babel}
\usepackage[numbers]{natbib}
\usepackage{graphicx}
\usepackage{placeins}
\usepackage{multirow}
\usepackage{color}
\usepackage[]{epstopdf}

\makeatletter
\def\ps@pprintTitle{%
 \let\@oddhead\@empty
 \let\@evenhead\@empty
 \def\@oddfoot{}%
 \let\@evenfoot\@oddfoot}
\makeatother

\begin{document}
\title{Electro-magneto-mechanically response of polycrystalline materials: Computational Homogenization via the Virtual Element Method}

\author[1]{Christoph B\"ohm\corref{cor1}}
\ead{boehm@ikm.uni-hannover.de}
\author[1]{Bla\v{z} Hudobivnik}
\author[1,2]{Michele Marino}
\author[1]{Peter Wriggers}

\cortext[cor1]{Corresponding author}

\address[1]{Institute of Continuum Mechanics, Leibniz University Hanover, An der Universit\"at 1, 30823 Garbsen, Germany}
\address[2]{Dep. of Civil Eng. and Computer Science, University Rome Tor Vergata, Via del Politecnico 1, 00133 Rome, Italy}

\begin{abstract}
This work presents a study on the computational homogenization of electro-magneto-mechanically coupled problems through the Virtual Element Method (VEM). VE-approaches have great potential for the homogenization of the physical properties of heterogeneous polycrystalline microstructures with anisotropic grains. The flexibility in element shapes can be exploited for creating VE-mesh with a significant lower number of degrees of freedom if compared to finite element (FE) meshes, while maintaining a high accuracy. Evidence that VE-approaches outperform FEM are available in the literature, but only addressing purely-mechanic problems (i.e. elastic properties) and transversely anisotropic materials. The aim of this work is twofold. On one hand, the study compares VE-and FE-based numerical homogenization schemes for electro-mechanically coupled problems for different crystal lattice structures and degrees of elastic anisotropy. Within all considered materials, the VE-approach outperforms the FE-approach for the same number of nodes. On the other hand a hybrid microstructure made up by both electro-mechanical and magneto-mechanical grains is investigated resulting in a electro-magneto-mechanically coupled microstructure. Again, VEM provides a more accurate solution strategy. 
\end{abstract}
\begin{keyword}
Computational Homogenization \sep Virtual Element Method (VEM) \sep Microstructure \sep Electro-Magneto-Mechanics
\end{keyword}

\maketitle

\section{Introduction}
\noindent The consideration of physical effects, occuring at microscopic length scales of polycrystalline aggregates, is well established within the framework of computational micromechanics. A wide range of information, gained from this length scale, is available on macroscopic environments resulting from homogenization techniques. Voigt, Reuss and Hill in for instance \cite{voigt28,reuss29,hil52} developed strategies to bridge from the micro- up to the macroscopic framework. As another prominent example, the work on $\text{FE}^{\mathrm{2}}$-Schemes in \cite{terada2000simulation,schroder2014numerical,vsolinc2015simple} is mentioned here. Moreover, a detailed illustration of the underlying micromechanical principles from the standard setup up to more advanced relations, can be found in \cite{zohdi2008introduction}.\\
When addressing the numerical homogenization of polycrystalline microstructures, the material symmetry of single grains is affected by its particular lattice structure. Complex behavior can arise when grains are endowed by lower symmetries such as trigonal or orthorhombic structures. In addition, the geometric properties of grains can lead to distorted finite element meshes, if not highly refined in certain regions. For these reasons, classical FE-approaches require either a large number of elements or a higher polynomial degree to obtain small homogenization errors in the framework of computational micromechanics. As outlined in \cite{mar19}, VEM demonstrates an outperformance in comparison with classical FE-approaches regarding the homogenization of purely elastic mechanical properties of polycrystalline assembles with transversely isotropic grains.\\
However, when considering a multiphysics framework with piezo-electricity and/or piezo-magnetism, the performance of VE-homogenization schemes remains an open question. In this regards, a comparison between VE- and FE-approaches will be conducted in 3D and addressing different crystal systems, extending the results in \cite{mar19}. It will be shown that a VEM strategy is a more efficient and more robust option with regards to the same level of accuracy. In fact, FEM performs well when addressing quasi-isotropic behavior, even if geometric properties are complex and the number of nodes is at a minimum level. Contrary, by an increase of anisotropy, the error of a FE-approach increases up to a not acceptable level. To improve this behaviour, one could choose either a finer mesh, or a higher polynomial degree for the ansatz-functions, or a combination of both. But this results in an increase of the simulation time. On the other hand, VEM outperforms FEM in this regime with the minimum level of nodes, by defining only one element for each grain. The study addresses piezo-electric grains for electro-mechanical polycrystalline assemblies, as well as hybrid microstructures, with both piezo-electrical and magneto-mechanical grain materials for electro-magneto-mechanically coupled problems.

\section{Materials and Methods}

\subsection{Constitutive Framework of the microscopic Boundary Value Problem}
\begin{figure}[htp!]
	\centering
	\includegraphics[width=0.6\textwidth]{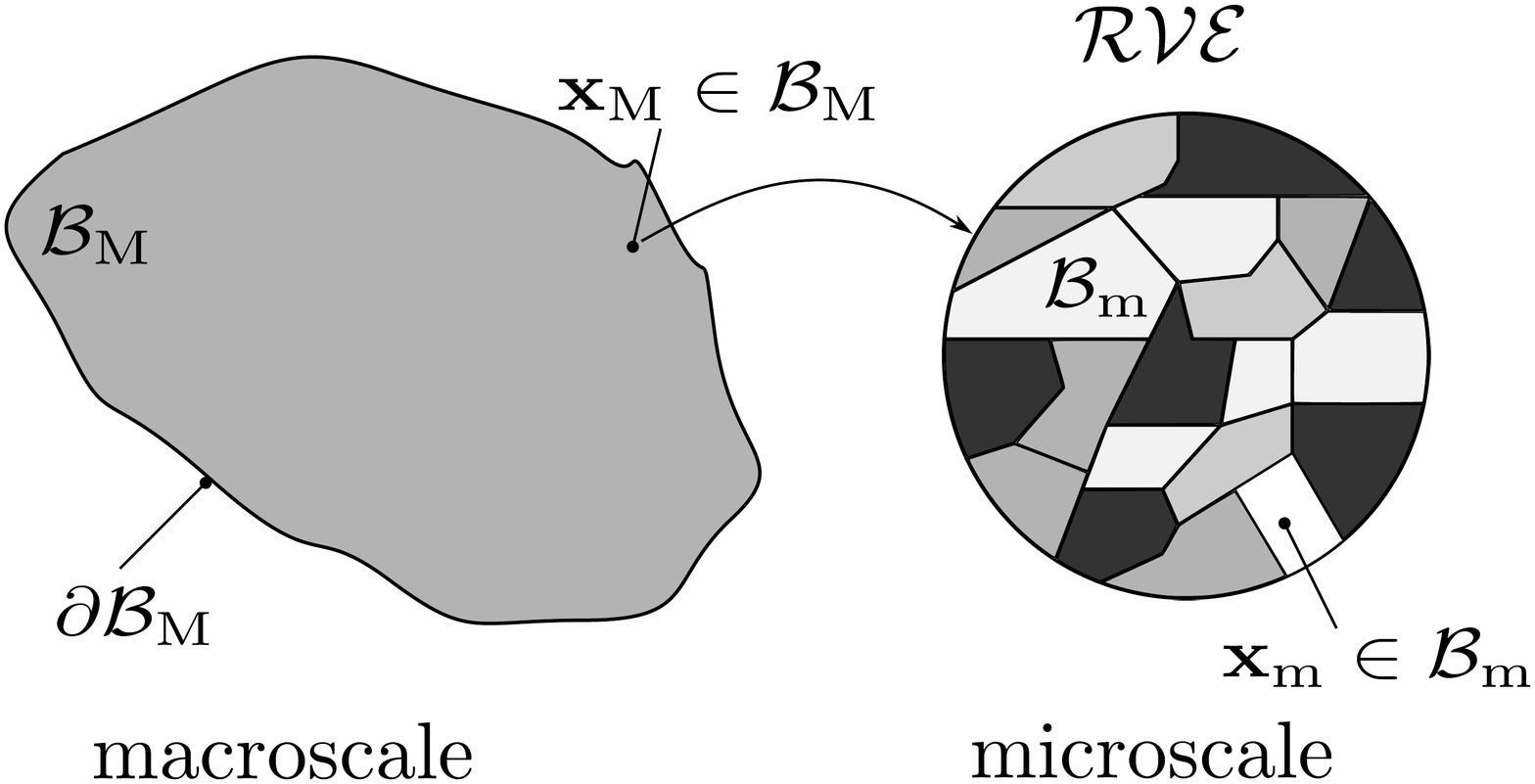}
	\caption{Schematic illustration of macroscopic continuum $\mathcal{B}_{\mathrm{M}}$ with material point $\mathbf{x}_{\mathrm{M}}$ and its related representative volume element
$\mathcal{RVE}$ at microscopic length scale}
	\label{fig:Config}
\end{figure}

\begin{figure}[htp!]
    \centering
    \includegraphics[width=0.5\textwidth]{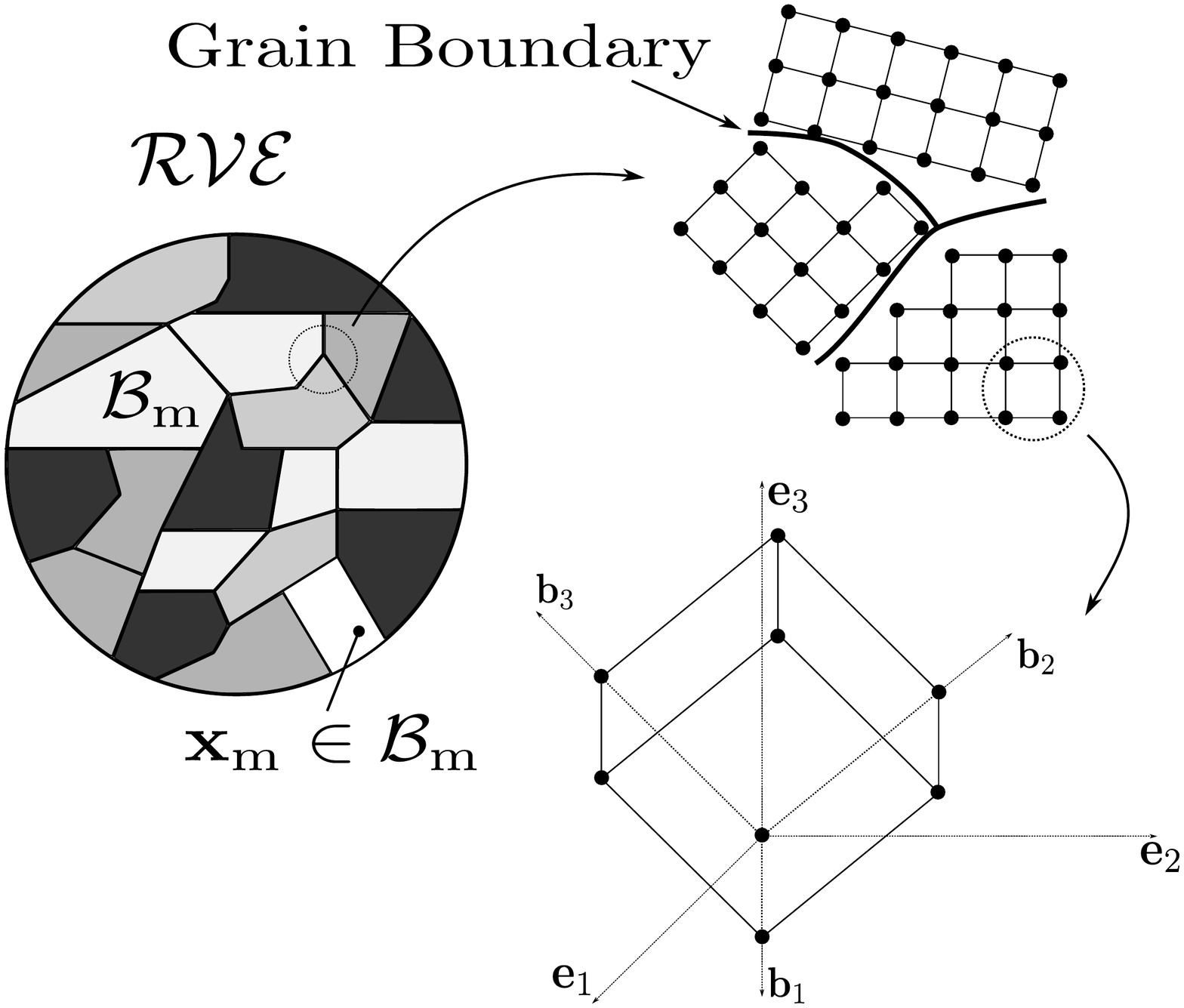}
    \caption{Schematic illustration of a cubic unit cell with lattice-aligned axes $\mathbf{b}_{i}$ rotated with respect to coordinate axes $\mathbf{e}_{i}$}
    \label{fig:RotatedLattice}
\end{figure}

\noindent Let $\mathcal{B}_{\mathrm{M}}\subset\mathbb{R}^{\mathrm{3}}$ be the continuum in three-dimensional space at macroscopic level. Then, $\mathbf{x}_{\mathrm{M}}$ denotes a material point at $\mathcal{B}_{\mathrm{M}}$, see Figure \ref{fig:Config}. Moreover, $\partial\mathcal{B}_{\mathrm{M}}\subset\mathbb{R}^{\mathrm{2}}$ depicts the surface of the macroscopic continuum, where appropriate boundary conditions are applied. 
\FloatBarrier
\noindent A representative volume element $\mathcal{RVE}$ at the microscale is attached to every macroscopic material point $\mathbf{x}_{\mathrm{M}}\in\mathcal{B}_{\mathrm{M}}$. Then, for every $\mathcal{RVE}$, let $\mathcal{B}_{\mathrm{m}}\subset\mathcal{B}_{\mathrm{M}}$ be the microscopic continuum. Thus, $\mathbf{x}_{\mathrm{m}}$ states a material point $\mathcal{B}_{\mathrm{m}}$ as well as $\partial\mathcal{B}_{\mathrm{m}}$ the surface at the microscopic level. The three primary fields, namely the mechanical displacement $\mathbf{u}$, the electric potential $\phi$ as well as the magnetic potential $\varphi$ in $\mathcal{B}_{\mathrm{m}}$, are introduced which are summarized in
\begin{equation}
    \mathcal{R}=\lbrace\mathbf{u},\phi,\varphi\rbrace.
\end{equation}
In line with the small strain assumption, their gradients are introduced as
\begin{equation}\label{eq:strain}
    \begin{aligned}
        \boldsymbol{\varepsilon}=\mathrm{sym}\left(\triangledown\mathbf{u}\right),\qquad \mathbf{E}=-\triangledown\phi,\qquad \mathbf{H}=-\triangledown\varphi,
    \end{aligned}
\end{equation}
with $\lbrace \boldsymbol{\varepsilon}, \mathbf{E}, \mathbf{H}\rbrace$ being the infinitesimal strain tensor $\boldsymbol{\varepsilon}$, the electric field vector $\mathbf{E}$, and the magnetic field vector $\mathbf{H}$ at microscopic length scale. The set of microscopic dual quantities to Eq. \eqref{eq:strain} is then given by $\lbrace \boldsymbol{\sigma}, \mathbf{D}, \mathbf{B}\rbrace$ being respectively the Cauchy stress tensor, the electric displacement vector and the magnetic flux density vector \cite{sch16}. The governing equations in the absence of body forces are the balance of linear momentum and Gauss' law of electrostatics as well as magneto-statics
\begin{equation}\label{eq:goveq}
    \begin{aligned}
        \mathrm{div}\left(\boldsymbol{\sigma}\right)=\mathbf{0},\qquad \mathrm{div}\left(\mathbf{D}\right)=Q,\qquad \mathrm{div}\left(\mathbf{B}\right)=0.
    \end{aligned}
\end{equation}
Here, $Q$ denotes the density of free charge carriers \cite{sch16}.\\
\noindent Microscopic constitutive relations depend on the type of crystal lattice characterizing grains in the polycrystalline assembly. 
Moreover, the material symmetry directions (i.e., crystal lattice) of each grain are rotated within the microstructure, see Figure \ref{fig:RotatedLattice}, resulting in a highly heterogeneous structure.
\FloatBarrier
\noindent Following \cite{sch16}, a linear relationship is assumed for each grain between the dual quantities at the microstructure, expressed in Voigt notation as:
\begin{equation}
\underbrace{
\begin{bmatrix}
\boldsymbol{\sigma}_{\mathrm{l}}\\
-\mathbf{D}_{\mathrm{l}}\\
-\mathbf{B}_{\mathrm{l}}
\end{bmatrix}
}_{\displaystyle \mathbf{L}_{\mathrm{l}}}=
\underbrace{
\begin{bmatrix}
\mathbb{C}_{\mathrm{l}} & -\mathbf{e}_{\mathrm{l}}^{\mathrm{T}} & -\mathbf{q}_{\mathrm{l}}^{\mathrm{T}}\\
-\mathbf{e}_{\mathrm{l}} & -\boldsymbol{\epsilon}_{\mathrm{l}} & -\boldsymbol{\alpha}_{\mathrm{l}}\\
-\mathbf{q}_{\mathrm{l}} & -\boldsymbol{\alpha}_{\mathrm{l}} & \boldsymbol{\mu}_{\mathrm{l}}
\end{bmatrix}
}_{\displaystyle \mathbb{G}_{\mathrm{l}}}
\underbrace{
\begin{bmatrix}
\boldsymbol{\varepsilon}_{\mathrm{l}}\\
\mathbf{E}_{\mathrm{l}}\\
\mathbf{H}_{\mathrm{l}}
\end{bmatrix}
}_{\displaystyle \mathbf{P}_{\mathrm{l}}}.
\label{eq:relation}
\end{equation}
Here $\left(\cdot\right)_{\mathrm{l}}$ denotes a quantity, related to the \textit{local} grain-specific coordinate system. $\lbrace\mathbb{C},\mathbf{e},\mathbf{q},\boldsymbol{\epsilon},\boldsymbol{\alpha},\boldsymbol{\mu}\rbrace$ are respectively the mechanical, piezoelectric, piezomagnetic, dielectric, electromagnetic and the magnetic modulus at grain-level. These moduli are collected in a generalized modulus $\mathbb{G}$, see \cite{sch16}. The constitutive relationship in Eq. \eqref{eq:relation} can be described by introducing a quadratic potential:
\begin{equation}\label{eq:grainpotential}
    \psi_{g}=\frac{1}{2}\mathbf{P}\cdot\mathbb{G}\cdot\mathbf{P}.
\end{equation}
Where $\mathbf{P}=\mathrm{vec}\lbrace\boldsymbol{\varepsilon},\mathbf{E},\mathbf{H}\rbrace$ and $\mathbb{G}$ are represented in a global coordinate system, common among all grains in the $\mathcal{RVE}$. Differentiation of Eq. \eqref{eq:grainpotential} with respect to the strains and the gradients of the magnetic and electro-static potentials yields 
\begin{equation}
    \begin{aligned}
        \boldsymbol{\sigma}=\frac{\partial\psi_{g}}{\partial\boldsymbol{\varepsilon}},\qquad \mathbf{D}=-\frac{\partial\psi_{g}}{\partial\mathbf{E}},\qquad \mathbf{B}=-\frac{\partial\psi_{g}}{\partial\mathbf{H}}.
    \end{aligned}
\end{equation}
\noindent A transformation of the grain specific coordinate system $\mathbb{G}_{\mathrm{l}}\to\mathbb{G}$ is obtained by a set of rotation tensors $\mathbb{II}_{\mathrm{\mathbf{L}}}$ and $\mathbb{II}_{\mathrm{\mathbf{P}}}$, extending the approach in \cite{kol03} to $\lbrace\mathbf{E},\mathbf{H}\rbrace$ and their dual quantities $\lbrace\mathbf{D},\mathbf{B}\rbrace$, which transforms $\mathbf{L}$ and $\mathbf{P}$, respectively, 
\begin{equation}\label{eq:coordtrafo}
\begin{aligned}
\mathbf{L}_{\mathrm{l}}=
\underbrace{
\begin{bmatrix}
\mathbf{T}^{\sigma} & \mathbf{0} & \mathbf{0}\\
\mathbf{0} & \mathbf{Q} & \mathbf{0}\\
\mathbf{0} & \mathbf{0} & \mathbf{Q}
\end{bmatrix}
}_{\displaystyle \mathbb{II}_{\mathbf{L}}}
\mathbf{L},\qquad \mathbf{P}_{\mathrm{l}}=
\underbrace{
\begin{bmatrix}
\mathbf{T}^{\varepsilon} & \mathbf{0} & \mathbf{0}\\
\mathbf{0} & \mathbf{Q} & \mathbf{0}\\
\mathbf{0} & \mathbf{0} & \mathbf{Q}
\end{bmatrix}
}_{\displaystyle \mathbb{II}_{\mathbf{P}}}
\mathbf{P},
\end{aligned}
\end{equation}
where $\mathbf{T}^{\lbrace\sigma,\varepsilon\rbrace}$ are coordinate transformations of the Cauchy stress tensor and of the infinitesimal strain tensor \cite{kol03} and $\mathbf{Q}$ denotes a rotation tensor (see Appendix \ref{subsec:TransformMatrices}). Modulus $\mathbb{G}$ is obtained from Eq. \eqref{eq:relation} and \eqref{eq:coordtrafo}, by
\begin{equation}
    \begin{aligned}
        \mathbb{G}=\mathbb{II}_{\mathbf{L}}^{\mathrm{-1}}\mathbb{G}_{\mathrm{l}}\mathbb{II}_{\mathbf{P}},\qquad \lbrace\mathbb{II}_{\mathbf{L}},\mathbb{II}_{\mathbf{P}}\rbrace\in\mathcal{SO}\left(3\right).
    \end{aligned}
\end{equation}

\subsection{Homogenization of the microscopic Boundary Value Problem}
\noindent Macroscopic homogenized quantities are obtained through homogenization of microscale fields within the $\mathcal{RVE}$. The macroscopic set $\lbrace\overline{\boldsymbol{\varepsilon}},\overline{\mathbf{E}},\overline{\mathbf{H}},\overline{\boldsymbol{\sigma}},\overline{\mathbf{D}},\overline{\mathbf{B}}\rbrace$ is linked to its set of microscopic counterparts $\lbrace\boldsymbol{\varepsilon},\mathbf{E},\mathbf{H},\boldsymbol{\sigma},\mathbf{D},\mathbf{B}\rbrace$ through averaged volume integrals \cite{sch16}, yielding
\begin{equation}\label{eq:VolumeAveragingQuantities}
    \begin{aligned}
        \overline{\boldsymbol{\varepsilon}}&=\frac{1}{V}\int_{\mathcal{B}_{\mathrm{m}}}\mathrm{sym}\left(\triangledown\mathbf{u}\right)\mathrm{dV}=\frac{1}{V}\int_{\partial\mathcal{B}_{\mathrm{m}}}\mathrm{sym}\left(\mathbf{u}\otimes\mathbf{n}_{\mathrm{m}}\right)\mathrm{dA},\quad \overline{\boldsymbol{\sigma}}=\frac{1}{V}\int_{\mathcal{B}_{\mathrm{m}}}\boldsymbol{\sigma}\mathrm{dV}=\frac{1}{V}\int_{\partial\mathcal{B}_{\mathrm{m}}}\mathrm{sym}\left(\mathbf{t}\otimes\mathbf{x}_{\mathrm{m}}\right)\mathrm{dA},\\
        \overline{\mathbf{E}}&=-\frac{1}{V}\int_{\mathcal{B}_{\mathrm{m}}}\triangledown\phi\mathrm{dV}=-\frac{1}{V}\int_{\partial\mathcal{B}_{\mathrm{m}}}\phi\mathbf{n}_{\mathrm{m}}\mathrm{dA},\quad \overline{\mathbf{D}}=-\frac{1}{V}\int_{\mathcal{B}_{\mathrm{m}}}\mathbf{D}\mathrm{dV}=-\frac{1}{V}\int_{\partial\mathcal{B}_{\mathrm{m}}}\Xi\mathbf{x}_{\mathrm{m}}\mathrm{dA},\\
          \overline{\mathbf{H}}&=-\frac{1}{V}\int_{\mathcal{B}_{\mathrm{m}}}\triangledown\varphi\mathrm{dV}=-\frac{1}{V}\int_{\partial\mathcal{B}_{\mathrm{m}}}\varphi\mathbf{n}_{\mathrm{m}}\mathrm{dA},\quad \overline{\mathbf{B}}=-\frac{1}{V}\int_{\mathcal{B}_{\mathrm{m}}}\mathbf{B}\mathrm{dV}=-\frac{1}{V}\int_{\partial\mathcal{B}_{\mathrm{m}}}\Theta\mathbf{x}_{\mathrm{m}}\mathrm{dA}.
    \end{aligned}
\end{equation}
The micro-macro transition criterion postulates the fulfillment of Hill's condition
\begin{equation}\label{eq:hill}
    \begin{aligned}
        \langle\boldsymbol{\sigma}\colon\boldsymbol{\varepsilon}\rangle-\langle\mathbf{D}\cdot\mathbf{E}\rangle-\langle\mathbf{B}\cdot\mathbf{H}\rangle=\overline{\boldsymbol{\sigma}}\colon\overline{\boldsymbol{\varepsilon}}-\overline{\mathbf{D}}\cdot\overline{\mathbf{E}}-\overline{\mathbf{B}}\cdot\overline{\mathbf{H}},
    \end{aligned}
\end{equation}
with $\langle\cdot\rangle=\frac{1}{V}\int\cdot\mathrm{d}V$ being the average operator in the $\mathcal{RVE}$. Eq. \eqref{eq:hill} shall hold true for every mechanical, electric and magnetic field applied to the $\mathcal{RVE}$, leading to
\begin{equation}\label{eq:hillparts}
    \begin{aligned}
      \langle\boldsymbol{\sigma}\colon\boldsymbol{\varepsilon}\rangle=\overline{\boldsymbol{\sigma}}\colon\boldsymbol{\varepsilon},\quad\langle\mathbf{D}\cdot\mathbf{E}\rangle=\overline{\mathbf{D}}\cdot\overline{\mathbf{E}},\quad\langle\mathbf{B}\cdot\mathbf{H}\rangle=\overline{\mathbf{B}}\cdot\overline{\mathbf{H}}.
    \end{aligned}
\end{equation}
Hence, the set of macroscopic, homogenized, moduli are given by:
\begin{equation}\label{eq:homogenModuli}
    \begin{aligned}
        \overline{\mathbb{C}}=\frac{\partial\langle\boldsymbol{\sigma}\rangle}{\partial\overline{\boldsymbol{\varepsilon}}},\qquad \overline{\mathbf{e}}=\frac{\partial\langle\mathbf{D}\rangle}{\partial\overline{\boldsymbol{\varepsilon}}},\qquad \overline{\mathbf{q}}=\frac{\partial\langle\mathbf{B}\rangle}{\partial\overline{\boldsymbol{\varepsilon}}},\\
        \overline{\boldsymbol{\epsilon}}=\frac{\partial\langle\mathbf{D}\rangle}{\partial\overline{\mathbf{E}}},\qquad \overline{\boldsymbol{\alpha}}=\frac{\partial\langle\mathbf{B}\rangle}{\partial\overline{\mathbf{E}}},\qquad \overline{\boldsymbol{\mu}}=\frac{\partial\langle\mathbf{B}\rangle}{\partial\overline{\mathbf{H}}}.
    \end{aligned}
\end{equation}
At the macroscopic level, a linear constitutive relation is introduced as in Eq. \eqref{eq:relation} at the microscopic length scale, defining the homogenized macroscopic modulus $\overline{\mathbb{G}}$:
\begin{equation}\label{eq:macroscopicModulus}
    \overline{\mathbb{G}}=\begin{bmatrix}
    \overline{\mathbb{C}} & -\overline{\mathbf{e}}^{\mathrm{T}} & -\overline{\mathbf{q}}^{\mathrm{T}}\\
    -\overline{\mathbf{e}} & -\overline{\boldsymbol{\epsilon}} & -\overline{\boldsymbol{\alpha}}^{\mathrm{T}}\\
    -\overline{\mathbf{q}} & -\overline{\boldsymbol{\alpha}} & \overline{\boldsymbol{\mu}}
    \end{bmatrix}.
\end{equation}
The homogenized quantities from Eq. \eqref{eq:homogenModuli} are obtained by employing 12 independent states of boundary conditions which lead to $\langle\mathbf{P}\rangle=\overline{\mathbf{P}}_{m}\text{, such that }\left[\overline{\mathbf{P}}_{m}\right]_{i}=\delta_{im}$ where $\delta_{im}$ is the Kronecker delta $(i,m=1,...,12)$. The boundary conditions, applied at the boundary $\partial\mathcal{RVE}$ of the $\mathcal{RVE}$, are of Dirichlet-type. Thus, the surface $\partial\mathcal{RVE}$ is constrained by either prescribed values for uniform displacements, electric potential or magnetic potential $\lbrace\mathbf{u},\phi,\varphi\rbrace\rvert_{\partial\mathcal{RVE}}$, as:
\begin{equation}\label{eq:bconditions}
\mathrm{For}
\begin{cases}
    \begin{aligned}
    m&\leq 6 && \mathbf{u}\rvert_{\partial\mathcal{RVE}}=\bar{\boldsymbol{\varepsilon}}_{m}\mathbf{x}_{m}, &&& \phi\rvert_{\partial\mathcal{RVE}}&=0, &&&& \varphi\rvert_{\partial\mathcal{RVE}}=0\\
    6<m&\leq 9 && \mathbf{u}\rvert_{\partial\mathcal{RVE}}=\mathbf{0}, &&& \phi\rvert_{\partial\mathcal{RVE}}&=\frac{x_{m}}{L}, &&&& \varphi\rvert_{\partial\mathcal{RVE}}=0\\
    9<m&\leq 12 && \mathbf{u}\rvert_{\partial\mathcal{RVE}}=\mathbf{0}, &&& \phi\rvert_{\partial\mathcal{RVE}}&=0, &&&& \varphi\rvert_{\partial\mathcal{RVE}}=\frac{x_{m}}{L}
    \end{aligned}
    \end{cases},
\end{equation}
with $\bar{\boldsymbol{\varepsilon}}_{m}$ such that
\begin{equation}\label{eq:epsilonm}
    \begin{aligned}
         \left[\mathrm{vec}\lbrace\bar{\boldsymbol{\varepsilon}}_{m}\rbrace\right]_{i}&=\delta_{im};\; \text{and} && x_{7}=x_{10}=x, &&& x_{8}=x_{11}=y, &&&& x_{9}=x_{12}=z.
    \end{aligned}
\end{equation}
These are variables with respect to the coordinate axes $\mathbf{e}_{i}$ in the $\mathcal{RVE}$ and $L$ denotes the $\mathcal{RVE}$ dimension (for simplicity it is assumed to have a cubic shape). By generalizing the average strain theorem, it is straightforward to verify from  Eq. \eqref{eq:VolumeAveragingQuantities} that the above definitions lead to $\left[\overline{\mathbf{P}}_{m}\right]_{i}=\delta_{im}$ as introduced before. Thereof, the first applied boundary condition yields to ${\overline{\mathbf{P}}_{1}=\left[1,0,0,0,0,0,0,0,0,0,0,0,0\right]}$ and the remaining set of boundary conditions are introduced in the same way. Hence, the macroscopic modulus of the problem is determined by the averaged stresses, electric displacements and magnetic flux densities, obtained from the numerical solution of a series of $m=1,...,12$ boundary value problems
\begin{equation}
    \begin{aligned}
        \left[\overline{\mathbb{G}}\right]_{im}=\left[\mathrm{vec}\lbrace\overline{\boldsymbol{\sigma}},\overline{\mathbf{D}},\overline{\mathbf{B}}\rbrace\right]_{im},\quad\overline{\boldsymbol{\sigma}}=\langle\boldsymbol{\sigma}|_{\overline{\mathbf{P}}_{m}}\rangle,\quad\overline{\mathbf{D}}=\langle\mathbf{D}|_{\overline{\mathbf{P}}_{m}}\rangle,\quad\overline{\mathbf{B}}=\langle\mathbf{B}|_{\overline{\mathbf{P}}_{m}}\rangle.
    \end{aligned}
\end{equation}

\subsection{Computational Approach: The Virtual Element Method}
\begin{figure}[htp!]
\centering
    \begin{minipage}{0.31\linewidth}
    \centering
        \includegraphics[width=\textwidth]{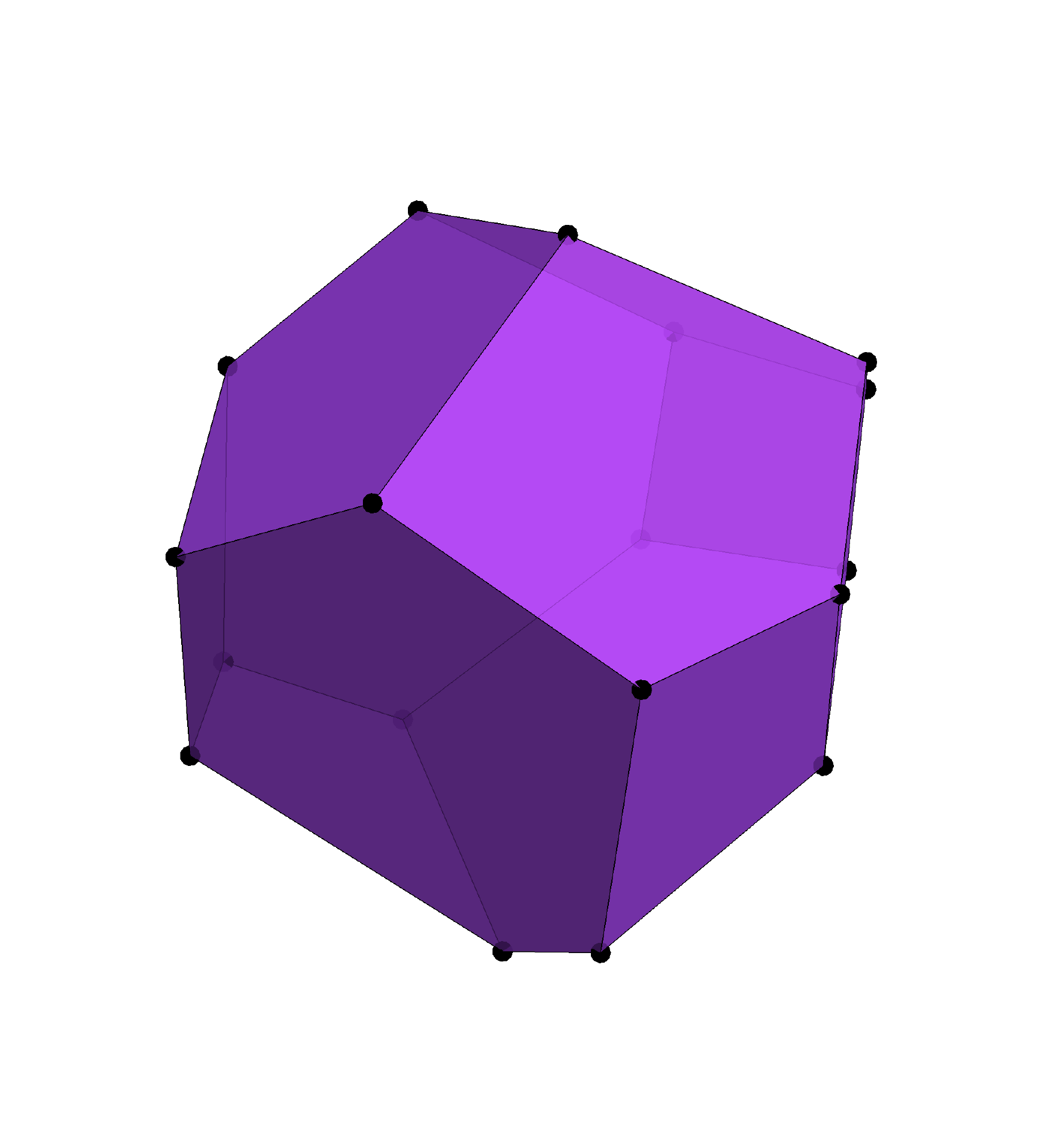}
        
        a)
    \end{minipage}
    \hfill
    \begin{minipage}{0.3\linewidth}
    \centering
        \includegraphics[width=\textwidth]{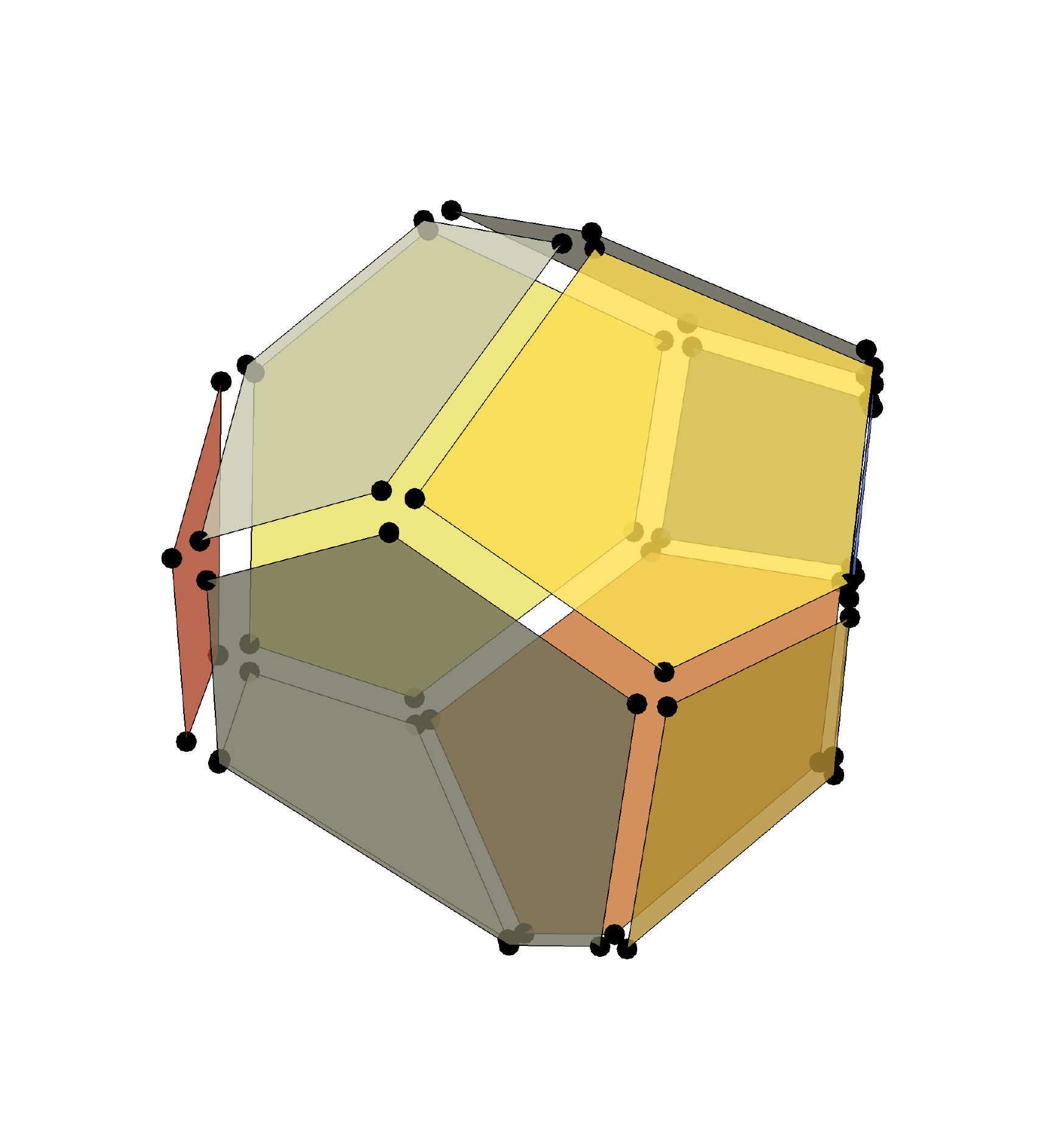}
        
        b)
    \end{minipage}
    \hfill
    \begin{minipage}{0.3\linewidth}
    \centering
        \includegraphics[width=\textwidth]{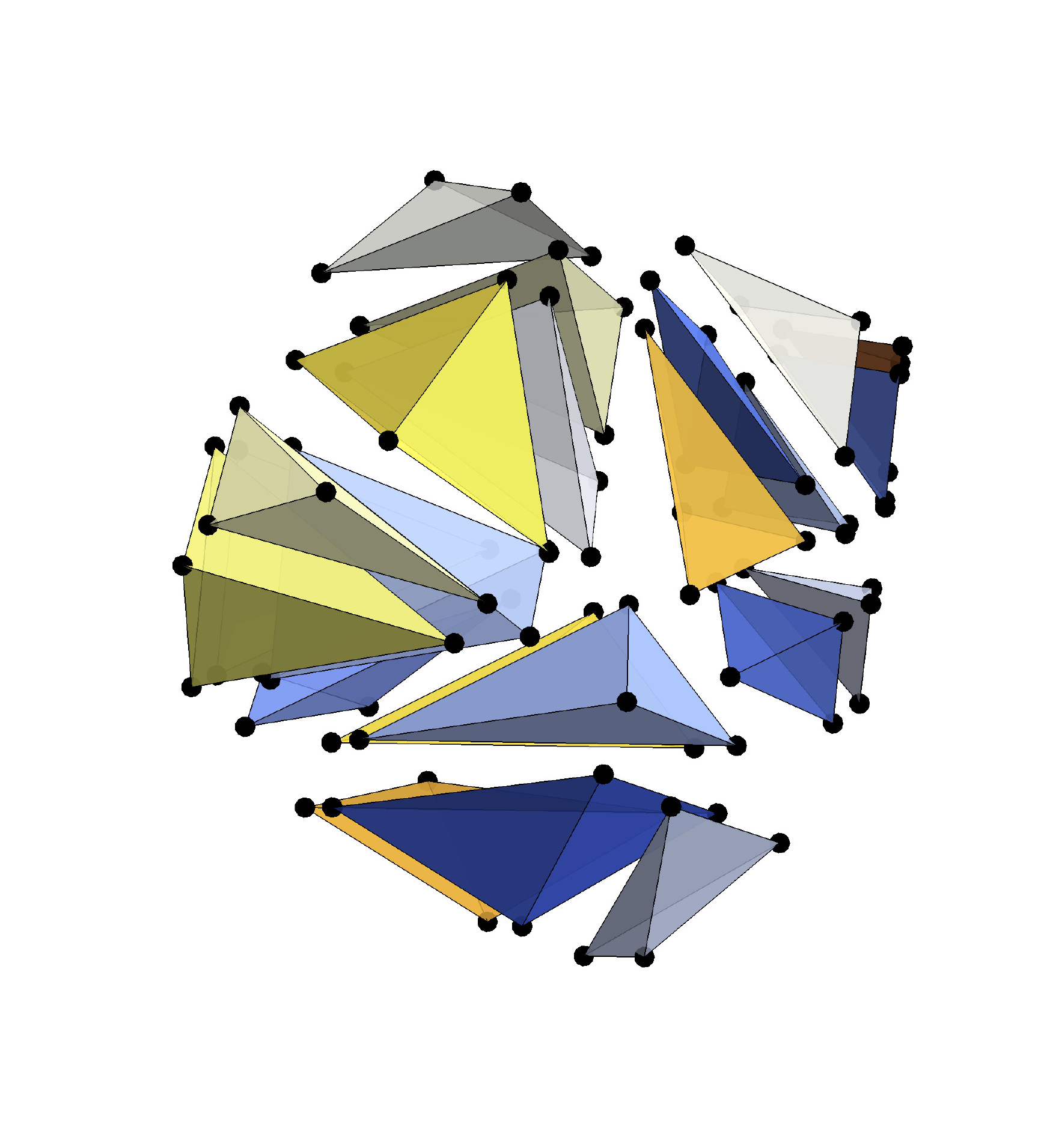}
        
        c)
    \end{minipage}
    \caption{Illustration of a) an artificial generated grain as a virtual element $\mathcal{G}\in\mathcal{B}_{\mathrm{m}}$, by Voronoi tesselation; b) its decomposition into faces $\mathcal{F}\in\partial\mathcal{G}$; c) its decomposition into a submesh with a minimum number of tetrahedrons $\mathfrak{T}\in\mathcal{G}$ by triangulation}
    \label{fig:GrainDecomposition}
\end{figure}

\noindent The virtual element method (VEM) has been introduced and highly developed in the last decade. From its general formulation in inter alia \cite{beirao2016virtual,beirao2013basic,beirao2014hitchhiker,da2014virtual} towards to recent works on homogenization \cite{mar19}, a framework regarding contact mechanics \cite{wriggers2016virtual} as well as elastodynamics \cite{cihan2020virtual} show its variety to multiple engineering fields as well as its performance and efficiency within. The advantage of VEM regarding the homogenization of polycrystalline microstructures is clearly the ability to match a grain of arbitrary convex/non-convex geometry perfectly \cite{mar19}.\\
Here, the particular grain structure at $\mathcal{RVE}$ is generated by a Voronoi tessellation, see Figure \ref{fig:GrainDecomposition}a). Hence, $\mathcal{B}_{\mathrm{m}}$ is discretized by non-overlapping elements of polyhedral type $\mathcal{G}\subset\mathbb{R}^{\mathrm{3}}$. The surface of a single polyhedron with volume $V_{\mathcal{G}}$ is composed of planar faces $\mathcal{F}\in\partial\mathcal{G}\subset\mathbb{R}^{\mathrm{2}}$, as illustrated in Figure \ref{fig:GrainDecomposition}b). These faces are characterized by linear edges $\mathfrak{e}\in\partial\mathcal{F}\subset\mathbb{R}^{\mathrm{1}}$ defined by two points, that represent the vertices $v$ of $\mathcal{G}_{i}\subset\mathcal{B}_{\mathrm{m}}$. Each virtual element, possibly non-convex, represents a single grain in the $\mathcal{RVE}$.
\FloatBarrier
\noindent A low-order virtual element formulation is here employed. As illustrated in \cite{mar19}, the element-wise defined virtual space $\mathcal{V}_{h}$ of primary fields $\mathcal{R}$ and their variations is then provided by
\begin{equation}\label{eq:VirtualSpaceG}
    \begin{aligned}
        \mathcal{V}_{h}\rvert_{\mathcal{G}}=\lbrace\mathcal{R}_{h}\in\left[H^{\mathrm{1}}\left(\mathcal{G}\right)\right]^{\mathrm{3}}:\bigtriangleup\mathcal{R}_{h}=0\in\mathcal{G}\land\mathcal{R}_{h}\rvert_{\mathcal{F}}\in\mathcal{V}_{\tau}\rvert_{\mathcal{F}},\forall\mathcal{F}\in\mathcal{G}\rbrace,
    \end{aligned}
\end{equation}
where
\begin{equation}\label{eq:VirtualSpaceF}
    \begin{aligned}
        \mathcal{V}_{\tau}\rvert_{\mathcal{F}}=\lbrace\mathcal{R}_{h}\in\left[H^{\mathrm{1}}\left(\mathcal{F}\right)\bigcap C^{\mathrm{0}}\left(\mathcal{F}\right)\right]^{\mathrm{3}}:\bigtriangleup_{\tau}\mathcal{R}_{h}=0\in\mathcal{F}\land\mathcal{R}_{h}\rvert_{\mathfrak{e}}\in\left[\mathcal{P}_{\mathrm{1}}\left(\mathfrak{e}\right)\right]^{\mathrm{3}},\forall\mathfrak{e}\in\partial\mathcal{F}\rbrace
    \end{aligned}
\end{equation}
defines the virtual space on each face $\mathcal{F}\in\partial\mathcal{G}$, with $\bigtriangleup,\bigtriangleup_{\tau}$ being the Laplacian operators in the global and local face (index $\tau$) variables. $\mathcal{V}_{\tau}\rvert_{\mathcal{F}}$ is the space of continuous harmonic functions, being piece-wise linear on $\partial\mathcal{F}$. $\mathcal{V}_{h}\rvert_{\mathcal{G}}$ is then the space of continuous harmonic functions at $\mathcal{G}$. At this point we note from \cite{beirao2014hitchhiker}, that with $\mathcal{V}_{h}\rvert_{\mathcal{G}}$ from Eq. \eqref{eq:VirtualSpaceG} a suitable set of degree of freedom can be introduced by means of point-wise values at the vertices of a polyhedron $v\in\mathcal{G}$.\\
The solution of the boundary value problem at hand can be achieved by minimization of a potential $U\rightarrow\text{min}$, where
\begin{equation}\label{eq:PseudoPotential}
    U:=\sum_{g}\int_{\mathcal{G}_{g}}\psi_{g}\mathrm{d}\mathcal{B}_{m}-\Pi_{ext}(\mathbf{u}),
\end{equation}
is obtained form the energy density function of each grain and the potential of external loads $\Pi_{ext}$. Therefore, the construction of this potential in a virtual element framework relies on a split of the primary fields $\mathcal{R}_{h}$ into a projection part and a remainder \cite{cihan2020virtual}
\begin{equation}\label{eq:SplitFields}
    \begin{aligned}
       \mathcal{R}_{h}=\Pi\mathcal{R}_{h}+\left(\mathcal{R}_{h}-\Pi\mathcal{R}_{h}\right),
    \end{aligned}
\end{equation}
with
\begin{equation}\label{eq:ProjectionOperator}
    \Pi:\mathcal{V}_{h}\mapsto\mathcal{P}_{1}
\end{equation}
being the projection operator on polynomial space of order one, defined such that:
\begin{equation}\label{eq:ProjectorOperatorDefinition}
    \begin{aligned}
    \int_{\mathcal{G}}\triangledown\Pi\mathcal{R}_{h}\mathrm{d}\mathcal{B}_{\mathrm{m}} \overset{!}{=} \int_{\mathcal{G}}\triangledown\mathcal{R}_{h}\mathrm{d}\mathcal{B}_{\mathrm{m}},\qquad \frac{1}{n_{v}}\sum_{k=1}^{n_{v}}\Pi\mathcal{R}_{h}\left(\mathbf{x}_{\mathrm{m}_{k}}\right)\overset{!}{=}\frac{1}{n_{v}}\sum_{k=1}^{n_{v}}\mathcal{R}_{h}\left(\mathbf{x}_{\mathrm{m}_{k}}\right).
    \end{aligned}
\end{equation}
Here $\lbrace\mathbf{x}_{\mathrm{m}}\in\mathbb{R}^{\mathrm{3}}:\mathbf{x}_{\mathrm{m}}\in\mathcal{B}_{\mathrm{m}}\subset\mathcal{B}_{\mathrm{M}}\rbrace$ denote the set of particular spatial coordinates at $v\in\mathcal{G}$ and $n_{v}$ its total number per polyhedron. In a linear formulation, the ansatz for $\Pi\mathcal{R}_{h}$ at each element is:
\begin{equation}\label{eq:linVEMansatz}
    \begin{aligned}
        \Pi\mathcal{R}_{h}=\sum_{i=1}^{3}\left(\mathbf{N}_{\Pi}\cdot\mathbf{a}_{i}\right)\mathbf{e}_{i},
    \end{aligned}
\end{equation}
with $\mathbf{N}_{\Pi}=\lbrace1,x,y,z\rbrace$ collecting constant $1$ and the coordinate variables with respect to $\mathbf{e}_{1}$, $\mathbf{e}_{2}$ and $\mathbf{e}_{3}$. Moreover, $\mathbf{a}_{i}=\lbrace a_{i1},...,a_{i6}\rbrace$ are the coefficients defining $\Pi\mathcal{R}_{h}$ in each element, i.e. $\Pi\mathcal{R}_{h}\rvert_{\mathcal{G}}$. In particular, the constant parts $a_{i\mathrm{1}}$ vanish through application of the gradient operator, and $\triangledown\Pi\mathcal{R}_{h}\rvert_{\mathcal{G}}=\lbrace\triangledown\Pi\mathbf{u}_{h}\rvert_{\mathcal{G}},\triangledown\Pi\phi_{h}\rvert_{\mathcal{G}},\triangledown\Pi\varphi_{h}\rvert_{\mathcal{G}}\rbrace$ is fully characterized by:
\begin{equation}\label{eq:GradientsOfProjectedParts}
    \begin{aligned}
    \triangledown\Pi\mathbf{u}_{h}\rvert_{\mathcal{G}}=\begin{bmatrix}
    a_{12} & a_{13} & a_{14}\\
    a_{22} & a_{23} & a_{24}\\
    a_{32} & a_{33} & a_{34}
    \end{bmatrix},\qquad
    \triangledown\Pi\phi_{h}\rvert_{\mathcal{G}}=\begin{bmatrix}
    a_{15}\\
    a_{25}\\
    a_{35}
    \end{bmatrix},\qquad
    \triangledown\Pi\varphi_{h}\rvert_{\mathcal{G}}=\begin{bmatrix}
    a_{16}\\
    a_{26}\\
    a_{36}
    \end{bmatrix}
    \end{aligned}.
\end{equation}
Due to the linear character of the ansatz, we obtain  $\triangledown\Pi\mathcal{R}_{h}\rvert_{\mathcal{G}}=const.\in\mathcal{G}$. Integration by parts of the right-hand side of the integral in Eq. \eqref{eq:ProjectorOperatorDefinition}, yields
\begin{equation}\label{eq:GradientsOfProjectedParts2}
    \begin{aligned}
    \triangledown\Pi\mathbf{u}_{h}\rvert_{\mathcal{G}}=\frac{1}{V_{\mathcal{G}}}\sum_{\mathcal{F\in\partial\mathcal{G}}}\int_{\mathcal{F}}\mathbf{u}_{h}\otimes\mathbf{n}_{\mathcal{F}}\mathrm{dA},\quad \lbrace\triangledown\Pi\phi_{h}\rvert_{\mathcal{G}}, \triangledown\Pi\varphi_{h}\rvert_{\mathcal{G}}\rbrace=\frac{1}{V_{\mathcal{G}}}\sum_{\mathcal{F\in\partial\mathcal{G}}}\int_{\mathcal{F}}\lbrace\phi_{h},\varphi_{h}\rbrace\mathbf{n}_{\mathcal{F}}\mathrm{dA},
    \end{aligned}
\end{equation}
with $\mathbf{n}_{\mathcal{F}}$ being the unit normal vector of $\mathcal{F}\in\partial\mathcal{G}$. Since integrals in Eq. \eqref{eq:GradientsOfProjectedParts2} can be computed from the values of primary variables at element vertices, the coefficients in Eq. \eqref{eq:GradientsOfProjectedParts} are univocally determined as function of nodal degrees of freedom. Since the primary fields are decomposed, so is the potential
\begin{equation}\label{eq:SplitPotential}
    U\left(\triangledown\mathcal{R}_{h}\right)=\underset{\mathcal{G}}{\textbf{\huge{\textsf{A}}}}\left[ \underbrace{U_{c}\left(\triangledown\Pi\mathcal{R}_{h}\rvert_{\mathcal{G}}\right)}_{\displaystyle\text{consistency}}-\underbrace{U_{s}\left(\triangledown\mathcal{R}_{h}\rvert_{\mathcal{G}}-\triangledown\Pi\mathcal{R}_{h}\rvert_{\mathcal{G}}\right)}_{\displaystyle\text{stabilization}}\right].
\end{equation}
While the consistency term is computable from Eq. \eqref{eq:linVEMansatz}, the stabilization term depends not only on the known projection term $\Pi\mathcal{R}_{h}$, but also on the total field $\mathcal{R}_{h}$, for which shape functions are not introduced. The construction of the stabilization term follows from \cite{wriggers2017efficient} by partitioning the potential $U\left(\mathcal{R}_{h}\right)$ into a fraction $\beta\in\left[0,1\right]$, based on the projected fields and another one based on the total field
\begin{equation}
    \begin{aligned}
         U\left(\triangledown\mathcal{R}_{h}\right)=\underset{\mathcal{G}}{\textbf{\huge{\textsf{A}}}}\left[\left(1-\beta\right) U\left(\triangledown\Pi\mathcal{R}_{h}\rvert_{\mathcal{G}}\right)+\beta U\left(\triangledown\mathcal{R}_{h}\rvert_{\mathcal{G}}\right)\right].
    \end{aligned}
\end{equation}
The stabilization part is computed by introducing for $\mathcal{R}_{h}$ a FE ansatz, based on a decomposition of $\mathcal{G}$ into a submesh of regular tetrahedrons by means of a triangulation with minimum number of internal elements, see Figure \ref{fig:GrainDecomposition}c). Here, also a linear ansatz is employed to obtain $\triangledown\mathcal{R}$. The element residual $\mathbf{R}_{\mathcal{G}}$ as well as the element stiffness tangent $\mathbf{K}_{\mathcal{G}}$ are then found by
\begin{equation}\label{eq:SKR}
    \begin{aligned}
        \mathbf{R}_{\mathcal{G}}=\left(1-\beta\right)\frac{\partial U\left(\triangledown\Pi\mathcal{R}_{h}\rvert_{\mathcal{G}}\right)}{\partial\mathbf{p}_{\mathcal{G}}}+\beta\frac{\partial U\left(\triangledown\mathcal{R}_{h}\rvert_{\mathcal{G}}\right)}{\partial\mathbf{p}_{\mathcal{G}}},\qquad \mathbf{K}_{\mathcal{G}}=\frac{\partial\mathbf{R}_{\mathcal{G}}}{\partial\mathbf{p}_{\mathcal{G}}},
    \end{aligned}
\end{equation}
with $\mathbf{p}_{\mathcal{G}}=\mathrm{vec}\left(\mathcal{R}_{h}\right)$ being the vector of nodal unknowns of the element $\mathcal{G}$. The computation is performed by using the software package AceGen/AceFEM as a symbolic mathematic software tool in Mathematica for deriving residual and tangent matrix through automatic differentiation tools \cite{korelc2016automation}.\\
It is worth noting, that, by following a low order virtual element formulation, the obtained set of microscopic quantities $\lbrace\boldsymbol{\varepsilon},\mathbf{E},\mathbf{H},\boldsymbol{\sigma},\mathbf{D},\mathbf{B}\rbrace=const.$ in each element, and thus grain, $\mathcal{G}$. Accordingly, the volume integral formulations in Eq. \eqref{eq:VolumeAveragingQuantities} simplify in general to
\begin{equation}\label{eq:VolAverageVEM}
    \begin{aligned}
        \int_{\mathcal{G}}\left(\cdot\right)\mathrm{d}\mathcal{B}_{\mathrm{m}}=V_{\mathcal{G}}\left(\cdot\right).
    \end{aligned}
\end{equation}

\section{Results}
\noindent A microstructure, consisting of 100 polyhedral grains, artificially generated, by a voronoi tessellation \cite{quey2011large}, is considered. To compare FE- and VE-approaches, a parametric study is conducted. 
\begin{figure}[htp!]
\centering
    \begin{minipage}{0.26\linewidth}
    \centering
        \includegraphics[width=\textwidth]{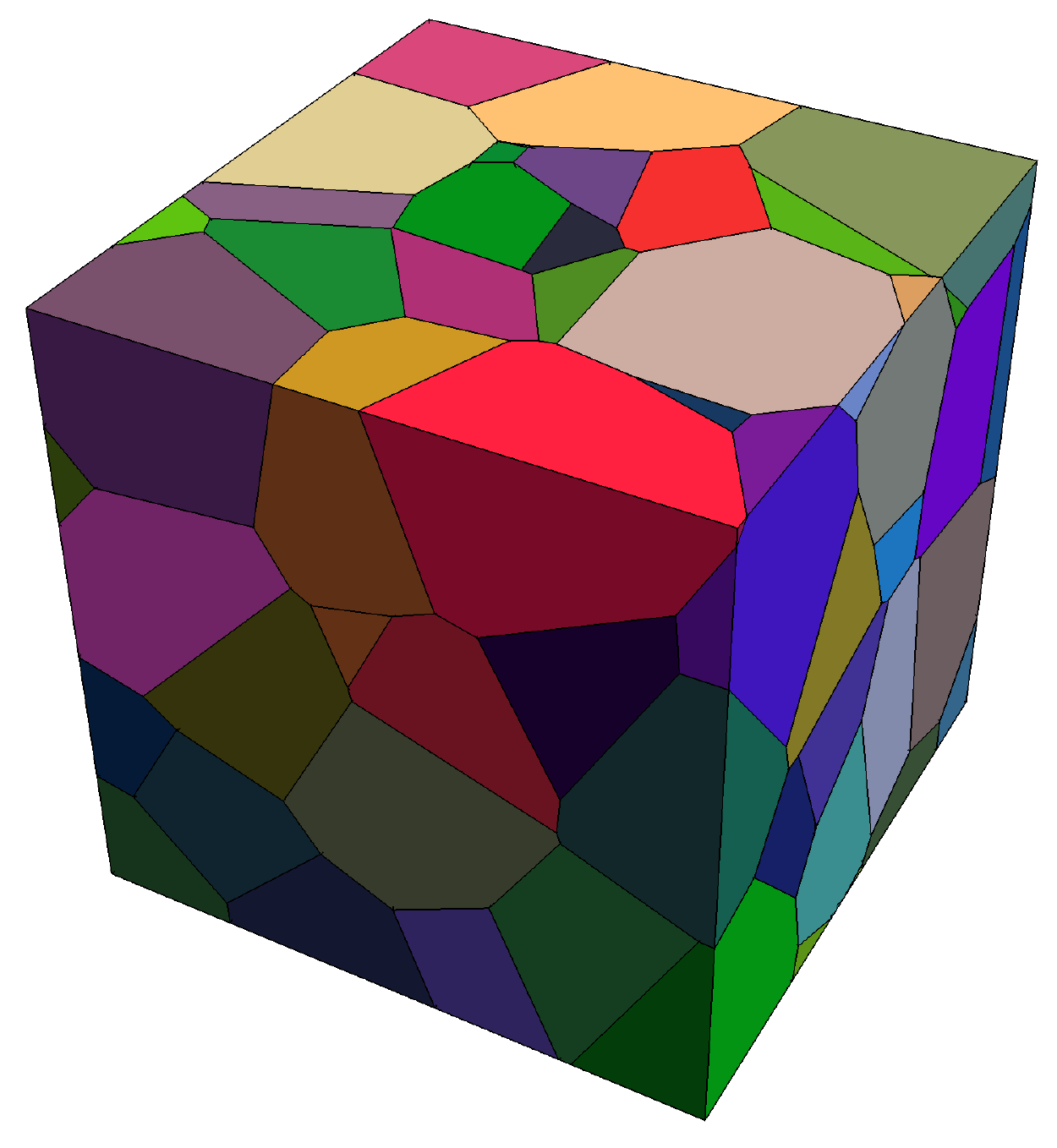}
        
       a)
    \end{minipage}
    \hfill
      \begin{minipage}{0.26\linewidth}
    \centering
        \includegraphics[width=\textwidth]{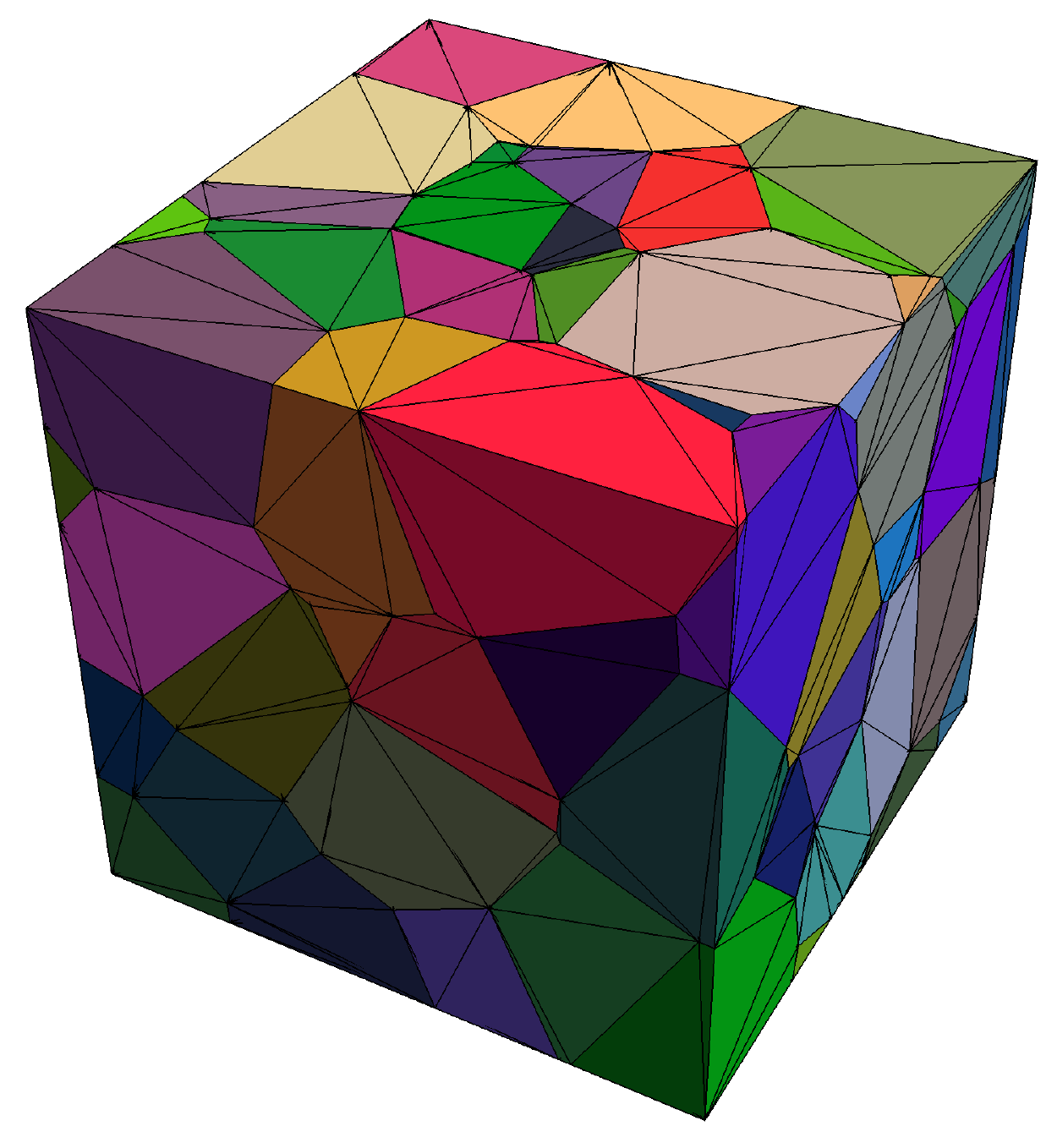}
        
        b)
    \end{minipage}
    \hfill
    \begin{minipage}{0.26\linewidth}
    \centering
        \includegraphics[width=\textwidth]{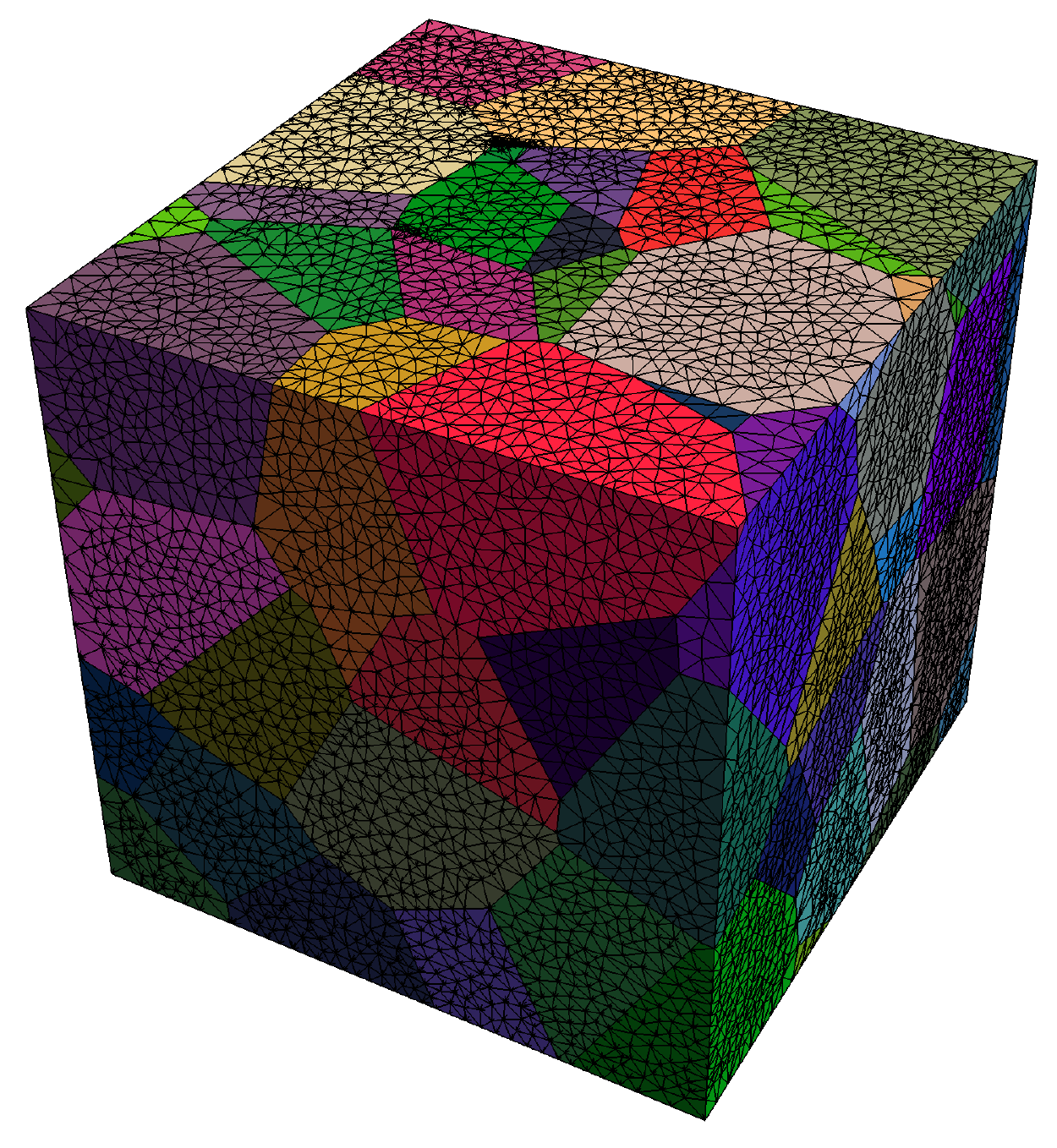}
        
        c)
    \end{minipage}
\caption{Illustration of an artificially generated $\mathcal{RVE}$ (100 grains), discretized by a) VEM-VO mesh, vertices of polyhedral elements are the nodes of the virtual elements (one polyhedral element equals one virtual element), b) coarse FEM-O1 mesh with same number of nodes as VEM-VO mesh (tetrahedralized polyhedral elements); c) fine FEM-O1 mesh (>350 k tetrahedral elements)}
    \label{fig:MeshedRVEs}
\end{figure}
\FloatBarrier
\noindent In the virtual element approach, each polyhedral grain is discretized by one virtual element (VEM-VO), see Figure \ref{fig:MeshedRVEs}a). By utilizing a tetrahedralization, the corresponding coarse FEM-O1 mesh with same number of nodes is designed. This coarse FE-mesh represents the submesh with which the stabilization term is computed. Benchmark results are obtained from a convergence study, where the FEM-O1 mesh is consequently refined up to >350 k elements, see Figure \ref{fig:MeshedRVEs}c). In addition to the coarse FEM-O1 mesh, illustrated in Figure \ref{fig:MeshedRVEs}b), where linear shape functions applied, also a mesh with the same number of elements, but interpolated by quadratic shape functions, is employed within the parametric study (coarse FEM-O2).

\subsection{Electro-Mechanically Coupled Problems}
\noindent First, the study is restricted to pure electro-mechanically coupled problems. Thus, the macroscopic effective modulus $\overline{\mathbb{G}}$ from Eq. \eqref{eq:macroscopicModulus} shrinks to
\begin{equation}\label{eq:electromechanicallyModulus}
    \begin{aligned}
        \overline{\mathbb{G}}=\begin{bmatrix}
        \overline{\mathbb{C}} & -\overline{\mathbf{e}}^{\mathrm{T}}\\
        -\overline{\mathbf{e}} & -\overline{\boldsymbol{\epsilon}}
        \end{bmatrix},\quad   \left[\overline{\mathbb{G}}\right]_{im}=\left[\mathrm{vec}\lbrace\overline{\boldsymbol{\sigma}},\overline{\mathbf{D}}\rbrace\right]_{im},\quad\overline{\boldsymbol{\sigma}}=\langle\boldsymbol{\sigma}|_{\overline{\mathbf{P}}_{m}}\rangle,\quad\overline{\mathbf{D}}=\langle\mathbf{D}|_{\overline{\mathbf{P}}_{m}}\rangle,
    \end{aligned}
\end{equation}
which requires 9 independent simulations with $\langle\mathbf{P}\rangle=\overline{\mathbf{P}}_{m}$ such that $\left[\overline{\mathbf{P}}_{m}\right]_{i}=\delta_{im}$ with $(i,m=1,...,9)$. Material properties stem from \cite{jain2013commentary,de2015database,de2015charting}, their corresponding IDs are listed in Table \ref{tab:MPErrorMats}.
\begin{table}[htp!]
    \centering
    \begin{tabular}{c|c|c|c|c}
        \textbf{Material} & \textbf{MP-ID} & \textbf{Lattice Structure} & $A^{\mathrm{U}}$ & \textbf{Point Group} \\\hline
       $\text{GaPO}_{\mathrm{4}}$  & mp-553932 & Orthorhombic & 1.31 & $222$\\
        $\text{AlPO}_{\mathrm{4}}$ & mp-4051 & Orthorhombic & 1.22 & $222$\\
        $\text{MoS}_{\mathrm{2}}$ & mp-1434 & Trigonal & 140.72 & $3m$\\
        $\text{BN}$ & mp-604884 & Hexagonal & 244.58 & $\bar{6}m2$ \\
        $\text{BaNiO}_{\mathrm{3}}$ & mp-19241 & Hexagonal & 14.49 & $6mm$\\\hline
    \end{tabular}
    \caption{Identification of materials used for studies on electro-mechanically coupled problems}
    \label{tab:MPErrorMats}
\end{table}
\FloatBarrier
\noindent For each material, $A^{\mathrm{U}}\geq 0$ (also given in Table \ref{tab:MPErrorMats}) denotes the index of elastic anisotropy, introduced by \cite{ranganathan2008universal} and is influenced by the differences of magnitude of entries related to main- and off-diagonal terms of $\mathbb{C}$, being directly proportional to the degree of anisotropy. Though the underlying lattice structure differs through applied materials, see Table \ref{tab:MPErrorMats}, the geometric properties of polycrystalline grain structure is hold constant, see Figure \ref{fig:MeshedRVEs}. Hence, the computations analyze the influence of different material response in an electro-mechanical framework related to the applied lattice structure, which affect the location of the non-zero entries in $\mathbb{G}_{l}$, see Appendix \ref{subsec:LatticeStructure}.

\subsubsection{Computational Error of macroscopic effective Properties}
\begin{figure}[htp!]
\centering
    \begin{minipage}{0.3\linewidth}
    \centering
        \includegraphics[width=\textwidth]{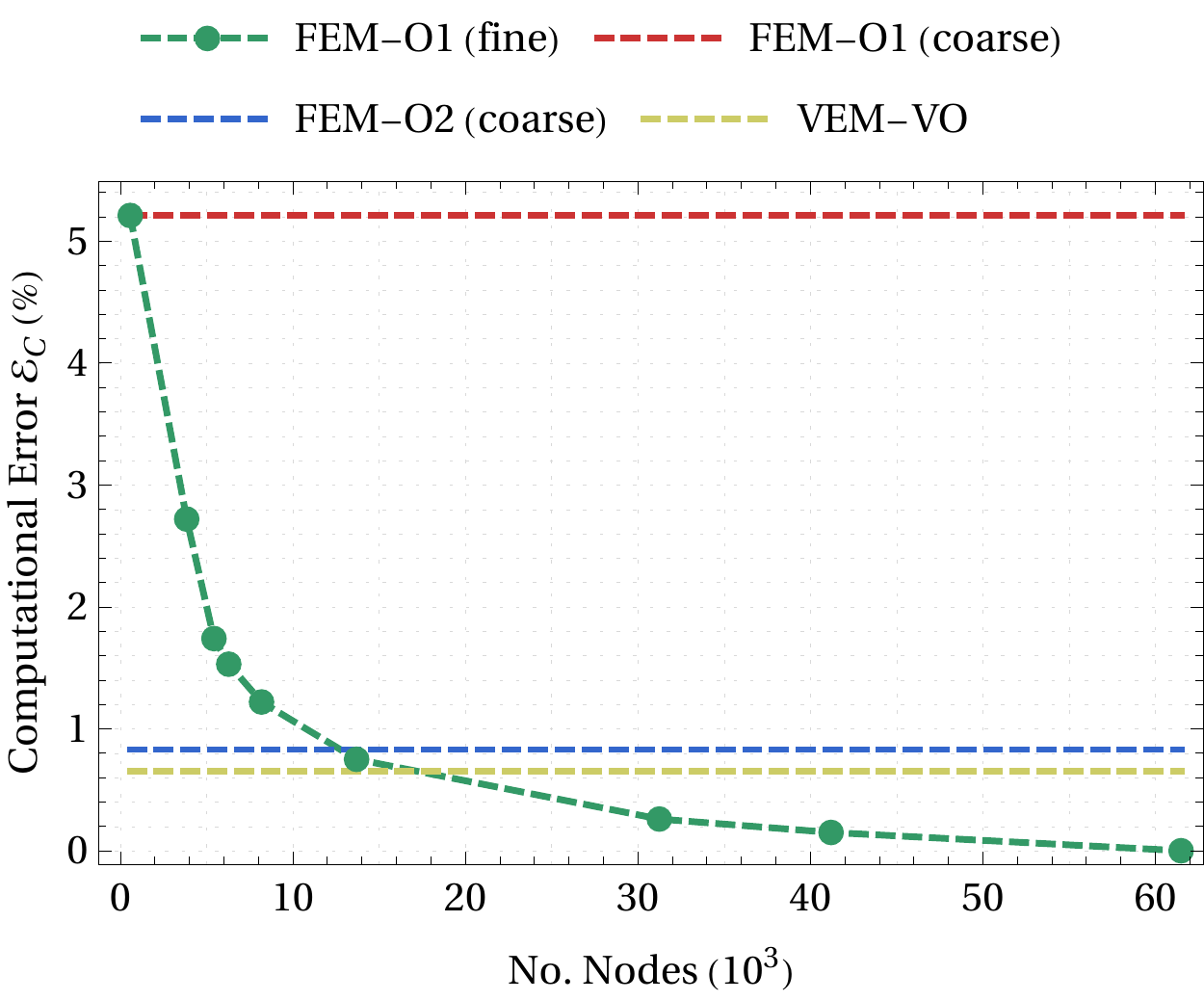}
        
        a)
    \end{minipage}
    \hfill
    \begin{minipage}{0.3\linewidth}
    \centering
        \includegraphics[width=\textwidth]{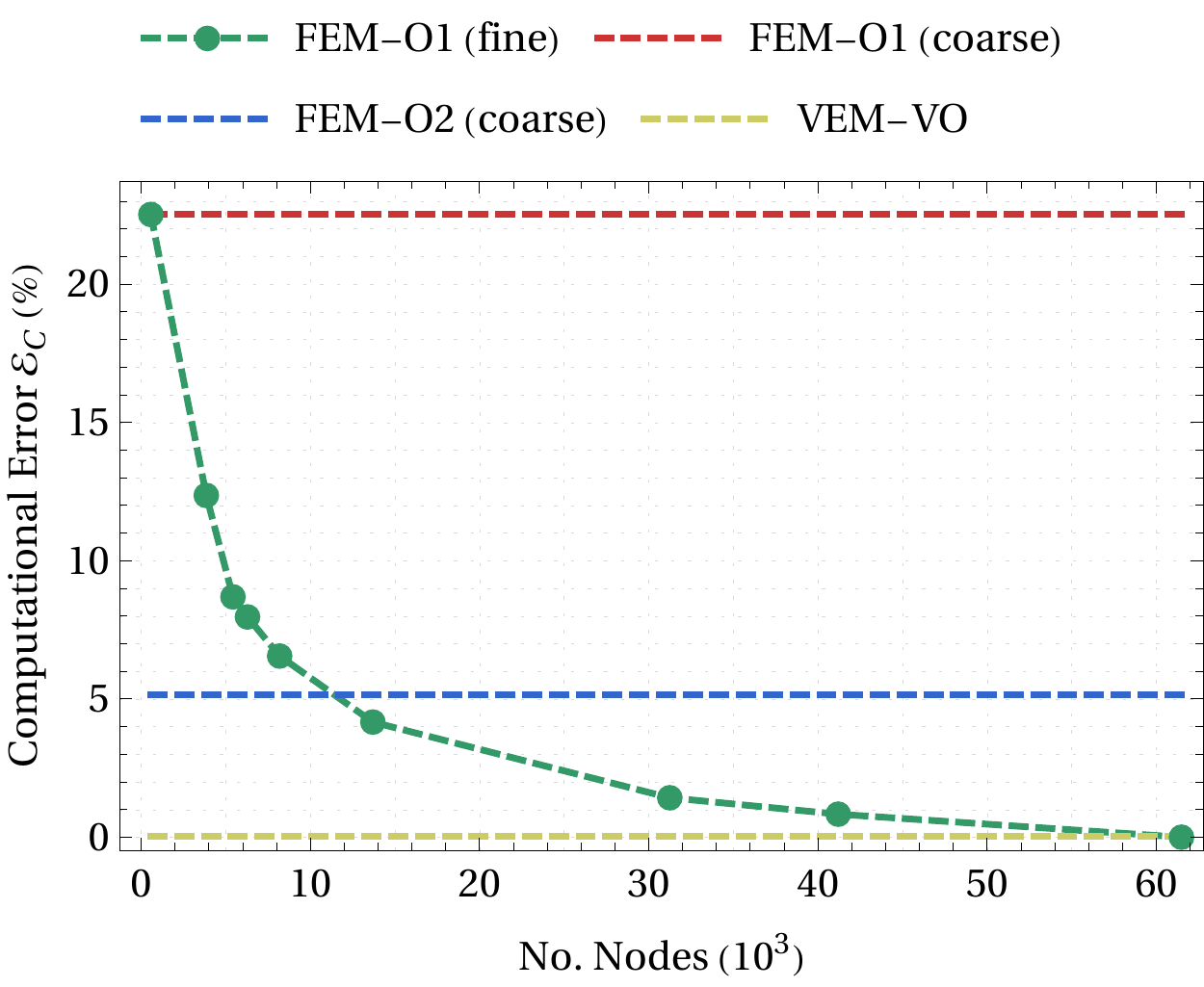}
        
        b)
    \end{minipage}
    \hfill
    \begin{minipage}{0.3\linewidth}
    \centering
        \includegraphics[width=\textwidth]{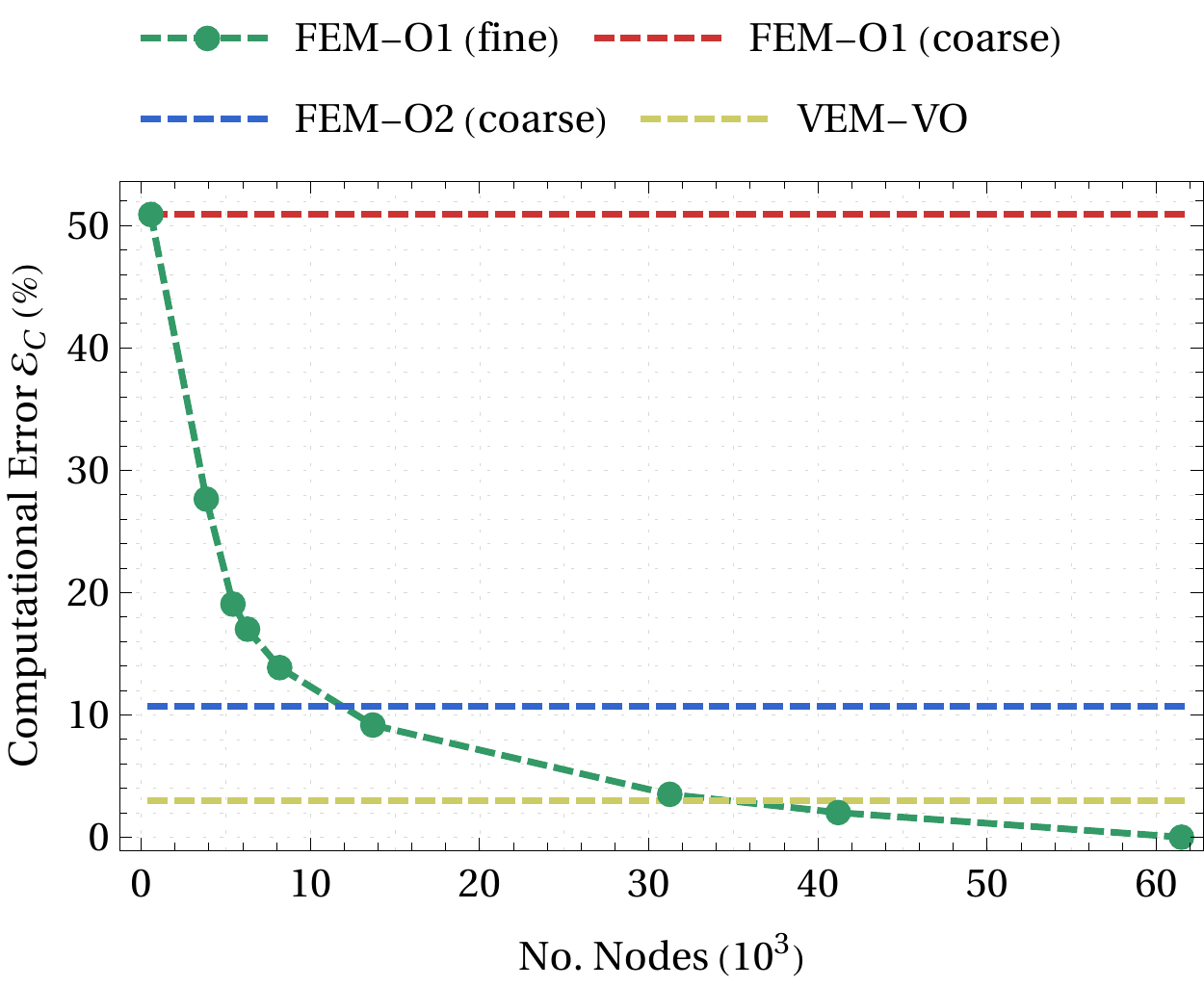}
        
        c)
    \end{minipage}
    \caption{Computational error $\mathcal{E}_{C}$ of $\overline{\mathbb{G}}$ of a) $\mathrm{AlPO_{4}}$, quasi-isotropic, orthorhombic unit cell; b) $\mathrm{BaNiO_{3}}$, mildly anisotropic, hexagonal unit cell; c) $\mathrm{MoS_{2}}$, highly anisotropic, trigonal unit cell}
    \label{fig:NormGPlots}
\end{figure}

\begin{figure}[htp!]
\centering
    \begin{minipage}{0.3\linewidth}
    \centering
        \includegraphics[width=\textwidth]{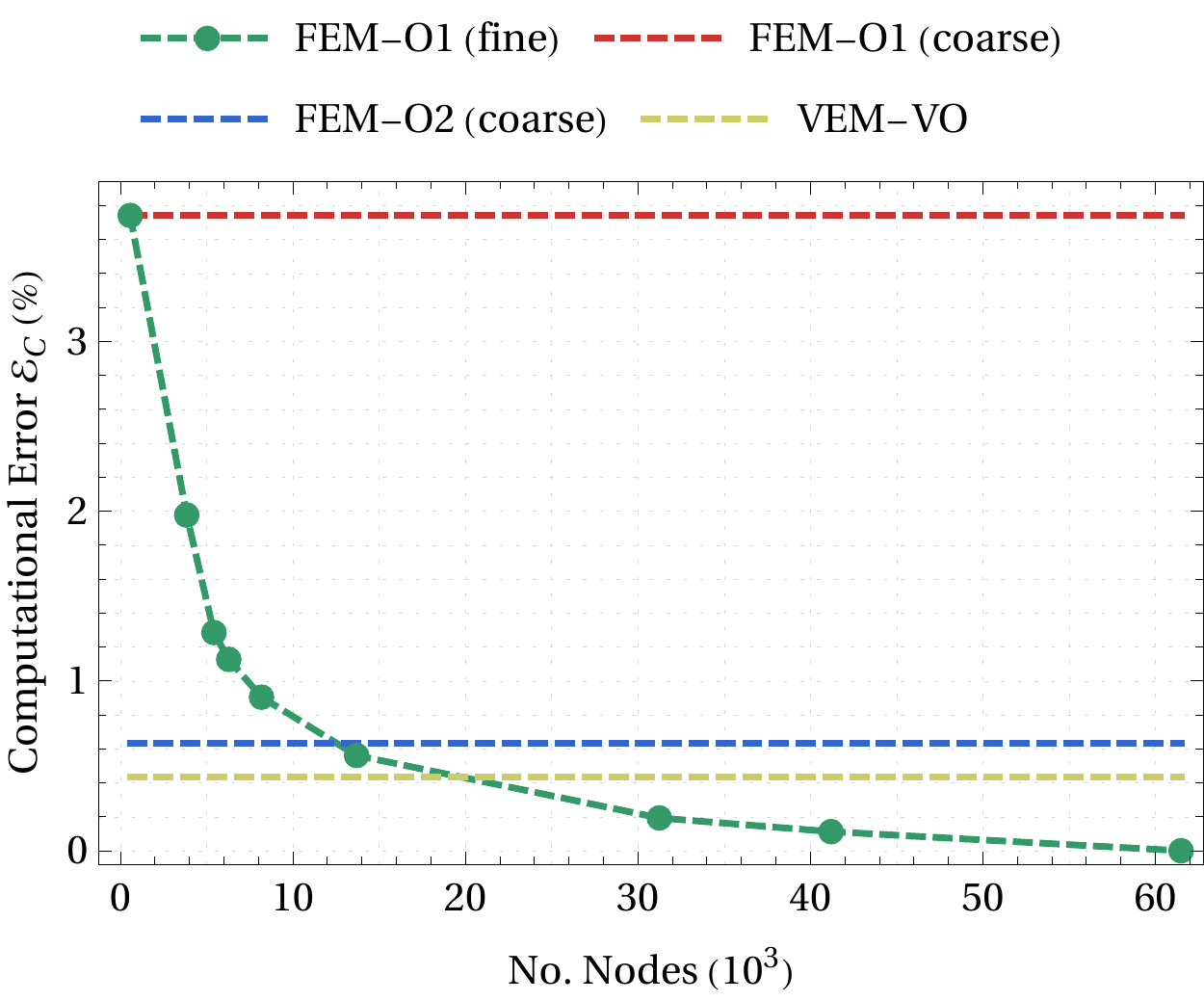}
        
        a)
    \end{minipage}
    \hfill
    \begin{minipage}{0.3\linewidth}
    \centering
        \includegraphics[width=\textwidth]{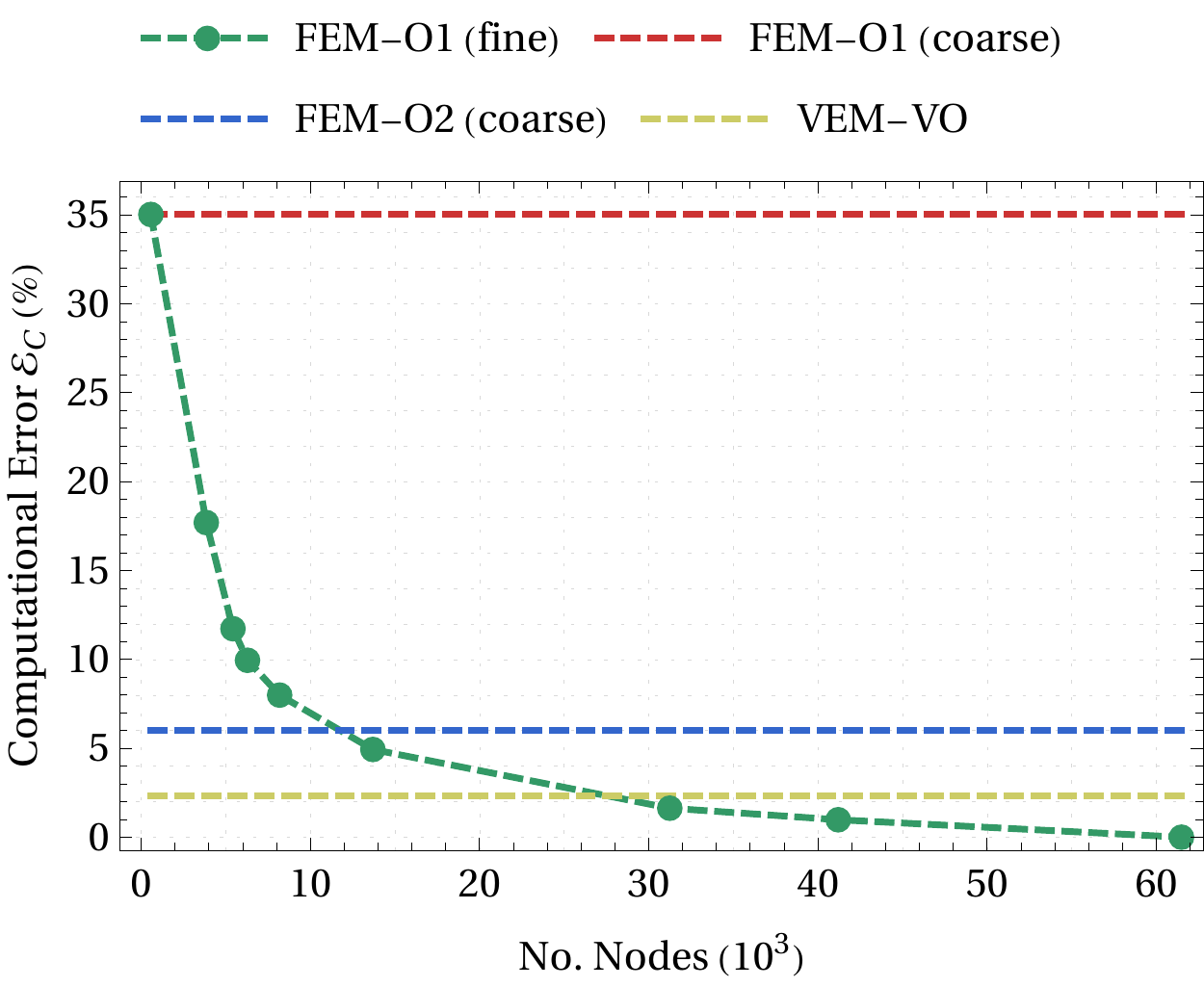}
        
        b)
    \end{minipage}
    \hfill
    \begin{minipage}{0.3\linewidth}
    \centering
        \includegraphics[width=\textwidth]{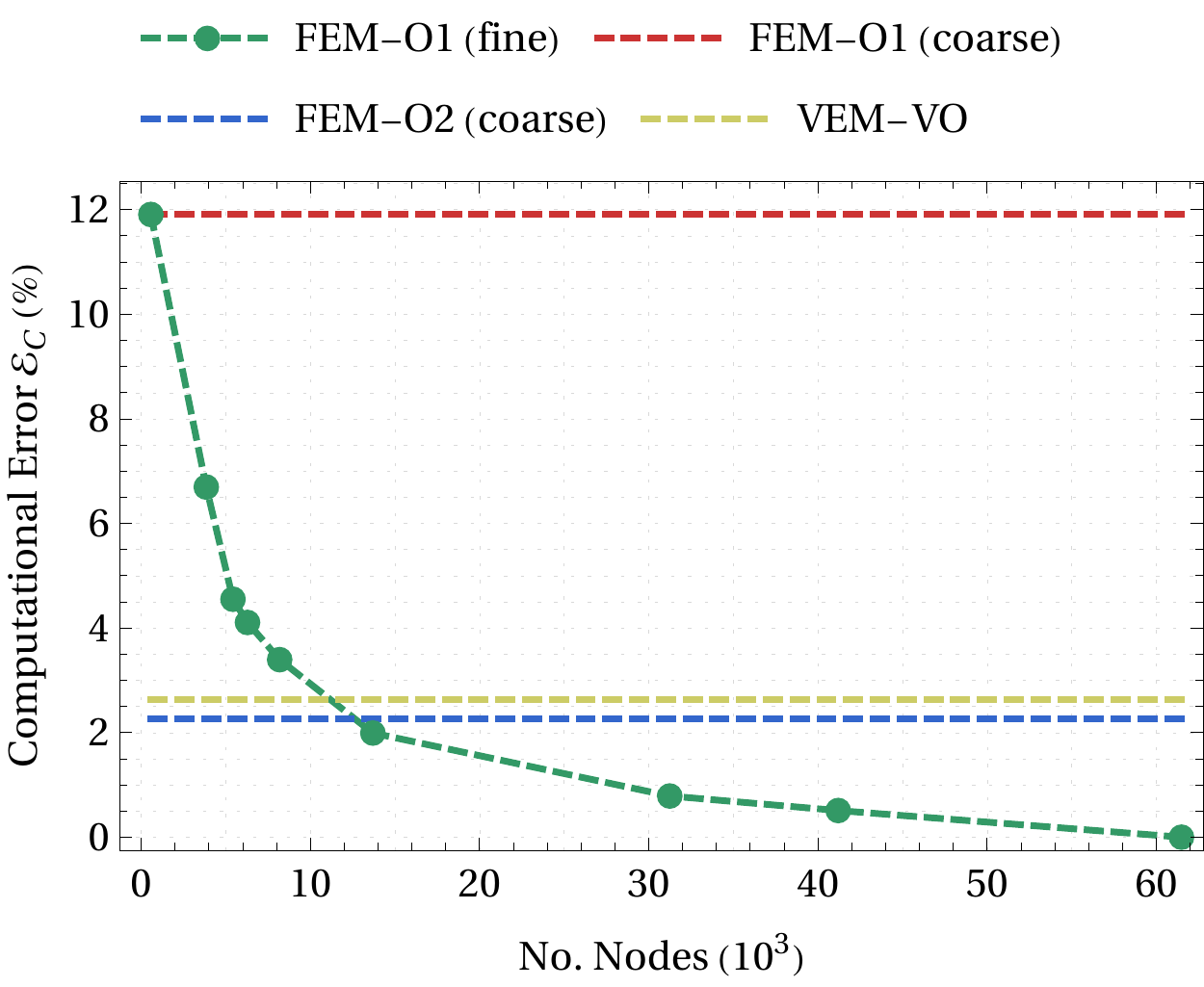}
        
        c)
    \end{minipage}
    \caption{Computational error $\mathcal{E}_{C}$ of $\mathrm{GaPO_{4}}$ with orthorhombic unit cell: 
    a) effective mechanical modulus $\overline{\mathbb{C}}$; b) effective electro-mechanical modulus $\overline{\mathbf{e}}$; c) effective dielectric modulus $\overline{\boldsymbol{\epsilon}}$}
    \label{fig:NormPlotsOrth3}
\end{figure}

\begin{figure}[htp!]
\centering
    \begin{minipage}{0.3\linewidth}
    \centering
        \includegraphics[width=\textwidth]{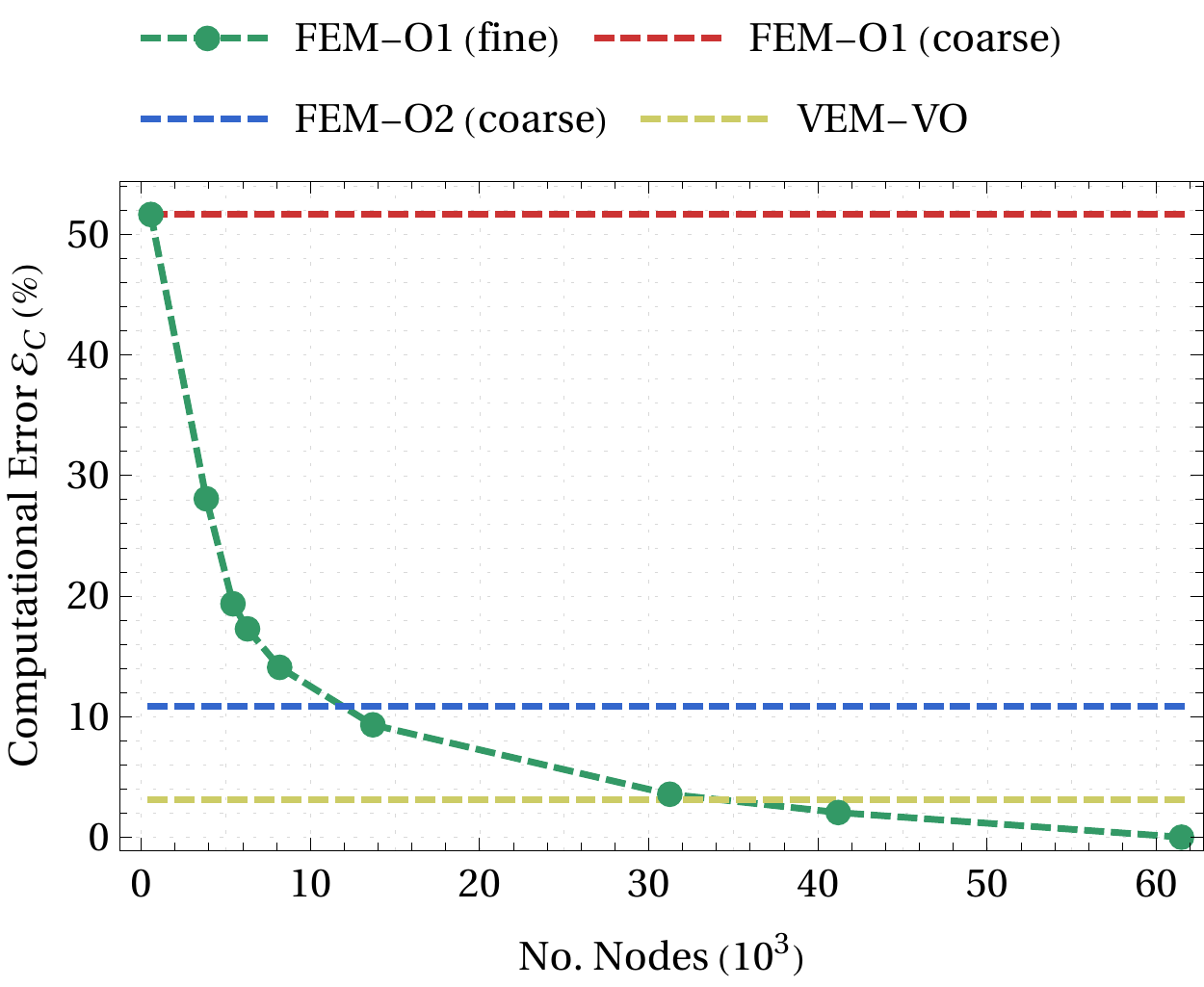}
        
        a)
    \end{minipage}
    \hfill
    \begin{minipage}{0.3\linewidth}
    \centering
        \includegraphics[width=\textwidth]{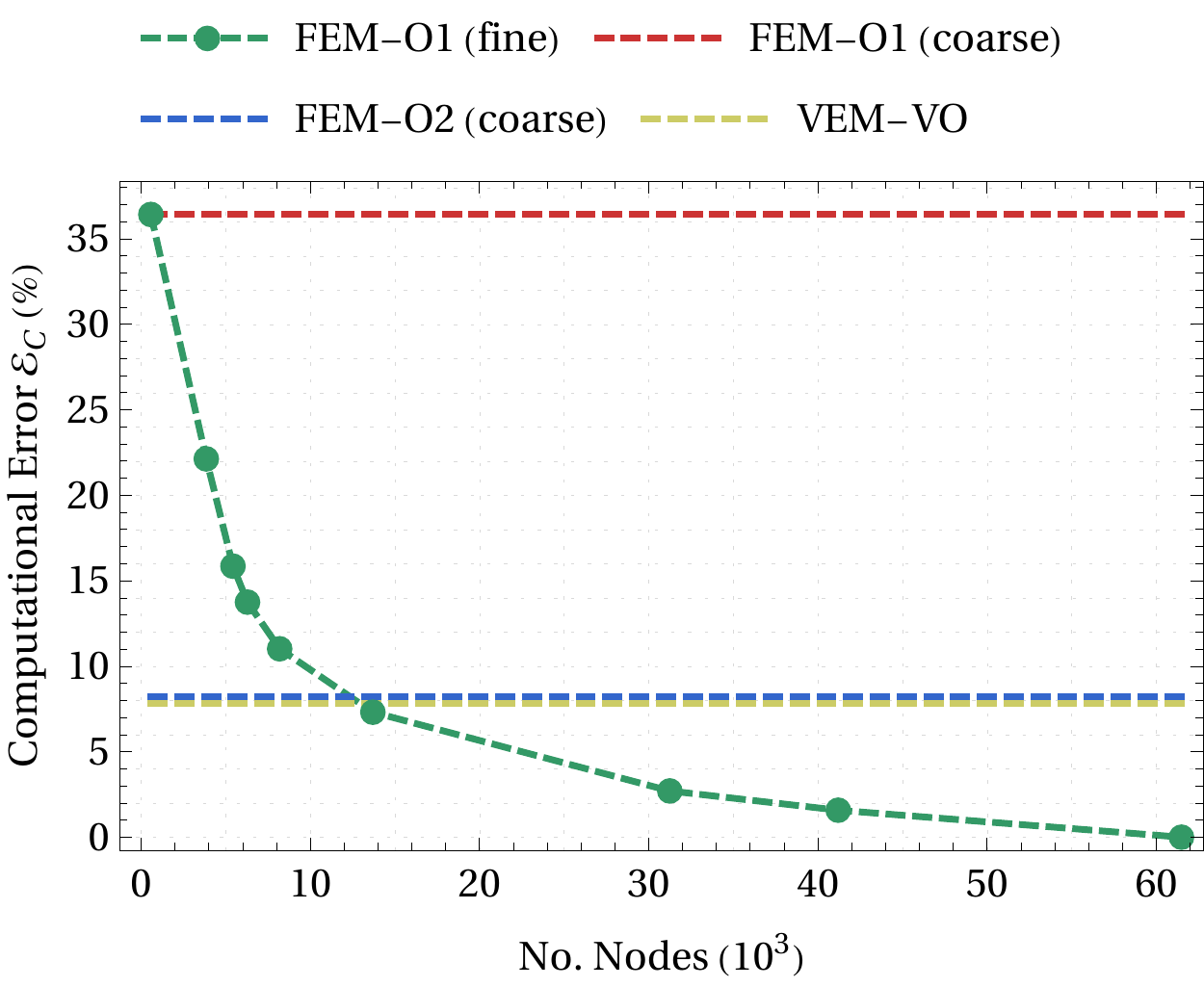}
        
        b)
    \end{minipage}
    \hfill
    \begin{minipage}{0.3\linewidth}
    \centering
        \includegraphics[width=\textwidth]{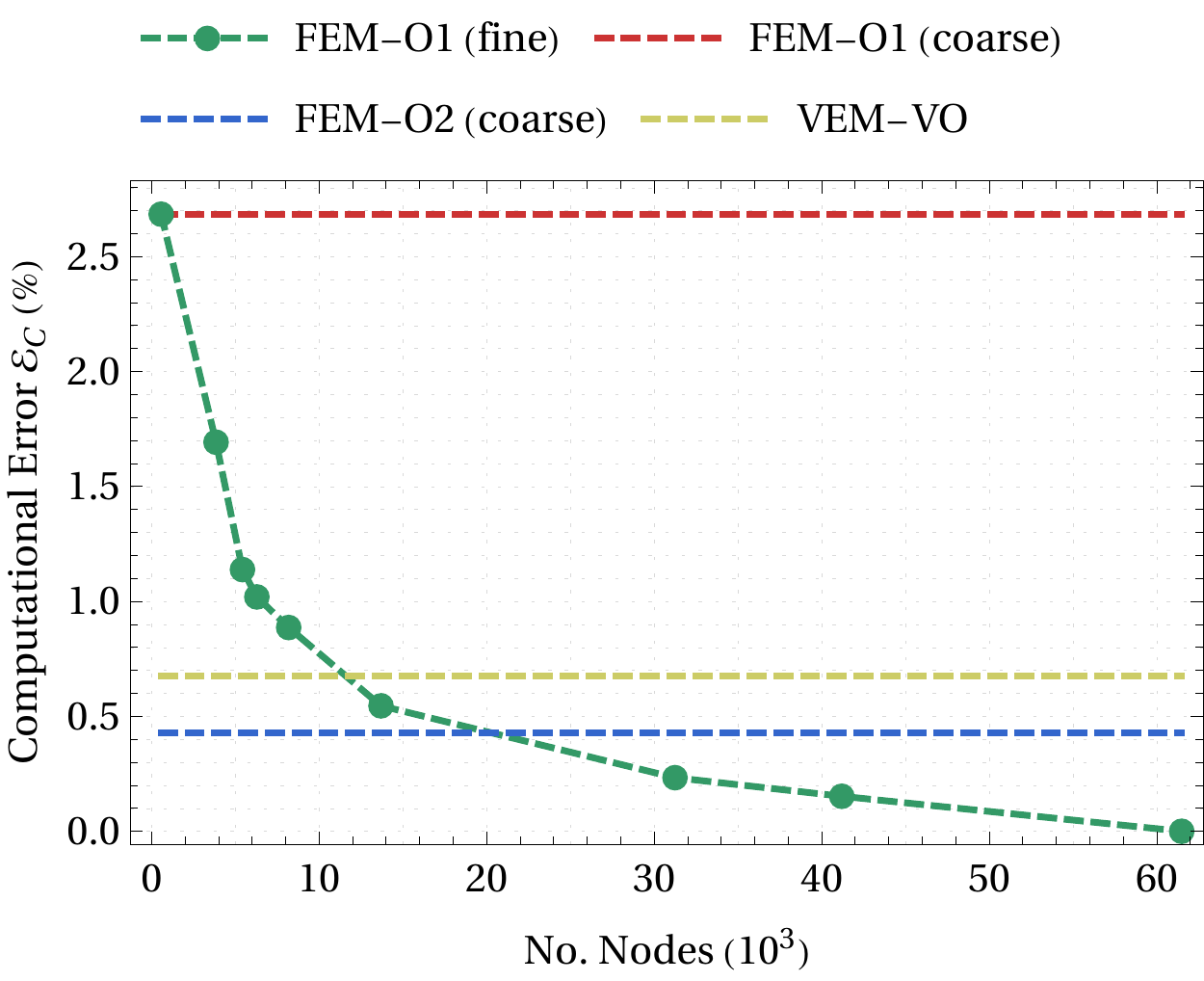}
        
        c)
    \end{minipage}
    \caption{Computational error $\mathcal{E}_{C}$ of $\mathrm{MoS_{2}}$ with trigonal unit cell: 
    a) effective mechanical modulus $\overline{\mathbb{C}}$; b) effective electro-mechanical modulus $\overline{\mathbf{e}}$; c) effective dielectric modulus $\overline{\boldsymbol{\epsilon}}$}
    \label{fig:NormPlotsTri3}
\end{figure}

\begin{figure}[htp!]
\centering
    \begin{minipage}{0.3\linewidth}
    \centering
        \includegraphics[width=\textwidth]{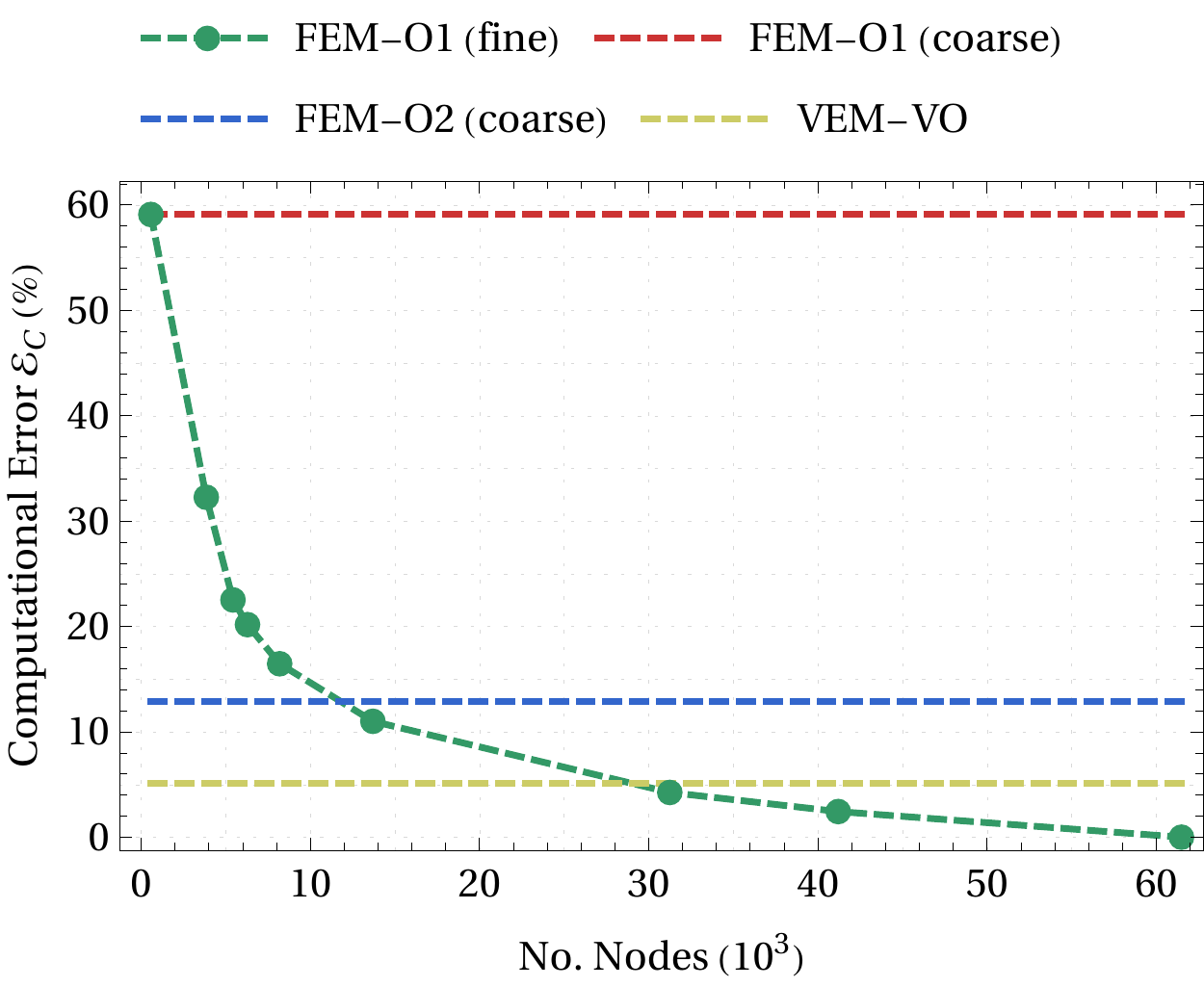}
        
        a)
    \end{minipage}
    \hfill
    \begin{minipage}{0.3\linewidth}
    \centering
        \includegraphics[width=\textwidth]{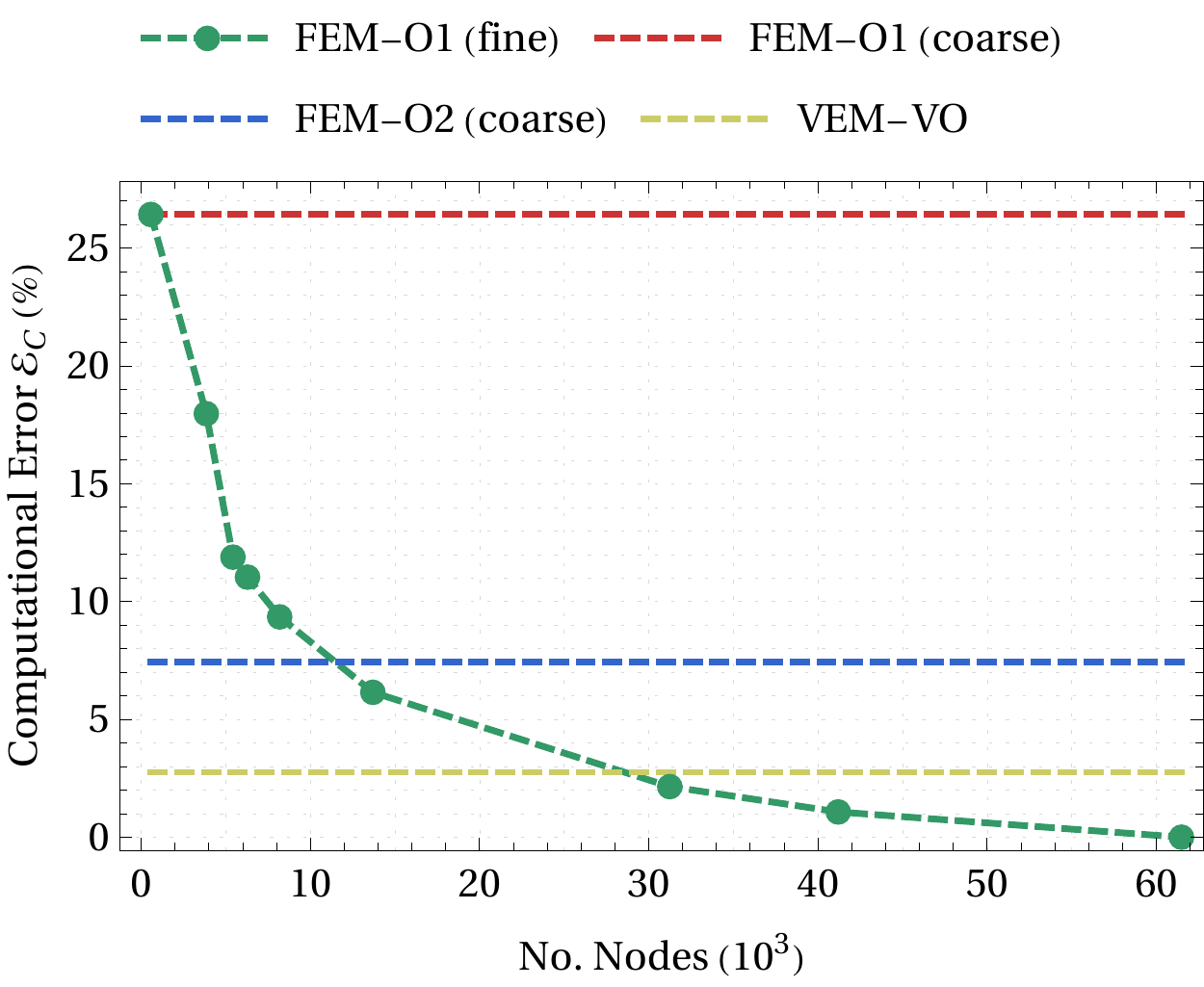}
        
        b)
    \end{minipage}
    \hfill
    \begin{minipage}{0.3\linewidth}
    \centering
        \includegraphics[width=\textwidth]{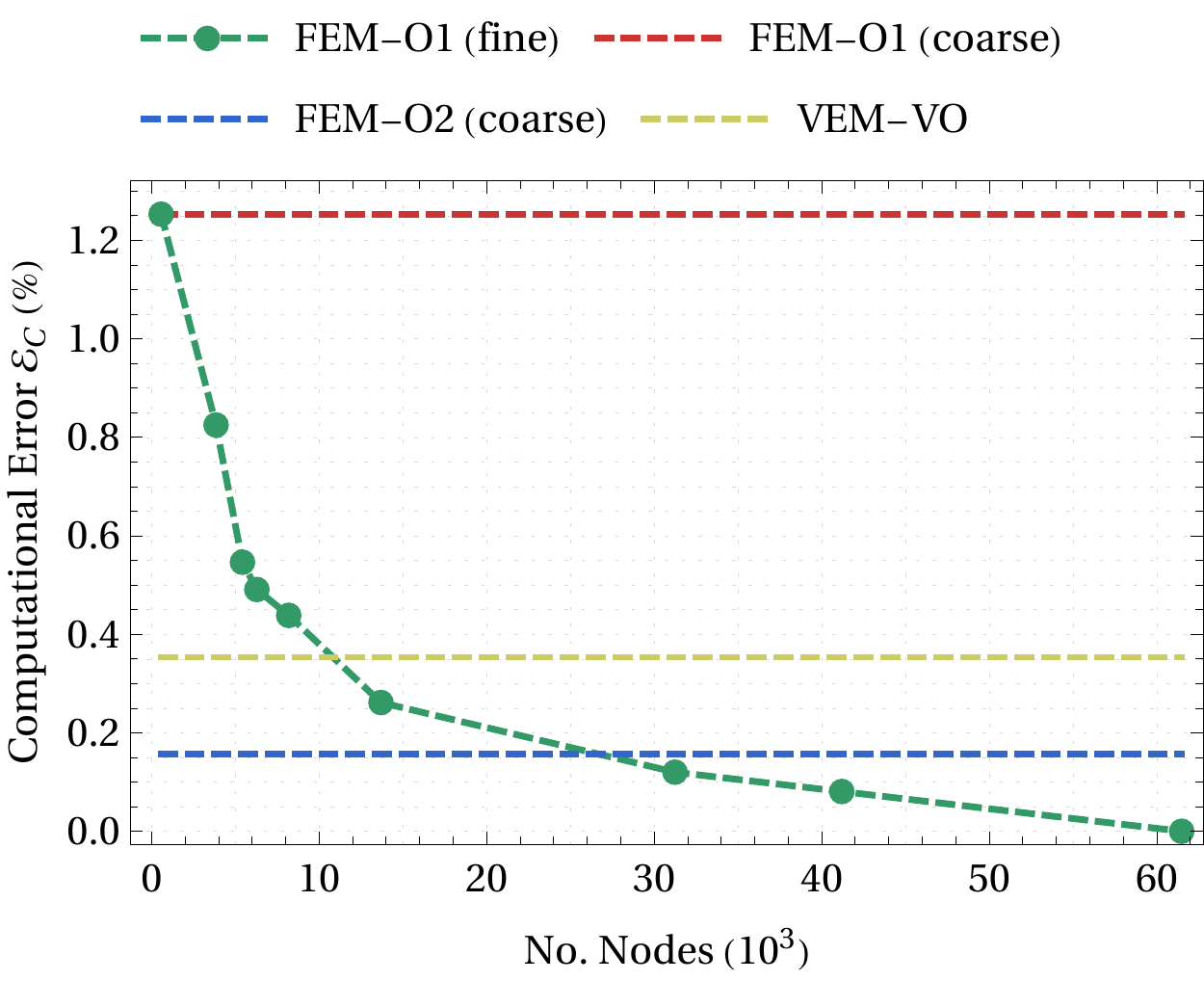}
        
        c)
    \end{minipage}
    \caption{Computational error $\mathcal{E}_{C}$ of $\mathrm{BN}$ with hexagonal unit cell: 
    a) effective mechanical modulus $\overline{\mathbb{C}}$; b) effective electro-mechanical modulus $\overline{\mathbf{e}}$; c) effective dielectric modulus $\overline{\boldsymbol{\epsilon}}$}
    \label{fig:NormPlotshexa3}
\end{figure}

\noindent An appropriate measure for comparison regarding performance of VE- and FE-approaches is introduced by an estimator of the computational error $\mathcal{E}_{C}$. In this regard, the Frobenius norm 
\begin{equation}\label{eq:FrobNorm}
    \begin{aligned}
        \lvert\lvert\left(\cdot\right)\rvert\rvert=\sqrt{\sum_{i=1}^{\mathrm{n}}\sum_{j=1}^{\mathrm{m}}\left(\cdot\right)_{ij}^{\mathrm{2}}}
    \end{aligned}
\end{equation}
is introduced to compare the effective macroscopic moduli in Eq. \eqref{eq:electromechanicallyModulus} obtained by different computational approaches. For the comparison, the computational error $\mathcal{E}_{C}$ related to the different discretizations
\begin{equation}\label{eq:error}
    \begin{aligned}
        \mathcal{E}_{C}&=\bigg\lvert \frac{\lvert\lvert\left(\cdot\right)\rvert\rvert^{\mathcal{M}}}{\lvert\lvert\left(\cdot\right)\rvert\rvert^{fine}}-1\bigg\rvert 10^{\mathrm{2}}, &&
        \mathcal{M}&\in\lbrace \text{FEM-O1},\;  \text{FEM-O2}, \; \text{VEM-VO}\rbrace, &&& (\cdot)\in\lbrace\overline{\mathbb{G}},\overline{\mathbb{C}},\overline{\mathbf{e}},\overline{\boldsymbol{\epsilon}}\rbrace,
    \end{aligned}
\end{equation}
is employed for different number of nodes for each approach. $\mathcal{E}_{C}$ in Eq. \eqref{eq:error} is defined as a relative quantity with respect to the result of the finest solution of the FEM-O1 approach using the largest number of elements, see Figure \ref{fig:MeshedRVEs}c), denoted by $\vert\vert(\cdot)\vert\vert^{fine}$. Thus it is assumed that the convergence of a FEM-O1 approach leads to decreasing computational error, with $\mathcal{E}_{C}=0$ at the finest mesh with largest number of nodes. In comparison to that convergence curve, the results, obtained by a FEM-O1 (coarse) and VEM-VO approach (having the same number of nodes, see Figure \ref{fig:MeshedRVEs}a) and b)), are illustrated. Moreover, this comparison includes also the result for Eq. \eqref{eq:error}, obtained by a FEM-O2 (coarse) approach, computed with the same number of elements as the FEM-O1 (coarse) approach (2769 elements). This provides a clear picture of the performance of the methods VEM-VO, FEM-O1 and FEM-O2 with respect to results, obtained by FEM-O1, undergoing mesh refinement. At first, we investigate on $\mathcal{E}_{C}\left(\overline{\mathbb{G}}\right)$ out of Eq. \eqref{eq:electromechanicallyModulus}, computed on varying degree of elastic anisotropy as well as different crystal lattice structures, see Figure \ref{fig:NormGPlots}. From the investigations in \cite{mar19}, we fix stabilization influence at VEM-VO computation to $\beta=0.1$.
\FloatBarrier
\noindent In Figure \ref{fig:NormGPlots}a) we obtain for a quasi-isotropic elastic response a computational error of 5\% versus <1\% regarding FEM-O1 versus VEM-VO at same number of nodes. When increasing the degree of anisotropy as well as varying material symmetry classes, as illustrated in Figure \ref{fig:NormGPlots}b) and \ref{fig:NormGPlots}c), VEM-VO demonstrates a robust performance, with $\mathcal{E}_{C}<5\%$ for the highest index of anisotropy while the computational error of FEM-O1 (coarse) increases up to $\approx$50\%. It has to be noted, that, even if orthorhombic lattice structures induce more non-zero entries in $\mathbb{C}$ than hexagonal structures, the computational error is mainly affected by the degree of elastic anisotropy.\\
\noindent As an example, the computational homogenization based on VEM-VO of a $\mathcal{RVE}$ made up by grains of $\mathrm{BaNiO_{3}}$ (the mildly-anisotropic material with an hexagonal lattice structure and $\mathcal{E}_{C}$ in Figure \ref{fig:NormGPlots}b)) yields the effective macroscopic modulus
\begin{equation}\label{eq:macroModulusVEM}
     \overline{\mathbb{G}}=\begin{bmatrix}
     \mathbf{124.361} & \mathbf{68.114} & \mathbf{58.489} & 2.090 & -0.676 & -0.232 & \mathbf{-1.039} & -0.244 & 0.130\\
      & \mathbf{122.194} & \mathbf{60.344} & -2.885 & 1.013 & -3.510 & \mathbf{-0.620} & -0.070 & 0.238\\
       & & \mathbf{129.372} & 1.942 & -0.102 & 1.808 & \mathbf{-0.349} & -0.151 & 0.391\\
        & & & \mathbf{53.906} & 6.096 & 2.126 & -0.001 & \mathbf{-0.117} & 0.0514\\
         & & & & \mathbf{51.975} & -0.741 & -0.105 & -0.051 & \mathbf{-0.047}\\
          & & \mathrm{sym.} & & & \mathbf{59.111} & 0.159 & 0.242 & -0.017\\
           & & & & & & \mathbf{145.883} & 28.358 & 6.017\\
            & & & & & & & \mathbf{123.173} & -11.078\\
             & & & & & & & & \mathbf{85.543}\\
     \end{bmatrix}.
\end{equation}
Here, the bold values represent the position of non-zero entries in the modulus $\mathbb{G}_{l}$ at grain-level of the $\mathcal{RVE}$, while other entries demonstrate the coupling given by the different grain orientations. At next, we investigate the performance of VEM-VO regarding particular moduli $\lbrace\overline{\mathbb{C}},\overline{\mathbf{e}},\overline{\boldsymbol{\epsilon}}\rbrace$, collected by $\overline{\mathbb{G}}$. Therefore, we employ again materials with three different crystal lattice structures, namely orthorhombic, trigonal and hexagonal structures in Figures \ref{fig:NormPlotsOrth3}, \ref{fig:NormPlotsTri3} and \ref{fig:NormPlotshexa3}. Again, a robust performance of VEM-VO solution is obtained with regard to orthorhombic lattice structure with quasi-isotropic moduli, see Figure \ref{fig:NormPlotsOrth3}. While the degree of elastic anisotropy is in range of quasi-isotropic and thus the FEM-O1 solution demonstrates a low computational error regarding $\overline{\mathbb{C}}$, the results for $\lbrace\overline{\mathbf{e}},\overline{\boldsymbol{\epsilon}}\rbrace$ can be highly inaccurate with FE-based approaches, see Figure \ref{fig:NormPlotsOrth3}b) and \ref{fig:NormPlotsOrth3}c). In particular, VEM-VO vs. FEM-O1 demonstrate $\approx$3\% vs. 35\% computational error. Also, VEM-VO leads only to half of the error when compared to FEM-O2 ($\approx$6\%).
\FloatBarrier
\FloatBarrier
\noindent In Figure \ref{fig:NormPlotsTri3}, a highly anisotropic behavior regarding $\mathbb{C}$ within trigonal lattice structures is investigated. For all solutions of the particular moduli $\lbrace\overline{\mathbb{C}},\overline{\mathbf{e}},\overline{\boldsymbol{\epsilon}}\rbrace$ a robust, low error producing, performance of VEM-VO is obtained. When addressing the mechanical modulus $\overline{\mathbb{C}}$, VEM-VO demonstrates approximately 12-times lower error $\mathcal{E}_{C}$ than FEM-O1 using the same number of nodes ($\approx$4\% vs. $\approx$51\%) and even when applying quadratic shape functions, the VEM-VO solution is still characterized by an error that is half of FEM-O2 ($\approx$4\% vs. $\approx$11\%). By investigating the piezoelectric modulus $\overline{\mathbf{e}}$ in Figure \ref{fig:NormPlotsTri3}b), VEM-VO shows four-times lower values of $\mathcal{E}_{C}$ as coarse FEM-O1 at same number of nodes ($\approx$8\% vs. $\approx$36\%). Here, VEM-VO and FEM-O2 provide approximately the same computational error. When investigating on the dielectric modulus $\overline{\boldsymbol{\epsilon}}$ in Figure \ref{fig:NormPlotsTri3}c), low values of $\mathcal{E}_{C}$ are obtained (FEM-O1 coarse $\approx$2.5\%). However, VEM-VO provides still better results than the FEM-O1 solution while FEM-O2 shows a slightly better performance as VEM-VO.
\FloatBarrier
\noindent In Figure \ref{fig:NormPlotshexa3}, we depict results for a highly anisotropic hexagonal lattice structure, where less non-zero entries exist in $\mathbb{C}$ with respect to a material with an orthorhombic or trigonal structure. However, such highly anisotropic environment leads to a high computational error for the effective mechanical modulus regarding the coarse FEM-O1 solution ($\approx$59\%), see Figure \ref{fig:NormPlotshexa3}a). As observed before, VEM-VO provides accurate computations with an error of $\mathcal{E}_{C}$ ($\approx$5\%), which is again approximately 12-times lower than FEM-O1 for coarse meshes  and approximately two-times lower than FEM-O2 for coarse meshes. Regarding the piezoelectric modulus in Figure \ref{fig:NormPlotshexa3}b), we obtain an approximately 8-times lower values of $\mathcal{E}_{C}$, when comparing VEM-VO to FEM-O1 ($\approx$3\% vs. $\approx$26\%) and approximately two-times lower error when comparing VEM-VO to coarse FEM-O2 solutions ($\approx$3\% vs. $\approx$7\%). The solutions for the dielectric modulus, illustrated in Figure \ref{fig:NormPlotshexa3}c), provide all a low amount of $\mathcal{E}_{C}$ and show similar behavior as observed before. The VEM-VO computation provides approximately three-times less computational error, compared to FEM-O1 solution on coarse mesh ($\approx$0.4\% vs. $\approx$1.2\%). Again, FEM-O2 leads to a slightly better result than VEM-VO, being anyway both very accurate ($\approx$0.4\% vs. 0.2\%).\\
Next, the error for the computation of the effective modulus of the coupled problem $\overline{\mathbb{G}}$ of orthorhombic and hexagonal lattice structures is illustrated in Figure \ref{fig:NormPlotsGort3Ghexa3}. 
\begin{figure}[htp!]
\centering
    \begin{minipage}{0.3\linewidth}
    \centering
        \includegraphics[width=\textwidth]{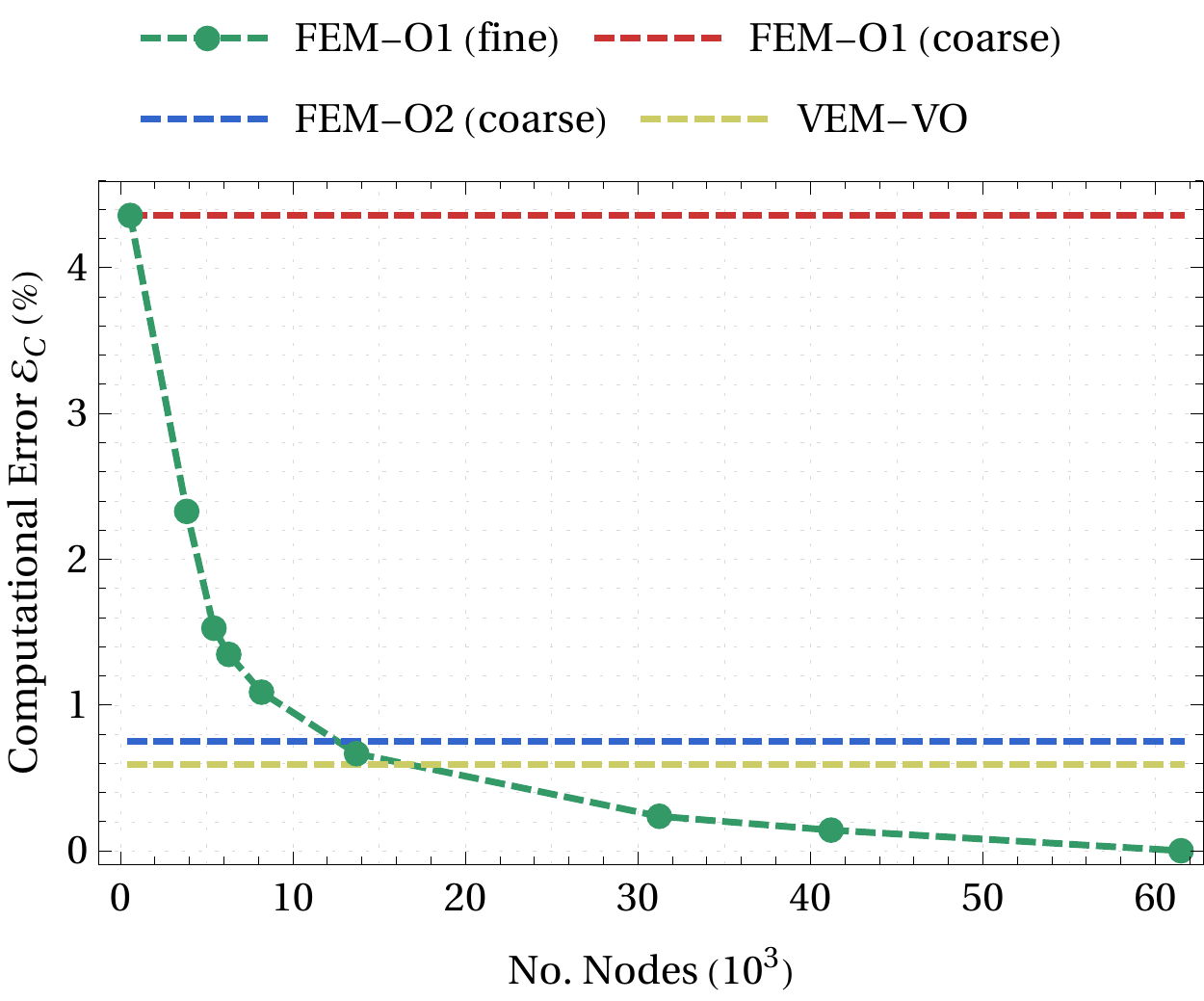}
        
        a)
    \end{minipage}
    \begin{minipage}{0.3\linewidth}
    \centering
        \includegraphics[width=\textwidth]{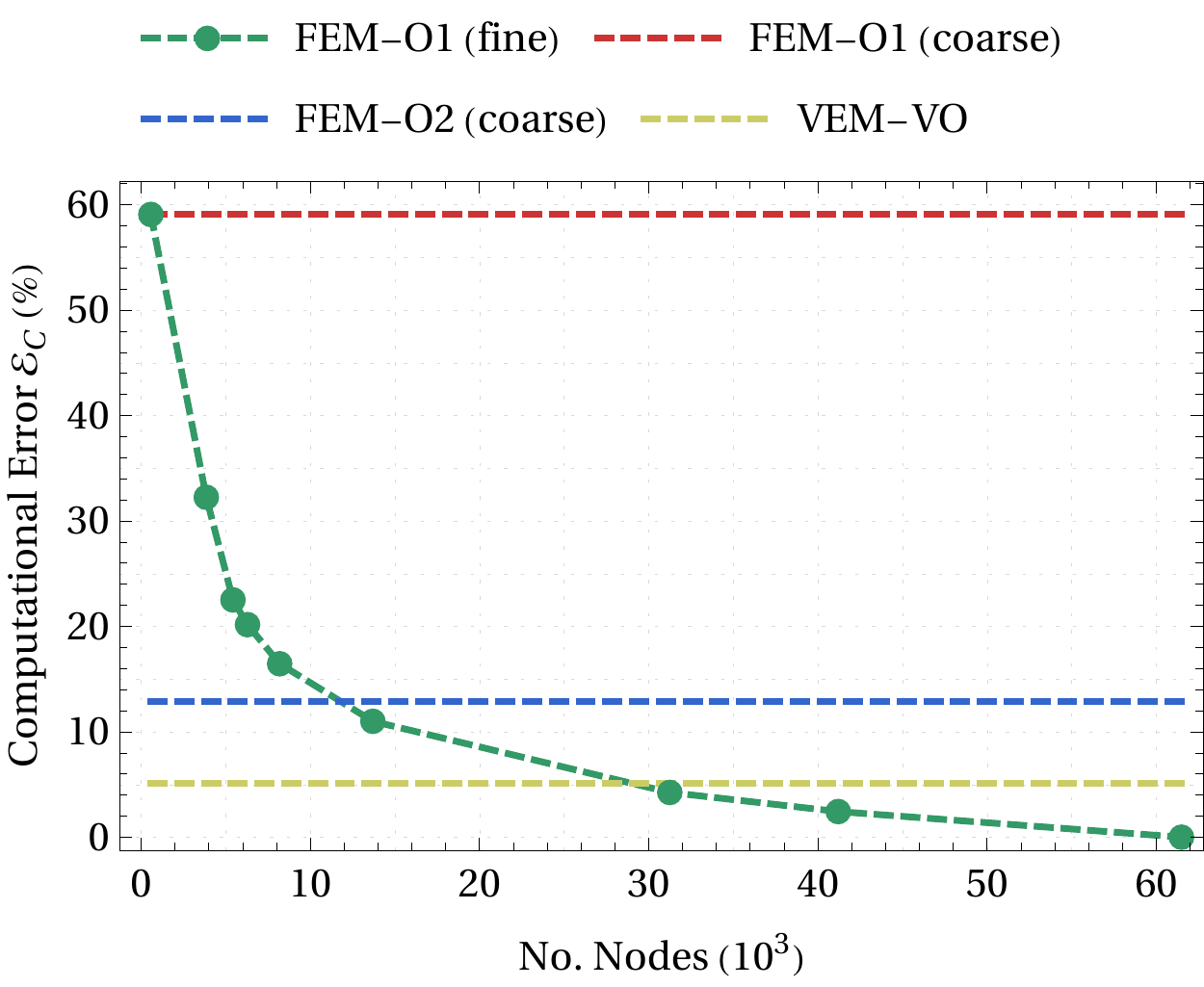}
        
        b)
    \end{minipage}
    \caption{Computational error $\mathcal{E}_{C}$ of $\overline{\mathbb{G}}$ of a) $\text{GaPO}_{\text{4}}$ with orthorhombic unit cell; b) $\text{BN}$ with hexagonal unit cell}
    \label{fig:NormPlotsGort3Ghexa3}
\end{figure}
\FloatBarrier
\noindent The results of the two $\overline{\mathbb{G}}$ show that the effective modulus of the macroscopic coupled problem is mainly influenced by $\overline{\mathbb{C}}$. The influence of $\overline{\boldsymbol{\epsilon}}$ on $\overline{\mathbb{G}}$ seems to be negligible. This effect might be caused by the high differences in magnitude of $\lbrace\mathbb{C},\mathbf{e},\boldsymbol{\epsilon}\rbrace$. A detailed overview for additional results on the computational error $\mathcal{E}_{C}$ is listed in the Appendix \ref{subsec:AdditionalResults}.

\subsubsection{Influence of Stabilization on relative Deviation}
\begin{figure}[htp!]
\centering
    \begin{minipage}{0.3\linewidth}
    \centering
        \includegraphics[width=\textwidth]{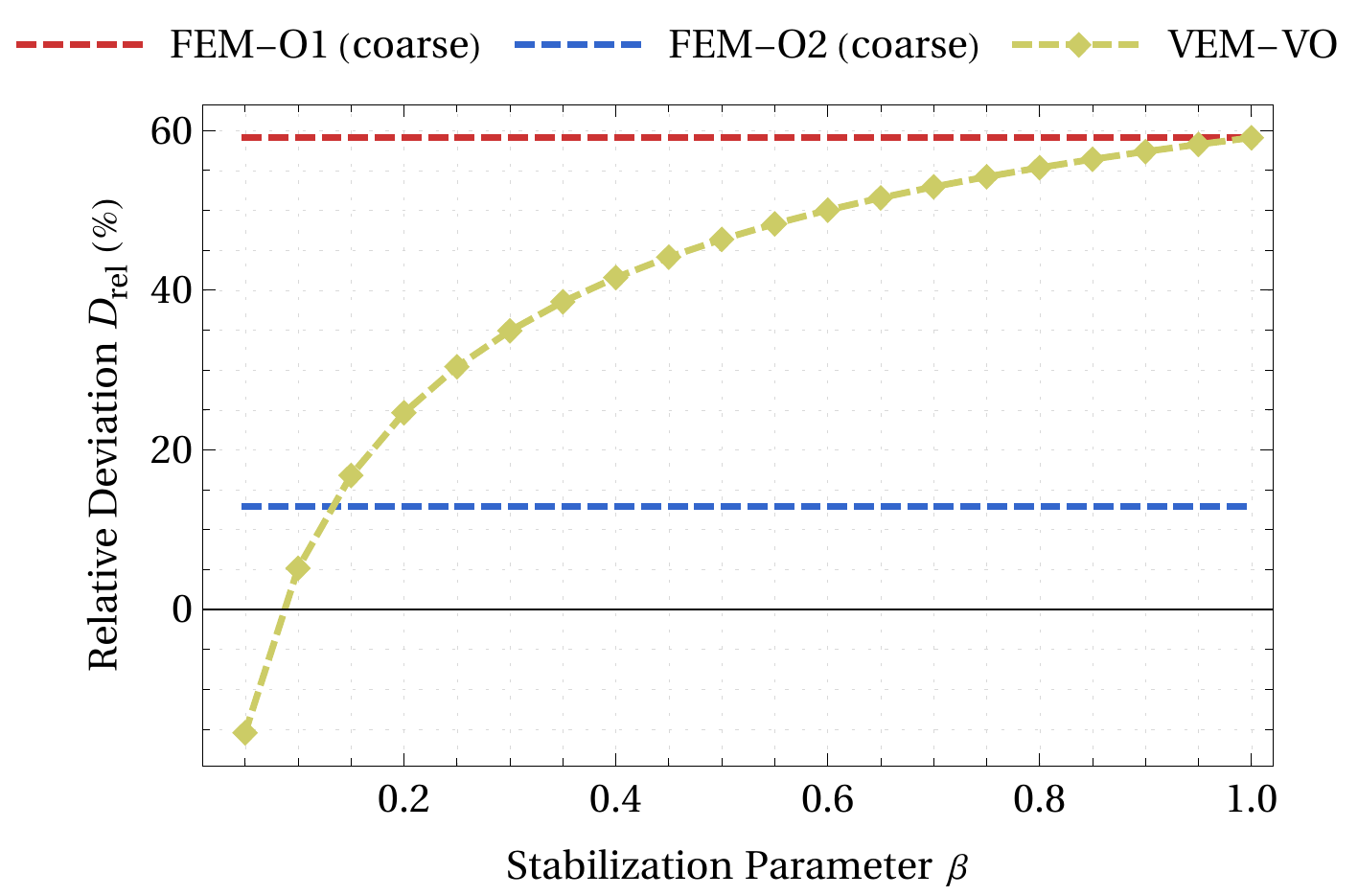}
        
        a)
    \end{minipage}
    \hfill
    \begin{minipage}{0.3\linewidth}
    \centering
        \includegraphics[width=\textwidth]{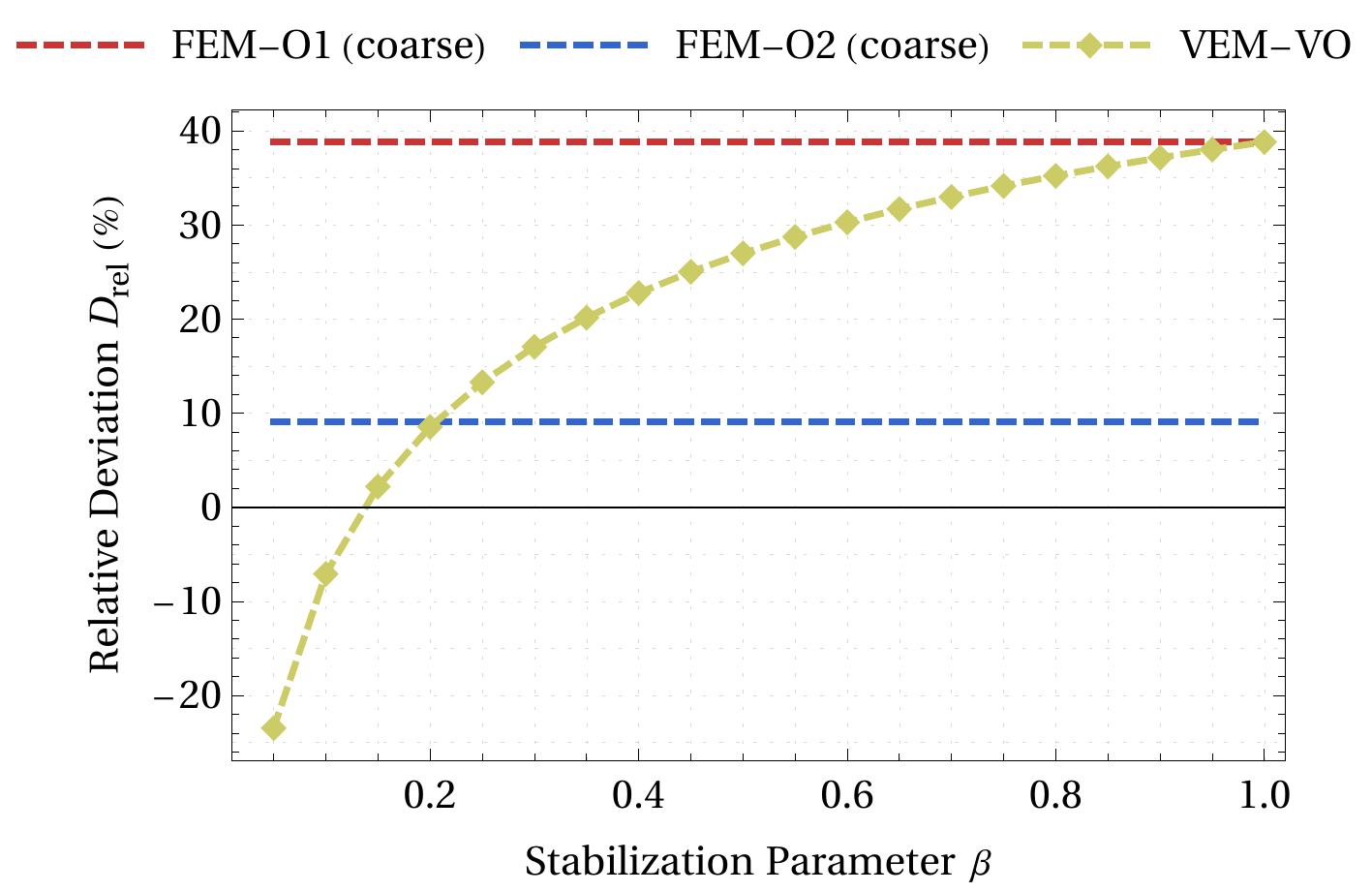}
        
        b)
    \end{minipage}
    \hfill
    \begin{minipage}{0.3\linewidth}
    \centering
        \includegraphics[width=\textwidth]{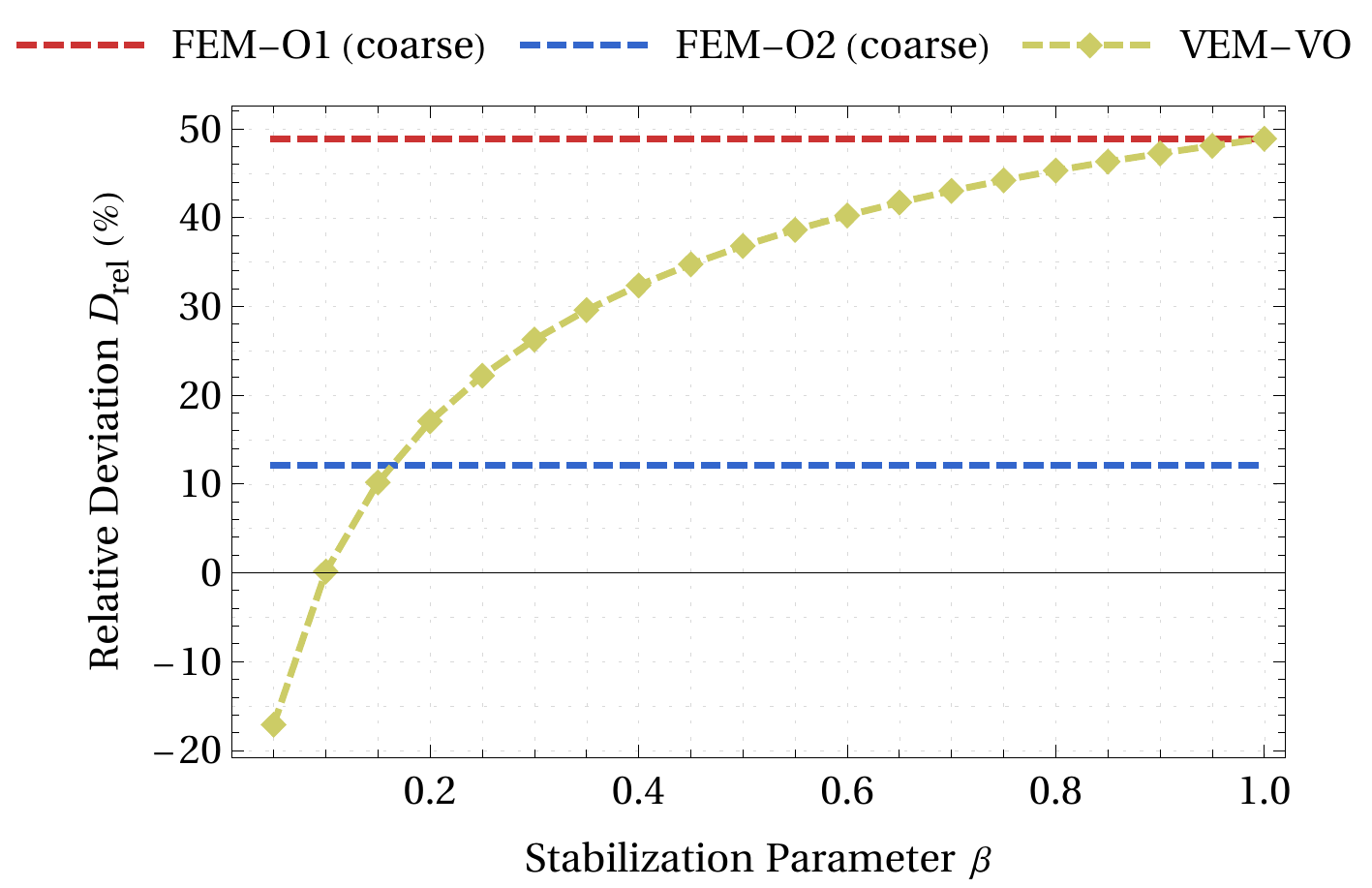}
        
        c)
    \end{minipage}
    \caption{Relative deviation $D_{\mathrm{rel}}$ of effective macroscopic moduli: a) mechanical modulus $\overline{\mathbb{C}}$, $\text{BN}$, hexagonal unit cell; b) electro-mechanical modulus $\overline{\mathbf{e}}$, $\text{BaNiO}_{\text{3}}$, hexagonal unit cell; c) dielectric modulus $\overline{\boldsymbol{\epsilon}}$, $\text{BaNiO}_{\text{3}}$, hexagonal unit cell}
    \label{fig:NormpartbetaPlots}
\end{figure}

\noindent The results of investigations on the computational error $\mathcal{E}_{C}$ were obtained by using a fixed value of the stabilization control parameter $\beta=0.1$ for the VE-approach. Next, the effect of this stabilization parameter in the range $\beta\in\left[0.05,\; 1\right]$ with $\Delta\beta=0.05$ is studied. Therefore, a relative deviation is introduced, reading
\begin{equation}\label{eq:reldeviation}
    \begin{aligned}
        D_{\mathrm{rel}}=10^{\mathrm{2}}\left(\frac{\lvert\lvert\left(\cdot\right)\rvert\rvert^{\mathcal{M}}-\lvert\lvert\left(\cdot\right)\rvert\rvert^{fine}}{\lvert\lvert\left(\cdot\right)\rvert\rvert^{fine}}\right),
    \end{aligned}
\end{equation}
which leads to a percentage of dismiss of the utilized approaches with respect to the solution with FEM-O1, computed at finest mesh, see Figure \ref{fig:MeshedRVEs}c). For $\beta=1$, the solution of the VE-approach degenerates to the coarse FEM-O1 one, since the VE-approach is computed by means of the stabilization part only. In addition, the case studies associated with the highest values of $\mathrm{D_{rel}}$ regarding particular effective moduli $\lbrace\overline{\mathbb{C}},\overline{\mathbf{e}},\overline{\boldsymbol{\epsilon}}\rbrace$ are collected in Figure \ref{fig:NormpartbetaPlots}. A detailed overview regarding additional results of relative deviation are provided in the Appendix \ref{subsec:AdditionalResults}.
\begin{figure}[htp!]
\centering
    \begin{minipage}{0.3\linewidth}
    \centering
        \includegraphics[width=\textwidth]{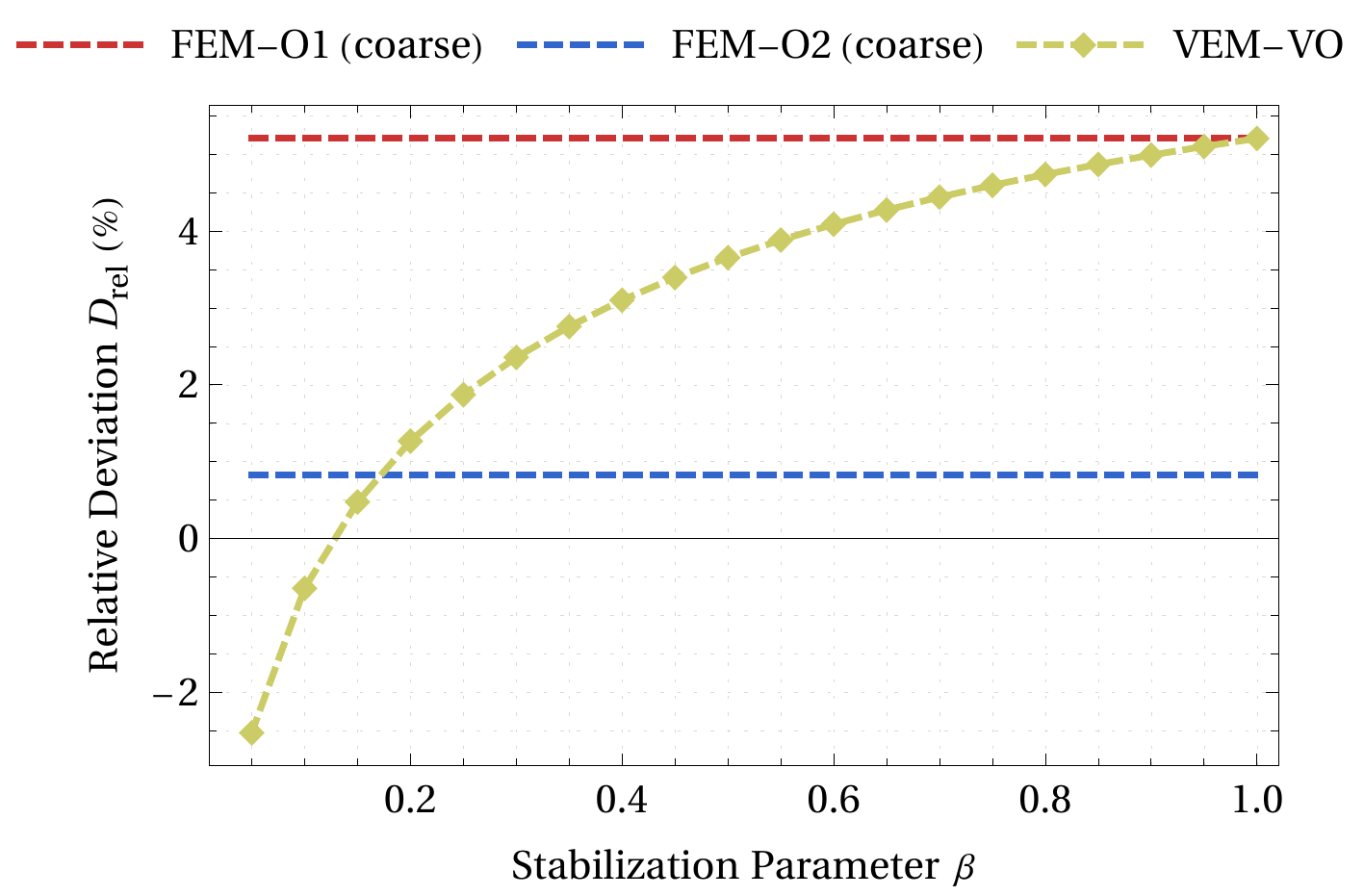}
        
        a)
    \end{minipage}
    \hfill
    \begin{minipage}{0.3\linewidth}
    \centering
        \includegraphics[width=\textwidth]{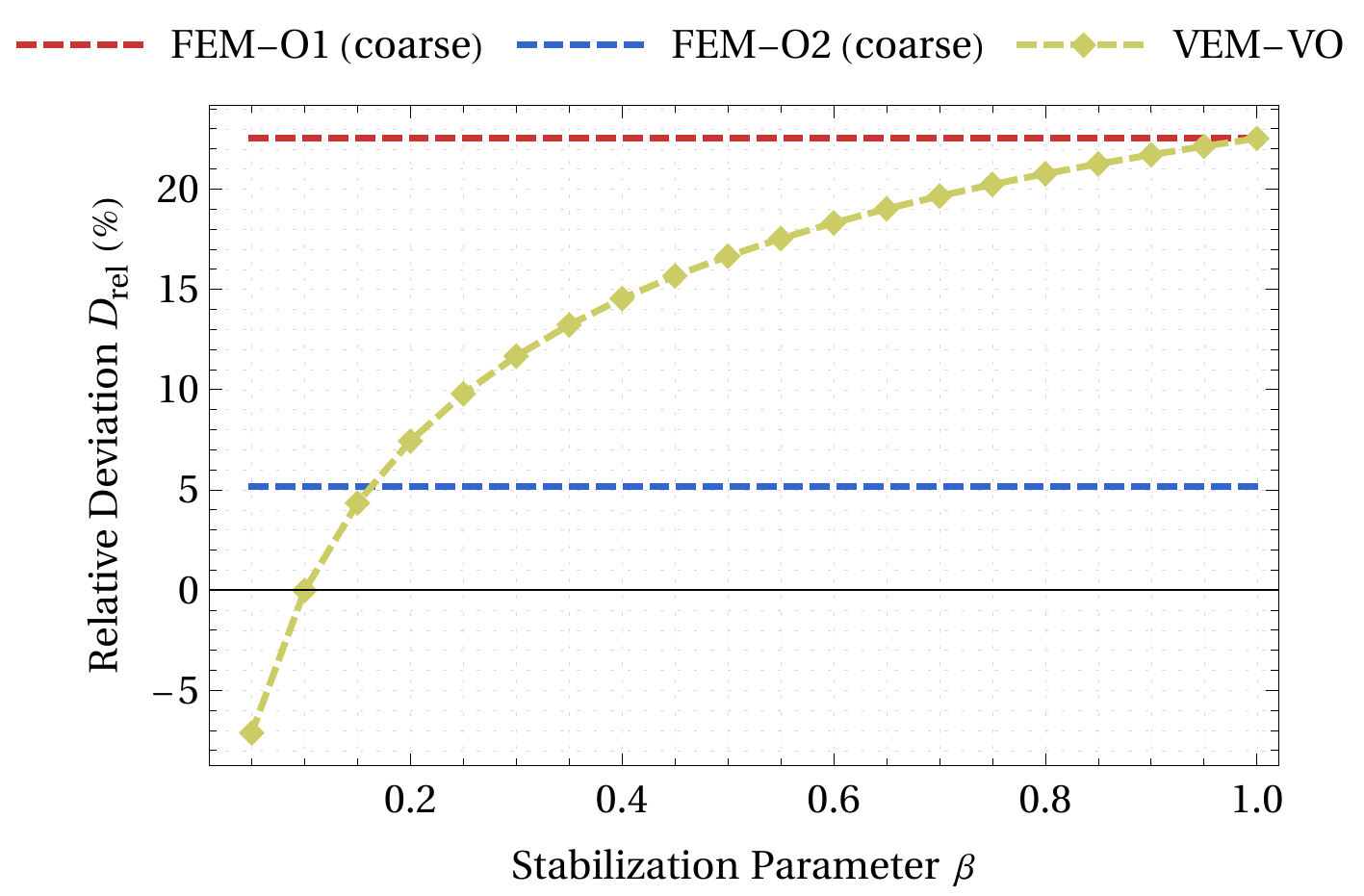}
        
        b)
    \end{minipage}
    \hfill
    \begin{minipage}{0.3\linewidth}
    \centering
        \includegraphics[width=\textwidth]{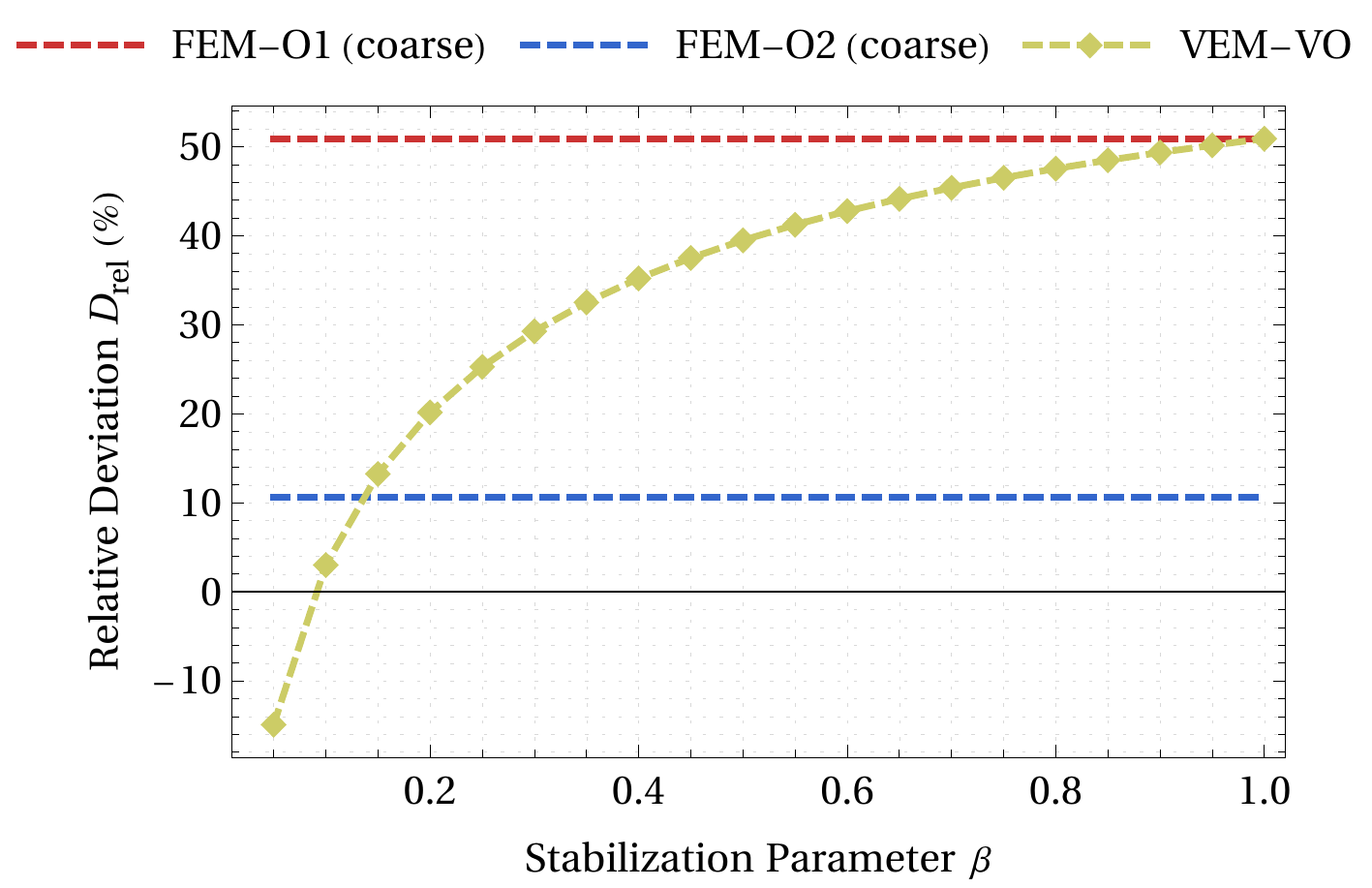}
        
        c)
    \end{minipage}
    \caption{Relative deviation $D_{\mathrm{rel}}$ of $\overline{\mathbb{G}}$ of a) $\text{AlPO}_{\text{4}}$, quasi-isotropic, orthorhombic unit cell; b) $\text{BaNiO}_{\text{3}}$, mildly anisotropic, hexagonal unit cell; c) $\text{MoS}_{\text{2}}$, highly anisotropic, trigonal unit cell}
    \label{fig:NormGbetaPlots}
\end{figure}
\FloatBarrier
\FloatBarrier
\noindent The solutions regarding $\overline{\mathbb{G}}$, illustrated in Figure \ref{fig:NormGbetaPlots}, show that VEM-VO in all three cases matches the finest solution at $D_\mathrm{{rel}}=0$ for low values of the stabilization parameter $\beta\in\left[0.1,0.15\right]$. Moreover, as expected, VEM-VO matches coarse FEM-O1 for $\beta=1$. The same behavior is observed for particular moduli $\lbrace\overline{\mathbb{C}},\overline{\mathbf{e}},\overline{\boldsymbol{\epsilon}}\rbrace$, illustrated in Figure \ref{fig:NormpartbetaPlots}. The fact that exact results are obtained with VEM-VO employing, in all cases, values of $\beta$ in the range $[0.1,0.15]$  demonstrates a performance of the virtual element method which is accurate and robust with respect to a variation of material properties. In fact, a constant value $\beta \approx 0.1$ can be employed for the computational homogenization of anisotropic materials in the wide range of anisotropic properties, obtaining accurate results ($\text{D}_{rel} \approx 0$). In all cases, FEM-O1, computed on the sub-mesh of VEM-VO, demonstrates the highest values of $\text{D}_{rel}$ (from $\approx 4\%$ up to $\approx 59\%$), always higher than VEM-VO. Also FEM-O2 shows higher relative deviation (from $\approx 0.9\%$ up to $\approx 12\%$) than VEM-VO with $\beta=0.1$.\\
The low optimal value of $\beta \approx 0.1$ suggests that the application at hand requires a small stabilization of the virtual element. Thus the stabilization term is only needed to eliminate the virtual elements rank-deficiency.

\subsection{Investigation on macroscopic electro-magneto-mechanical hybrid material}
\begin{figure}[htp!]
\centering
    \begin{minipage}{0.3\linewidth}
    \centering
            \includegraphics[width=\textwidth]{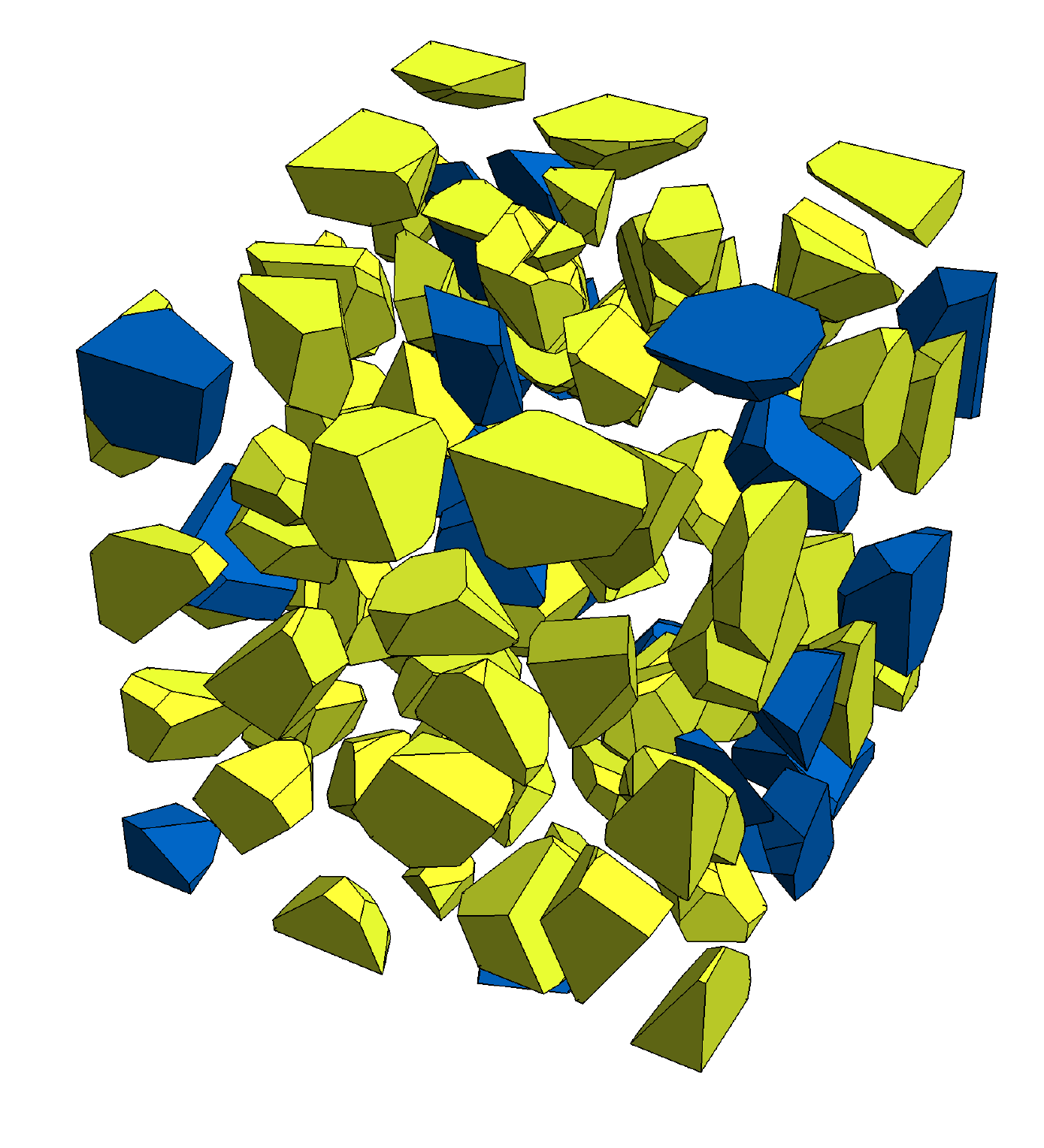}
        
        a)
    \end{minipage}
    \hfill
    \begin{minipage}{0.3\linewidth}
    \centering
        \includegraphics[width=\textwidth]{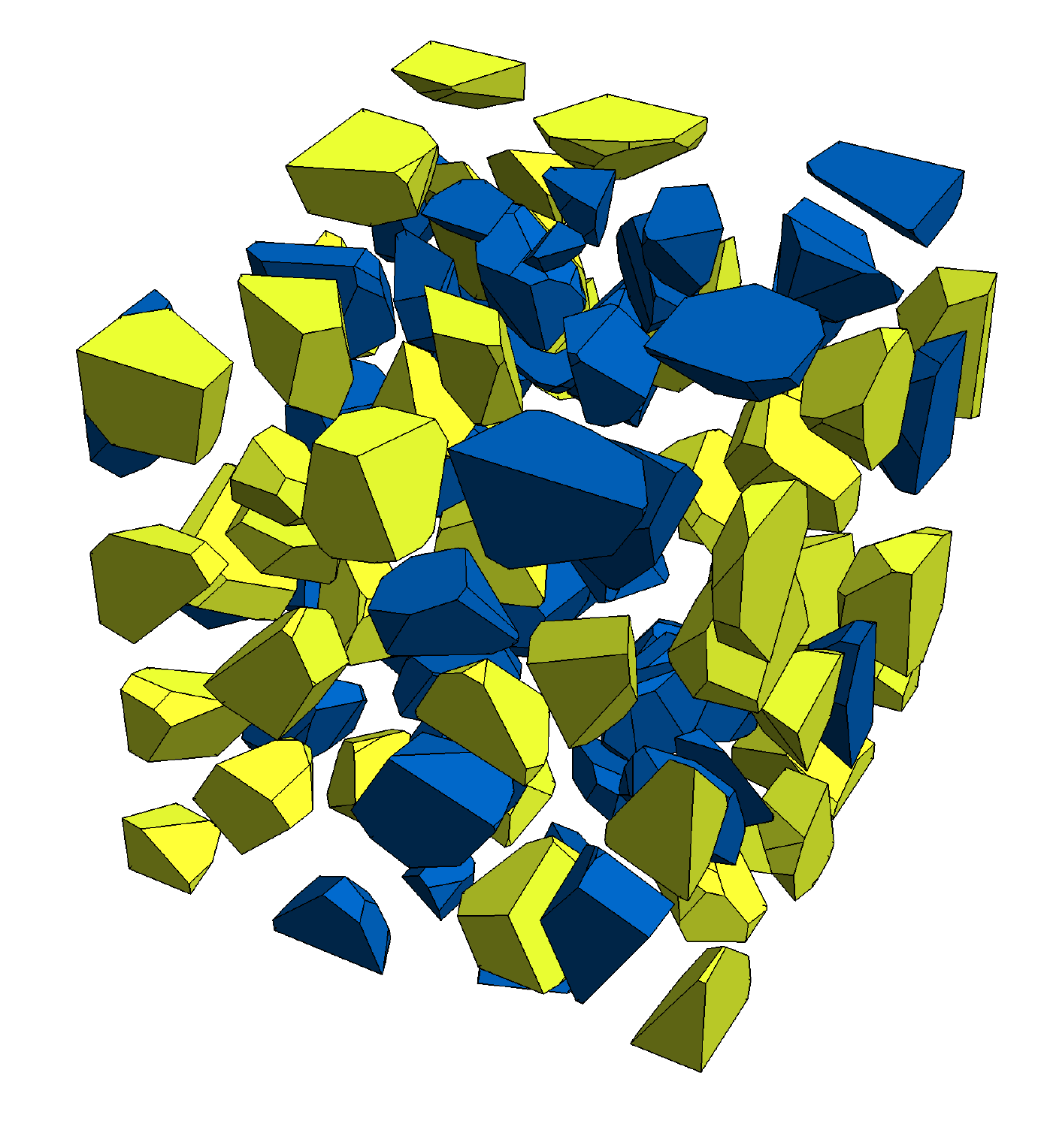}
        
        b)
    \end{minipage}
    \hfill
    \begin{minipage}{0.3\linewidth}
    \centering
        \includegraphics[width=\textwidth]{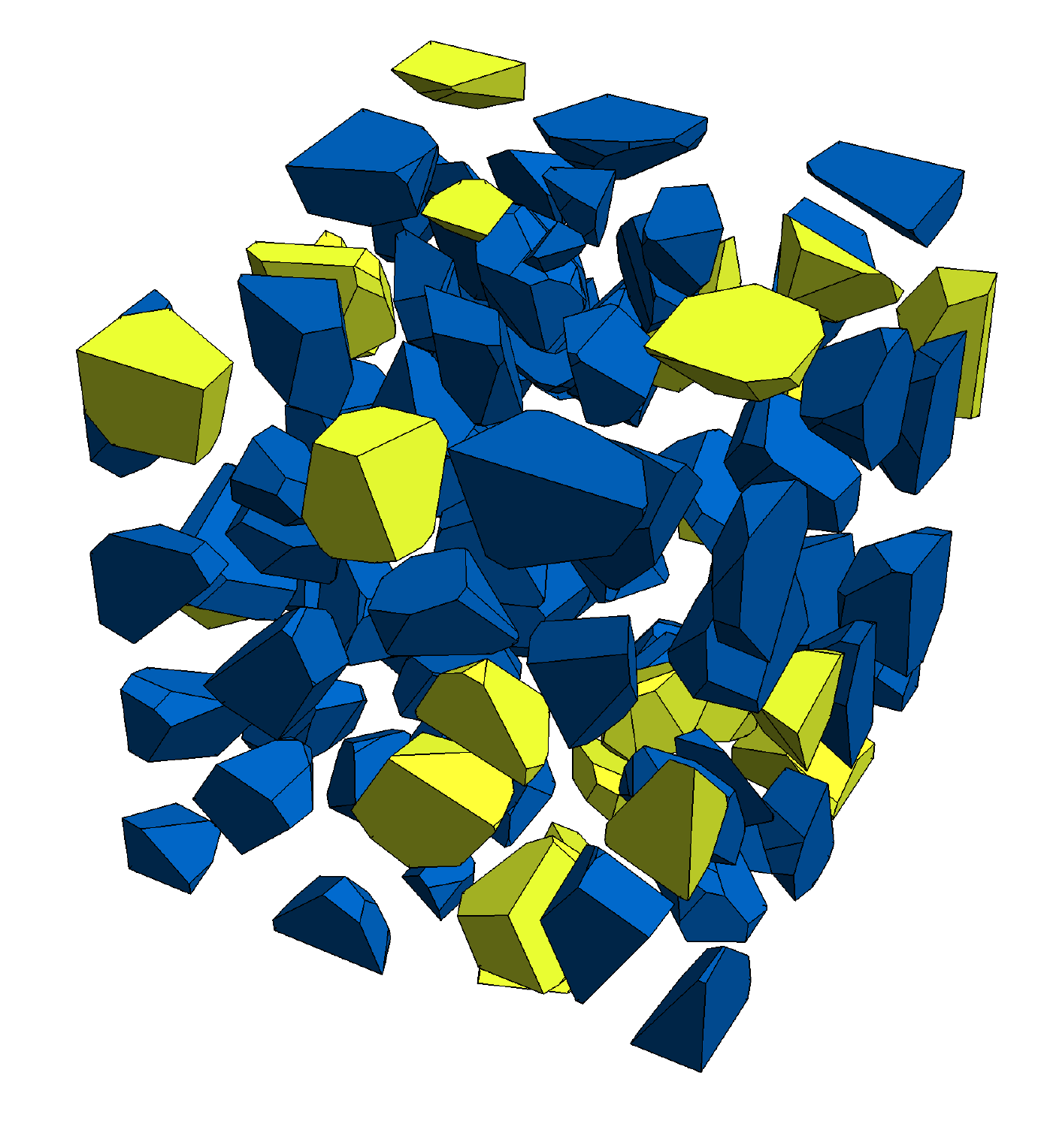}
        
        c)
    \end{minipage}
    \caption{Artificially generated hybrid $\mathcal{RVE}$ (exploded view) illustrating a composite microstructure of $\mathrm{BaTiO_{3}}$ (green) and $\mathrm{CoFe_{2}O_{4}}$ (blue) with fraction of a) $\mathcal{P}=0.25$; b) $\mathcal{P}=0.5$; c) $\mathcal{P}=0.75$}
    \label{fig:HybridRVEs}
\end{figure}

\noindent Additionally the virtual element framework is applied to a $\mathcal{RVE}$ consisting of a hybrid structure made up by two different polycrystalline materials. One of them behaves electro-mechanically and the other one magneto-mechanically. Thus, an electro-magneto-mechanical fully coupled problem is obtained, characterized by the effective macroscopic modulus
\begin{equation}\label{eq:macroscopicModulushybridMat}
    \overline{\mathbb{G}}=\begin{bmatrix}
    \overline{\mathbb{C}} & -\overline{\mathbf{e}}^{\mathrm{T}} & -\overline{\mathbf{q}}^{\mathrm{T}}\\
    -\overline{\mathbf{e}} & -\overline{\boldsymbol{\epsilon}} & \mathbf{0}\\
    -\overline{\mathbf{q}} & \mathbf{0} & \overline{\boldsymbol{\mu}}
    \end{bmatrix}.
\end{equation}
Note that there is no electro-magnetic coupling and thus $\overline{\boldsymbol{\alpha}}=\mathbf{0}$. The material properties $\mathbb{G}$ of piezo-electric ($\mathrm{BaTiO_{3}}$) and magneto-mechanical ($\mathrm{CoFe_{2}O_{4}}$) single grains are taken from \cite{sch16} and listed in Table \ref{tab:MatParaBaCo}.
\begin{table}[htp!]
    \centering
    \begin{tabular}{ccc||ccc}
        Parameter & $\mathrm{BaTiO_{3}}$ & $\mathrm{CoFe_{2}O_{4}}$ & Parameter & $\mathrm{BaTiO_{3}}$ & $\mathrm{CoFe_{2}O_{4}}$\\\hline
        $\mathbb{C}_{11}$ & $166$ & $212.1$ & $\mathbf{e}_{31}$ & $-4.4$ & 0\\
        $\mathbb{C}_{12}$ & $76.6$ & $74.5$ & $\mathbf{e}_{33}$ & $18.6$ & 0\\
        $\mathbb{C}_{13}$ & $77.5$ & $74.5$ & $\mathbf{e}_{15}$ & $11.6$ & 0\\
        $\mathbb{C}_{33}$ & $162$ & $212.1$ & $\mathbf{q}_{31}$ & $0$ & $580.3$\\
        $\mathbb{C}_{44}$ & $42.9$ & $68.8$ & $\mathbf{q}_{33}$ & $0$ & $-699.7$\\
        $\boldsymbol{\epsilon}_{11}$ & $0.0112$ & $8\cdot 10^{-5}$ & $\mathbf{q}_{15}$ & $0$ & $550$\\
        $\boldsymbol{\epsilon}_{33}$ & $0.0126$ & $9.3\cdot 10^{-5}$ & $\boldsymbol{\alpha}_{11}$ & $0$ & $0$\\
        $\boldsymbol{\mu}_{11}$ & $1.26$ & 157 & \multirow{2}{*}{$\boldsymbol{\alpha}_{33}$} & \multirow{2}{*}{$0$} & \multirow{2}{*}{$0$}\\
         $\boldsymbol{\mu}_{33}$ & $1.26$ & 157\\\hline
    \end{tabular}
    \caption{Material Parameters of $\mathrm{BaTiO_{3}}$ (piezo-electric) and $\mathrm{CoFe_{2}O_{4}}$ (magneto-mechanical) single grains. Units: $\mathbb{C}_{ij}(\mathrm{GPa})$, $\mathbf{e}_{ij}(\mathrm{C/m^{2}}), \; \mathbf{q}_{ij}(\mathrm{N/Am}), \; \boldsymbol{\epsilon}_{ij}(\mathrm{mC/kVm}), \; \boldsymbol{\mu}_{ij}(\mathrm{N/kA^{2}}),$ $\boldsymbol{\alpha}_{ij}(\mathrm{s/m})$}
    \label{tab:MatParaBaCo}
\end{table}
\FloatBarrier
\noindent Since the hybrid $\mathcal{RVE}$ consists of two materials, $\mathcal{P}$ is introduced as the volume fraction of $\mathrm{CoFe_{2}O_{4}}$ grains. The $\mathcal{RVE}$ from Figure \ref{fig:HybridRVEs} with 100 grains is used here.
\FloatBarrier
\subsubsection{Constitutive model specialization}
\noindent The material symmetry of $\mathrm{BaTiO_{3}}$ and $\mathrm{CoFe_{2}O_{4}}$ is transversely isotropic. The potential energy function is introduced in a convenient invariant space\footnote{Prominent works on the constitutive modelling of anisotropic multiphysics materials via potential energy functions, embedded in an invariant-formulation, are done by \textsc{J. Schr\"oder} and co-workers in for example \cite{sch03, sch04, sch10a, sch10b}.} from literature \cite{sch16}. Hence, a split into field- and coupling related parts is used. The energetic contribution of underlying anisotropy is added by introducing a vector $\mathbf{a}_{g}$ such that $\vert\vert\mathbf{a}_{g}\vert\vert=1$, for any $g\in\mathcal{RVE}$, that refers to the preferred direction of the solid grain. The energy density functional of each grain $g$ is additivly decomposed as
\begin{equation}
    \begin{aligned}
    \psi_{g}&=\psi_{el,g}+\psi_{em,g}+\psi_{mm,g}+\psi_{diel,g}+\psi_{mag,g},\\
    \psi_{el,g}&=\frac{\lambda}{2}I_{1}^{2}+\mu I_{2}+\omega_{1}I_{5,g}+\omega_{2}I_{4,g}^{2}+\omega_{3}I_{1}I_{4,g},\\
    \psi_{em,g}&=\beta_{1}I_{1}J_{2,g}^{e}+\beta_{2}I_{4,g}J_{2,g}^{e}+\beta_{3}K_{1,g}^{e},\\
    \psi_{mm,g}&=\kappa_{1}I_{1}J_{2,g}^{m}+\kappa_{2}I_{4,g}J_{2,g}^{m}+\kappa_{3}K_{1,g}^{m},\\
    \psi_{diel,g}&=\gamma_{1}J_{1}^{e}+\gamma_{2}(J_{2,g}^{e})^{2},\\
    \psi_{mag,g}&=\xi_{1}J_{1}^{m}+\xi_{2}(J_{2,g}^{m})^{2},
    \end{aligned}
\end{equation}
representing, respectively, the pure elastic, electro-mechanically, magneto-mechanically, dielectric as well as magnetic contributions. The invariants of the considered energy density functions are given by
\begin{equation}
    \begin{aligned}
        I_{1}&=\mathrm{Tr}\left[\boldsymbol{\varepsilon}\right], & I_{2}&=\mathrm{Tr}\left[\boldsymbol{\varepsilon}^{2}\right], && I_{4,g}=\mathrm{Tr}\left[\boldsymbol{\varepsilon}\mathbf{m}_{g}\right], &&& I_{5,g}=\mathrm{Tr}\left[\boldsymbol{\varepsilon}^{2}\mathbf{m}_{g}\right],\\
        J_{1}^{e}&=\mathrm{Tr}\left[\mathbf{E}\otimes\mathbf{E}\right], & J_{2,g}^{e}&=\mathrm{Tr}\left[\mathbf{E}\otimes\mathbf{a}_{g}\right], &&  J_{1}^{m}=\mathrm{Tr}\left[\mathbf{H}\otimes\mathbf{H}\right], &&&  J_{2,g}^{m}=\mathrm{Tr}\left[\mathbf{H}\otimes\mathbf{a}_{g}\right],\\
        K_{1,g}^{e}&=\mathrm{Tr}\left[\boldsymbol{\varepsilon}\left(\mathbf{E}\otimes\mathbf{a}_{g}\right)\right], & K_{1,g}^{m}&=\mathrm{Tr}\left[\boldsymbol{\varepsilon}\left(\mathbf{H}\otimes\mathbf{a}_{g}\right)\right],
    \end{aligned}
\end{equation}
with $\mathbf{m}_{g}=\mathbf{a}_{g}\otimes\mathbf{a}_{g}$ being the structural tensor, identifying the contribution of underlying anisotropy to the energy density functional of each grain \cite{sch16}. The coefficients are obtained from material parameters (see Table \ref{tab:MatParaBaCo}) as
\begin{equation}
    \begin{aligned}
        \lambda&=\mathbb{C}_{12},\quad \mu=\frac{1}{2}\left(\mathbb{C}_{11}-\mathbb{C}_{12}\right),\quad \omega_{1}=2\mathbb{C}_{44}+\mathbb{C}_{12}-\mathbb{C}_{11},\quad \omega_{2}=\frac{1}{2}\left(\mathbb{C}_{11}+\mathbb{C}_{33}\right)-2\mathbb{C}_{44}-\mathbb{C}_{13},\\
        \omega_{3}&=\mathbb{C}_{13}-\mathbb{C}_{12},\quad \beta_{1}=-\mathbf{e}_{31},\quad \beta_{2}=\mathbf{e}_{31}-\mathbf{e}_{33}+2\mathbf{e}_{15},\quad\beta_{3}=-2\mathbf{e}_{15},\quad \kappa_{1}=-\mathbf{q}_{31},\\
        \kappa_{2}&=\mathbf{q}_{31}-\mathbf{q}_{33}+2\mathbf{q}_{15},\quad \kappa_{3}=-2\mathbf{q}_{15},\quad\gamma_{1}=-\frac{1}{2}\boldsymbol{\epsilon}_{11},\quad\gamma_{2}=\frac{1}{2}\left(\boldsymbol{\epsilon}_{11}-\boldsymbol{\epsilon}_{33}\right),\\
        \xi_{1}&=-\frac{1}{2}\boldsymbol{\mu}_{11},\quad\xi_{2}=\frac{1}{2}\left(\boldsymbol{\mu}_{11}-\boldsymbol{\mu}_{33}\right).
    \end{aligned}
\end{equation}
\subsubsection{Computational Error and Stabilization Influence}
\begin{figure}[htp!]
   \centering
    \begin{minipage}{0.47\linewidth}
    \centering
            \includegraphics[width=\textwidth]{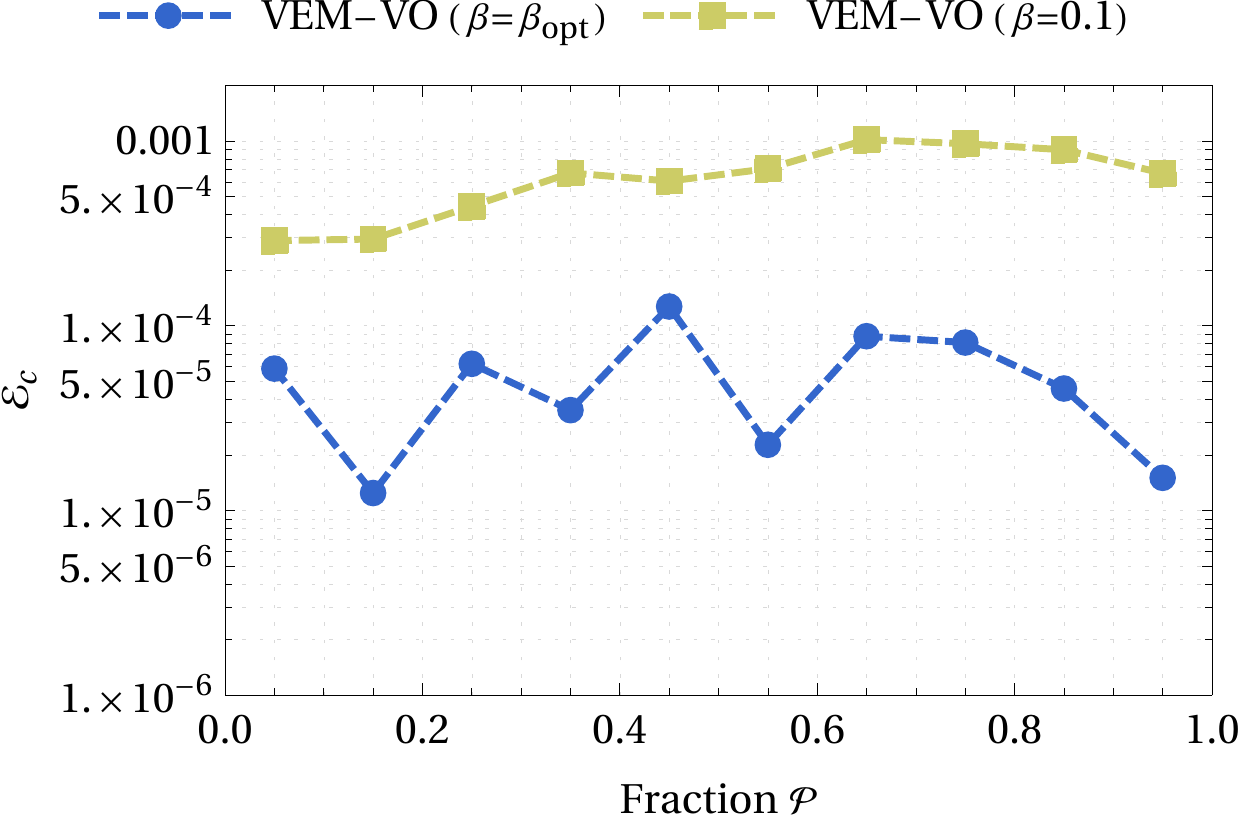}
        
        a)
    \end{minipage}
    \hfill
    \begin{minipage}{0.47\linewidth}
    \centering
        \includegraphics[width=\textwidth]{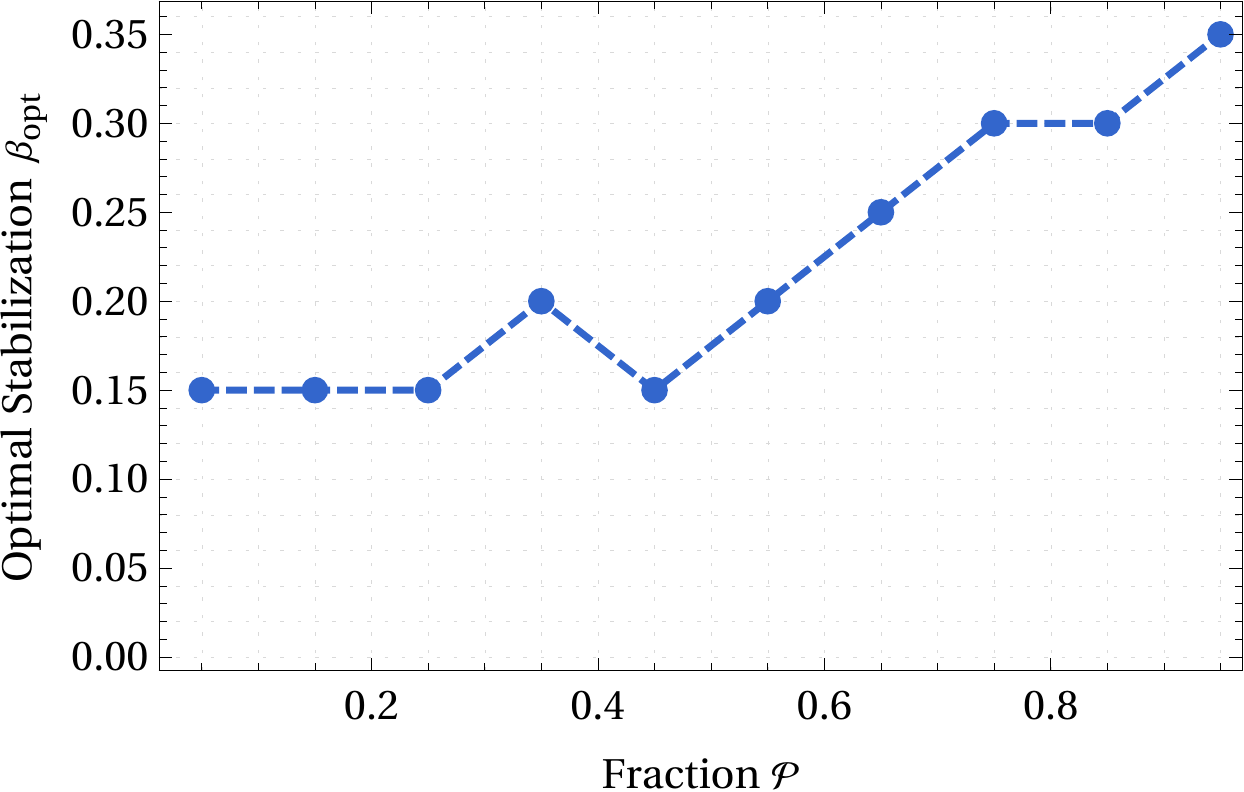}
        
        b)
    \end{minipage}
    \caption{a) computational error of the effective macroscopic modulus $\overline{\mathbb{G}}$, computed by VEM-VO for $(\beta=\beta_{opt})$ and $(\beta=0.1)$ with respect to a volume fraction $\mathcal{P}$ of $\mathrm{CoFe_{2}O_{4}}$; b) Influence of the volume fraction $\mathcal{P}$ of $\mathrm{CoFe_{2}O_{4}}$ on the optimal stabilization parameter $\beta_{opt}$}
    \label{fig:fracinfluence}
\end{figure}

\noindent The computational error $\mathcal{E}_{C}$ obtained by means of a VEM-based homogenization (see Eq. \eqref{eq:error}) is represented in Figure \ref{fig:fracinfluence}, employing a stabilization $\beta=0.1$, for different volume fractions $\mathcal{P}$ of $\mathrm{CoFe_{2}O_{4}}$ grains. 
\FloatBarrier
\noindent In order to investigate on the influence of the stabilization parameter in the case of a hybrid microstructure, the optimal value for $\beta$ regarding the solution for the effective macroscopic modulus $\overline{\mathbb{G}}$ is introduced as
\begin{equation}
    \beta_{opt}:=\underset{\beta}{min}\lbrace\mathcal{E}_{C}\left[\text{VEM-VO}(\mathcal{P,\beta})\right]\rbrace.
\end{equation}
The considered volume fractions are in the range $\mathcal{P}\in\left[0.05,\;0.95\right]$ with step-size being $\Delta\mathcal{P}=0.1$. Both results, the $\mathcal{E}_{C}\rvert_{\beta=0.1}$ and $\mathcal{E}_{C}\rvert_{\beta=\beta_{opt}}$ of the effective macroscopic modulus $\overline{\mathbb{G}}$, computed by the VEM-VO approach, are illustrated in Figure \ref{fig:fracinfluence}a). An overall increase of $\mathcal{E}_{C}$ is observed for increasing values of $\mathcal{P}$ when the stabilization influence is hold constant. A peak of $\mathcal{E}_{C}$ is reached at $\mathcal{P}=0.65$. However, the range of $\mathcal{E}_{C}$ of a VE-approach, acting in an anisotropic and heterogeneous framework remains low with a maximum at $\mathcal{E}_{C}=0.001$, which illustrates the accuracy of the computational approach. The results, depicted in Figure \ref{fig:fracinfluence}b), demonstrate an increase of the optimal stabilization parameter $\beta_{opt}$ with $\mathcal{P}$. Although the computational error can be optimized by tuning $\beta$ (by obtaining values of $\mathcal{E}_{C}\rvert_{\beta=\beta_{opt}}$ which are one order of magnitude lower than $\mathcal{E}_{C}\rvert_{\beta=0.1}$), the low absolute values obtained for $\mathcal{E}_{C}\rvert_{\beta=0.1}$ prove that the accuracy in the case under investigation is high and robust also by employing a constant value of $\beta$ for different microstructures (cf., Figure \ref{fig:fracinfluence}).

\section{Conclusion}
\noindent In this work, the authors investigated the performance of a low order virtual element homogenization scheme regarding polycrystalline microstructures. The underlying microstructure was modeled by artificially generated polyhedral elements, computed by voronoi tesselation. The advantage of a VE-approach is related to a perfect fit of grain geometries. Firstly, piezo-electric materials with different crystal lattice structures (orthorhombic, hexagonal and trigonal unit cells) were investigated. The effects of different degree of elastic anisotropy were analyzed. The applied virtual element scheme (VEM-VO) demonstrated a high and robust performance regarding the computational error, computed with respect to an overkilled FEM-based scheme. VEM-VO outperformes for all lattice structures and for every degree of elastic anisotropy a linear FEM-O1 approach with the same number of nodes by up to 12-times lower error values. In addition, VEM-VO shows also better performance of FEM-based approaches with quadratic shape functions.\\
\noindent In the second part of the work, the authors employed a hybrid microstructure, consisting of two different polycrystalline materials, having either piezo-electric or magneto-mechanical properties. Here, the obtained macroscopic behavior is fully coupled and exhibits an electro-magneto-mechanical behavior. The VEM-based homogenization scheme demonstrated an accurate behavior also in this case, independently from the composition of the microstructure (i.e., for different volume fractions of one grain type). VEM results are accurate by employing the same value of the stabilization parameter for all materials and microstructural properties investigated in this work. It might be considered that a stabilized method like VEM cannot be applied to different classes of problems easily since the stabilization parameter needs to be adjusted. However, in this paper and in \cite{mar19} it was shown that all results were obtained with the same stability parameter of $\beta=0.1$. Hence the application of VEM in homogenization can rely on a fixed stabilization parameter and thus needs no adjustment.\\
In conclusion, the results clearly demonstrate the applicability and the advantage of a VEM-VO scheme in comparison to classical FE-approaches for the homogenization of polycrystalline microstructures with grains characterized by anisotropic and multiphysical behaviors.

\section*{Acknowledgement}
\noindent C.B. and P.W. gratefully acknowledge the German Research Foundation (DFG, Deutsche Forschungsgemeinschaft) for financial support to this work with the Collaborative Research Centre 1153 (CRC 1153) \lq\lq{}\textit{Process chain for the production of hybrid high-performance components through tailored forming}\rq\rq{} with the subproject C4 \lq\lq{}\textit{Modelling and Simulation of the Joining Zone}\rq\rq{}, project number 252662854. M.M. acknowledges the Italian Ministry of Education, University and Research (MIUR) for funding in the framework of the Rita Levi Montalcini Program.

 \renewcommand*\thesection{\Alph{section}}
 \renewcommand*\thesubsection{\thesection.\arabic{subsection}}
 \setcounter{section}{0}
 \newpage
\section{Appendix}
\subsection{Transformation Matrices}
\label{subsec:TransformMatrices}
\noindent The rotation matrix $\mathbf{Q}$ is multiplicative decomposed to $\mathbf{Q}=\mathbf{Q}_{1}\mathbf{Q}_{2}\mathbf{Q}_{3}$ with its particular parts as
\begin{equation*}
    \begin{aligned}
        \mathbf{Q}_{\mathrm{1}}=\begin{bmatrix}
        1 & 0 & 0\\
        0 & \cos\vartheta_{\mathrm{1}} & -\sin\vartheta_{\mathrm{1}}\\
        0 & \sin\vartheta_{\mathrm{1}} & \cos\vartheta_{\mathrm{1}}
        \end{bmatrix},\quad
        \mathbf{Q}_{\mathrm{2}}=\begin{bmatrix}
        \cos\vartheta_{\mathrm{2}} & 0 & \sin\vartheta_{\mathrm{2}}\\
        0 & 1 & 0\\
        -\sin\vartheta_{\mathrm{2}} & 0 & \cos\vartheta_{\mathrm{2}}
        \end{bmatrix},\quad
         \mathbf{Q}_{\mathrm{3}}=\begin{bmatrix}
        \cos\vartheta_{\mathrm{3}} & -\sin\vartheta_{\mathrm{3}} & 0\\
        \sin\vartheta_{\mathrm{3}} & \cos\vartheta_{\mathrm{3}} & 0\\
        0 & 0 & 1
        \end{bmatrix},
    \end{aligned}
\end{equation*}
with $\vartheta_{i}\in\left[0,2\pi\right],\forall i\in\lbrace 1,2,3\rbrace$ as the angles of rotation with respect to origin main axes of the $\mathcal{RVE}$, aligned to unit (global) basis vectors $\mathbf{e}_{i}$ in spatial dimension $\mathbb{R}^{\mathrm{3}}$. The transformations $\lbrace\mathbf{T}^{\sigma},\mathbf{T}^{\varepsilon}\rbrace=\lbrace\mathbf{T}^{\sigma}_{1}\mathbf{T}^{\sigma}_{2}\mathbf{T}^{\sigma}_{3},\mathbf{T}^{\varepsilon}_{1}\mathbf{T}^{\varepsilon}_{2}\mathbf{T}^{\varepsilon}_{3}\rbrace$ of stresses and strains are also multiplicative decomposed and related to the same main axes of the $\mathcal{RVE}$. The single components read \cite{kol03}:
\begin{equation*}
    \begin{aligned}
        \mathbf{T}_{\mathrm{1}}^{\sigma}=\begin{bmatrix}
        1 & 0 & 0 & 0 & 0 & 0\\
        0 & \cos^{\mathrm{2}}\vartheta_{\mathrm{1}} & \sin^{\mathrm{2}}\vartheta_{\mathrm{1}} & 2\cos\vartheta_{\mathrm{1}}\sin\vartheta_{\mathrm{1}} & 0 & 0\\
        0 & \sin^{\mathrm{2}}\vartheta_{\mathrm{1}} & \cos^{\mathrm{2}}\vartheta_{\mathrm{1}} & -2\cos\vartheta_{\mathrm{1}}\sin\vartheta_{\mathrm{1}} & 0 & 0\\
        0 & -\cos\vartheta_{\mathrm{1}}\sin\vartheta_{\mathrm{1}} & \cos\vartheta_{\mathrm{1}}\sin\vartheta_{\mathrm{1}} & \cos^{\mathrm{2}}\vartheta_{\mathrm{1}}-\sin^{\mathrm{2}}\vartheta_{\mathrm{1}} & 0 & 0\\
        0 & 0 & 0 & 0 & \cos\vartheta_{\mathrm{1}} & -\sin\vartheta_{\mathrm{1}}\\
        0 & 0 & 0 & 0 & \sin\vartheta_{\mathrm{1}} & \cos\vartheta_{\mathrm{1}}
        \end{bmatrix},\\
        \mathbf{T}_{\mathrm{2}}^{\sigma}=\begin{bmatrix}
        \cos^{\mathrm{2}}\vartheta_{\mathrm{2}} & 0 & \sin^{\mathrm{2}}\vartheta_{\mathrm{2}} & 0 & 2\cos\vartheta_{\mathrm{2}}\sin\vartheta_{\mathrm{2}} & 0\\
        0 & 1 & 0 & 0 & 0 & 0\\
        \sin^{\mathrm{2}}\vartheta_{\mathrm{2}} & 0 & \cos^{\mathrm{2}}\vartheta_{\mathrm{2}} & 0 & -2\cos\vartheta_{\mathrm{2}}\sin\vartheta_{\mathrm{2}} & 0\\
        0 & 0 & 0 & \cos\vartheta_{\mathrm{2}} & 0 & -\sin\vartheta_{\mathrm{2}}\\
        -\cos\vartheta_{\mathrm{2}}\sin\vartheta_{\mathrm{2}} & 0 & \cos\vartheta_{\mathrm{2}}\sin\vartheta_{\mathrm{2}} & 0 & \cos^{\mathrm{2}}\vartheta_{\mathrm{2}}-\sin^{\mathrm{2}}\vartheta_{\mathrm{2}} & 0\\
        0 & 0 & 0 & \sin\vartheta_{\mathrm{2}} & 0 & \cos\vartheta_{\mathrm{2}}
        \end{bmatrix},\\
        \mathbf{T}_{\mathrm{3}}^{\sigma}=\begin{bmatrix}
        \cos^{\mathrm{2}}\vartheta_{\mathrm{3}} & \sin^{\mathrm{2}}\vartheta_{\mathrm{3}} & 0 & 0 & 0 & 2\cos\vartheta_{\mathrm{3}}\sin\vartheta_{\mathrm{3}}\\
        \sin^{\mathrm{2}}\vartheta_{\mathrm{3}} & \cos^{\mathrm{2}}\vartheta_{\mathrm{3}} & 0 & 0 & 0 & -2\cos\vartheta_{\mathrm{3}}\sin\vartheta_{\mathrm{3}}\\
        0 & 0 & 1 & 0 & 0 & 0\\
        0 & 0 & 0 & \cos\vartheta_{\mathrm{3}} & -\sin\vartheta_{\mathrm{3}} & 0\\
        0 & 0 & 0 & \sin\vartheta_{\mathrm{3}} & \cos\vartheta_{\mathrm{3}} & 0\\
        -\cos\vartheta_{\mathrm{3}}\sin\vartheta_{\mathrm{3}} & \cos\vartheta_{\mathrm{3}}\sin\vartheta_{\mathrm{3}} & 0 & 0 & 0 & \cos^{\mathrm{2}}\vartheta_{\mathrm{3}}-\sin^{\mathrm{2}}\vartheta_{\mathrm{3}}
        \end{bmatrix},\\
        \mathbf{T}_{\mathrm{1}}^{\varepsilon}=\begin{bmatrix}
         1 & 0 & 0 & 0 & 0 & 0\\
        0 & \cos^{\mathrm{2}}\vartheta_{\mathrm{1}} & \sin^{\mathrm{2}}\vartheta_{\mathrm{1}} & \cos\vartheta_{\mathrm{1}}\sin\vartheta_{\mathrm{1}} & 0 & 0\\
        0 & \sin^{\mathrm{2}}\vartheta_{\mathrm{1}} & \cos^{\mathrm{2}}\vartheta_{\mathrm{1}} & -\cos\vartheta_{\mathrm{1}}\sin\vartheta_{\mathrm{1}} & 0 & 0\\
        0 & -2\cos\vartheta_{\mathrm{1}}\sin\vartheta_{\mathrm{1}} & 2\cos\vartheta_{\mathrm{1}}\sin\vartheta_{\mathrm{1}} & \cos^{\mathrm{2}}\vartheta_{\mathrm{1}}-\sin^{\mathrm{2}}\vartheta_{\mathrm{1}} & 0 & 0\\
        0 & 0 & 0 & 0 & \cos\vartheta_{\mathrm{1}} & -\sin\vartheta_{\mathrm{1}}\\
        0 & 0 & 0 & 0 & \sin\vartheta_{\mathrm{1}} & \cos\vartheta_{\mathrm{1}}
        \end{bmatrix},\\
        \mathbf{T}_{\mathrm{2}}^{\varepsilon}=\begin{bmatrix}
        \cos^{\mathrm{2}}\vartheta_{\mathrm{2}} & 0 & \sin^{\mathrm{2}}\vartheta_{\mathrm{2}} & 0 & \cos\vartheta_{\mathrm{2}}\sin\vartheta_{\mathrm{2}} & 0\\
        0 & 1 & 0 & 0 & 0 & 0\\
        \sin^{\mathrm{2}}\vartheta_{\mathrm{2}} & 0 & \cos^{\mathrm{2}}\vartheta_{\mathrm{2}} & 0 & -\cos\vartheta_{\mathrm{2}}\sin\vartheta_{\mathrm{2}} & 0\\
        0 & 0 & 0 & \cos\vartheta_{\mathrm{2}} & 0 & -\sin\vartheta_{\mathrm{2}}\\
        -2\cos\vartheta_{\mathrm{2}}\sin\vartheta_{\mathrm{2}} & 0 & 2\cos\vartheta_{\mathrm{2}}\sin\vartheta_{\mathrm{2}} & 0 & \cos^{\mathrm{2}}\vartheta_{\mathrm{2}}-\sin^{\mathrm{2}}\vartheta_{\mathrm{2}} & 0\\
        0 & 0 & 0 & \sin\vartheta_{\mathrm{2}} & 0 & \cos\vartheta_{\mathrm{2}}
        \end{bmatrix},\\
        \mathbf{T}_{\mathrm{3}}^{\varepsilon}=\begin{bmatrix}
        \cos^{\mathrm{2}}\vartheta_{\mathrm{3}} & \sin^{\mathrm{2}}\vartheta_{\mathrm{3}} & 0 & 0 & 0 & \cos\vartheta_{\mathrm{3}}\sin\vartheta_{\mathrm{3}}\\
        \sin^{\mathrm{2}}\vartheta_{\mathrm{3}} & \cos^{\mathrm{2}}\vartheta_{\mathrm{3}} & 0 & 0 & 0 & -\cos\vartheta_{\mathrm{3}}\sin\vartheta_{\mathrm{3}}\\
        0 & 0 & 1 & 0 & 0 & 0\\
        0 & 0 & 0 & \cos\vartheta_{\mathrm{3}} & -\sin\vartheta_{\mathrm{3}} & 0\\
        0 & 0 & 0 & \sin\vartheta_{\mathrm{3}} & \cos\vartheta_{\mathrm{3}} & 0\\
        -2\cos\vartheta_{\mathrm{3}}\sin\vartheta_{\mathrm{3}} & 2\cos\vartheta_{\mathrm{3}}\sin\vartheta_{\mathrm{3}} & 0 & 0 & 0 & \cos^{\mathrm{2}}\vartheta_{\mathrm{3}}-\sin^{\mathrm{2}}\vartheta_{\mathrm{3}}
        \end{bmatrix}.
    \end{aligned}
\end{equation*}
\subsection{Shape of piezo-electric Moduli in Lattice Structures}
\label{subsec:LatticeStructure}
\noindent An hexagonal crystal system (point groups $\lbrace\bar{6}m2,\;6mm\rbrace$) is characterized by the following moduli $\mathbb{C},\; \boldsymbol{\epsilon},\; \mathbf{e}$ \cite{de2015database, mouhat2014necessary}:
\begin{equation*}
    \begin{aligned}
    \mathbb{C}_{hex}&=\begin{bmatrix}
    C_{11} & C_{12} & C_{13} & 0 & 0 & 0\\
     & C_{11} & C_{13} & 0 & 0 & 0\\
     & &  C_{33} & 0 & 0 & 0\\
     & & & C_{44} & 0 & 0\\
     & \text{sym.} & & & C_{44} & 0\\
     & & & & & \frac{1}{2}(C_{11}-C_{12})
    \end{bmatrix}, & \boldsymbol{\epsilon}_{hex}=\begin{bmatrix}
    \epsilon_{11} & 0 & 0\\
     & \epsilon_{11} & 0\\
     \text{sym.} & & \epsilon_{33}
    \end{bmatrix},\\
    \mathbf{e}_{hex}^{\bar{6}m2}&=\begin{bmatrix}
    0 & 0 & 0 & 0 & 0 & -e_{22}\\
    -e_{22} & e_{22} & 0 & 0 & 0 & 0\\
    0 & 0 & 0 & 0 & 0 & 0
    \end{bmatrix}, & \mathbf{e}_{hex}^{6mm}=\begin{bmatrix}
    0 & 0 & 0 & 0 & e_{15} & 0\\
    0 & 0 & 0 & e_{15} & 0 & 0\\
    e_{31} & e_{31} & e_{33} & 0 & 0 & 0
    \end{bmatrix}
    \end{aligned}
\end{equation*}.
A trigonal crystal system (point group $3m$) is characterized by the following moduli $\mathbb{C},\; \boldsymbol{\epsilon},\; \mathbf{e}$ \cite{de2015database, mouhat2014necessary}:
\begin{equation*}
    \begin{aligned}
    \mathbb{C}_{trig}&=\mathbb{C}_{hex},\quad \boldsymbol{\epsilon}_{trig}=\boldsymbol{\epsilon}_{hex},\\
     \mathbf{e}_{trig}&=\begin{bmatrix}
    0 & 0 & 0 & 0 & e_{15} & -e_{22}\\
    -e_{22} & e_{22} & 0 & e_{15} & 0 & 0\\
    e_{31} & e_{31} & e_{33} & 0 & 0 & 0
    \end{bmatrix}.
    \end{aligned}
\end{equation*}
Materials with trigonal crystal system, used in this work, demonstrate $e_{31}=0$.\\
An orthorhombic crystal system (point group $222$) is characterized by the following moduli $\mathbb{C},\; \boldsymbol{\epsilon},\; \mathbf{e}$ \cite{de2015database, mouhat2014necessary}:
\begin{equation*}
    \begin{aligned}
    \mathbb{C}_{orth}&=\begin{bmatrix}
    C_{11} & C_{12} & C_{13} & 0 & 0 & 0\\
     & C_{22} & C_{23} & 0 & 0 & 0\\
     & &  C_{33} & 0 & 0 & 0\\
     & & & C_{44} & 0 & 0\\
     & \text{sym.} & & & C_{55} & 0\\
     & & & & & C_{66}
    \end{bmatrix},\quad\boldsymbol{\epsilon}_{orth}=\begin{bmatrix}
    \epsilon_{11} & 0 & 0\\
     & \epsilon_{22} & 0\\
     \text{sym.} & & \epsilon_{33}
    \end{bmatrix},\\
    \mathbf{e}_{orth}&=\begin{bmatrix}
    0 & 0 & 0 & e_{14} & 0 & 0\\
    0 & 0 & 0 & 0 & e_{25} & 0\\
    0 & 0 & 0 & 0 & 0 & e_{36}
    \end{bmatrix}.
    \end{aligned}
\end{equation*}
\subsection{Additional Results}
\label{subsec:AdditionalResults}
\begin{figure}[htp!]
\centering
    \begin{minipage}{0.3\linewidth}
    \centering
        \includegraphics[width=\textwidth]{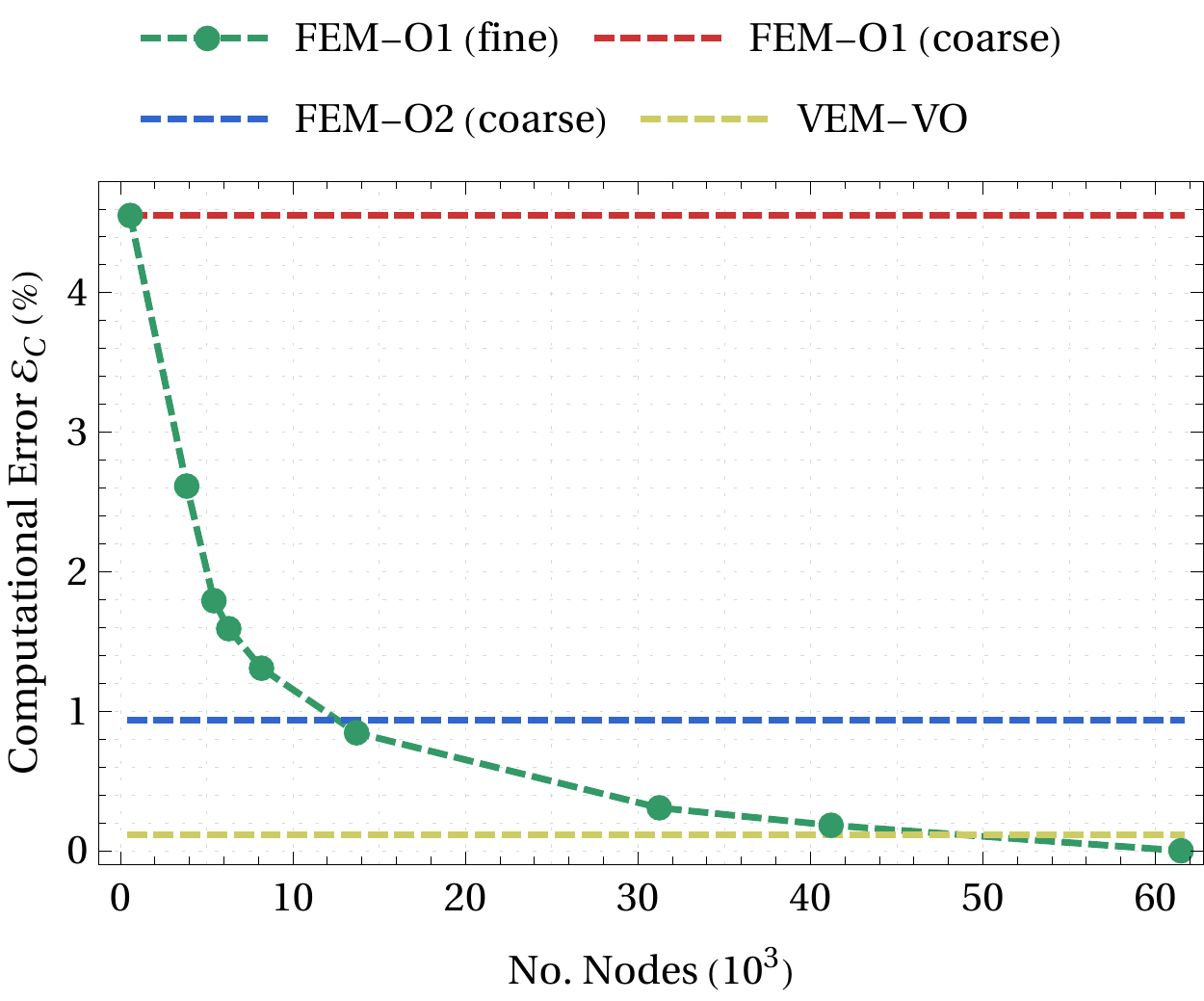}
        
        a)
    \end{minipage}
    \hfill
    \begin{minipage}{0.3\linewidth}
    \centering
        \includegraphics[width=\textwidth]{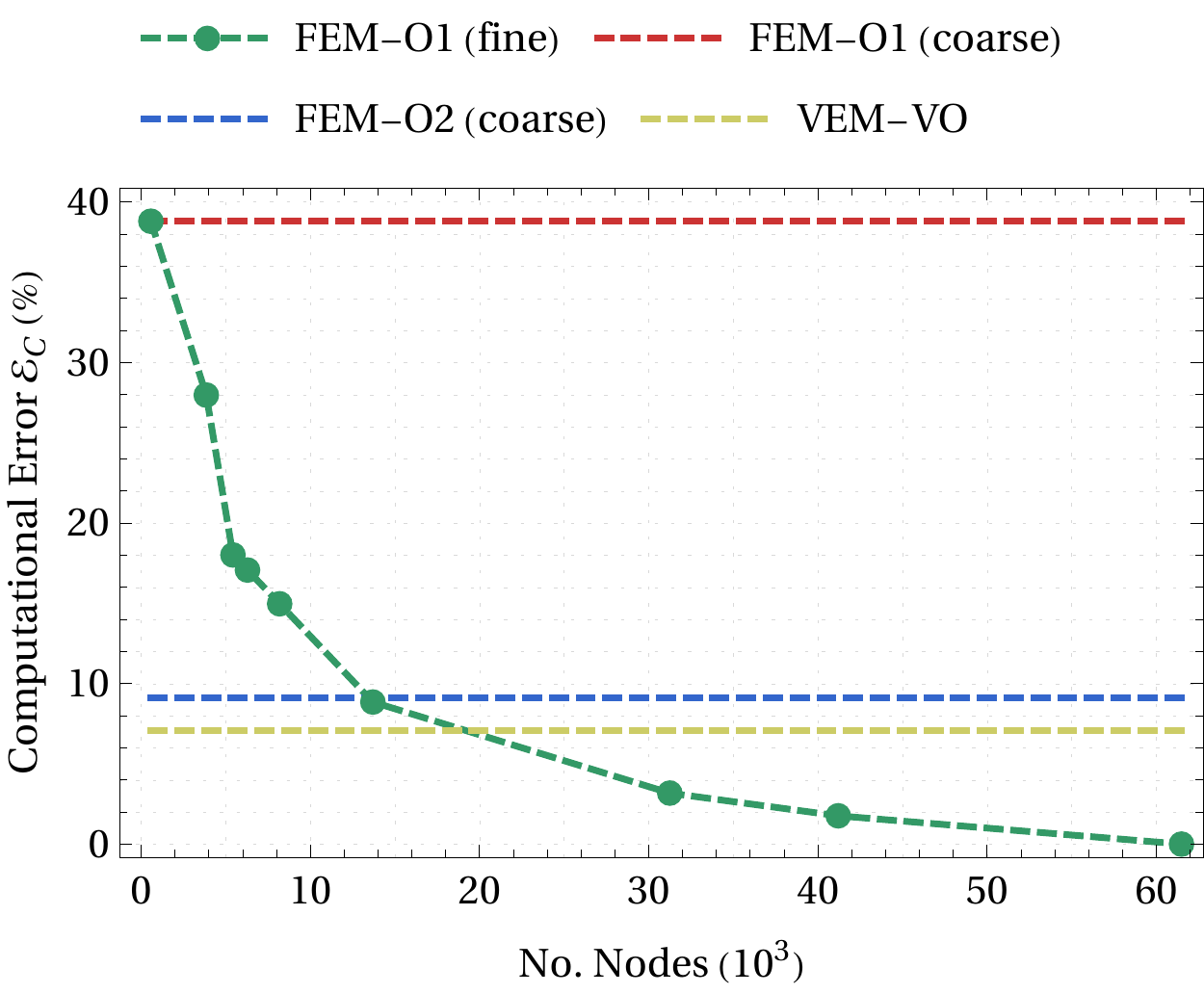}
        
        b)
    \end{minipage}
    \hfill
    \begin{minipage}{0.3\linewidth}
    \centering
        \includegraphics[width=\textwidth]{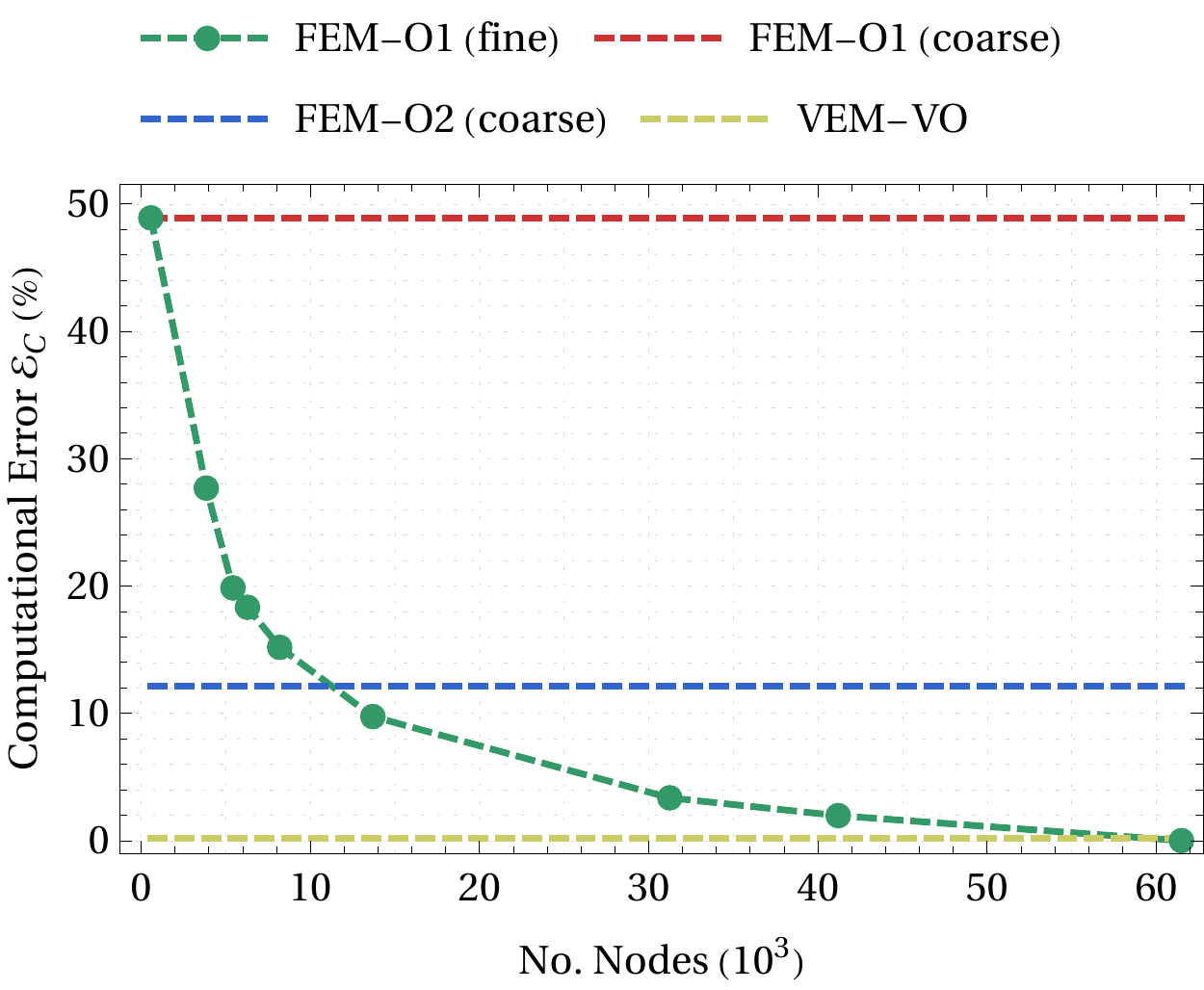}
        
        c)
    \end{minipage}
    \caption{Computational error $\mathcal{E}_{C}$ of $\mathrm{BaNiO_{3}}$ with hexagonal unit cell: 
    a) effective mechanical modulus $\overline{\mathbb{C}}$; b) effective electro-mechanical modulus $\overline{\mathbf{e}}$; c) effective dielectric modulus $\overline{\boldsymbol{\epsilon}}$}
    \label{fig:NormPlotshexa2}
\end{figure}
\FloatBarrier
\begin{figure}[htp!]
\centering
    \begin{minipage}{0.3\linewidth}
    \centering
        \includegraphics[width=\textwidth]{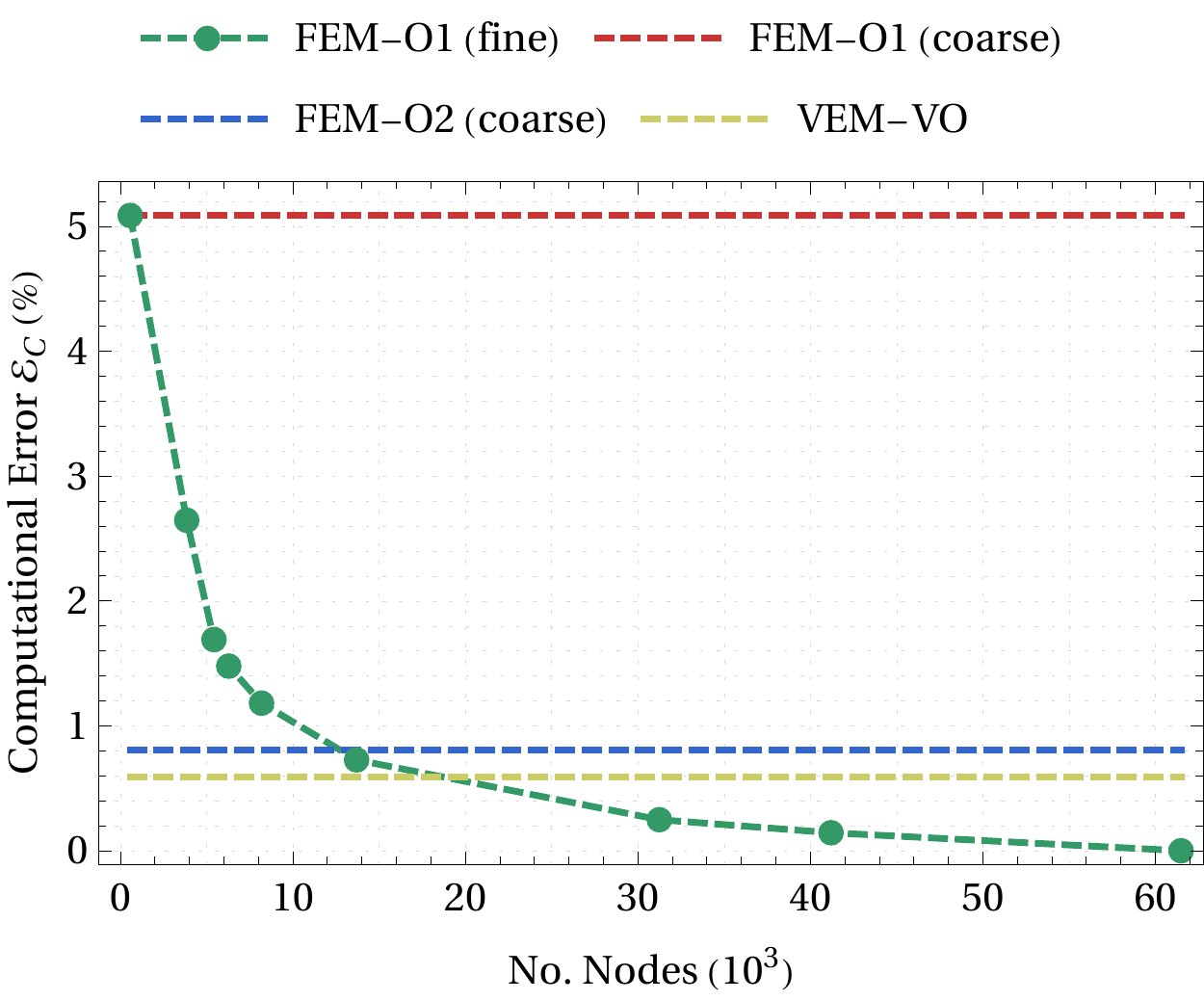}
        
        a)
    \end{minipage}
    \hfill
    \begin{minipage}{0.3\linewidth}
    \centering
        \includegraphics[width=\textwidth]{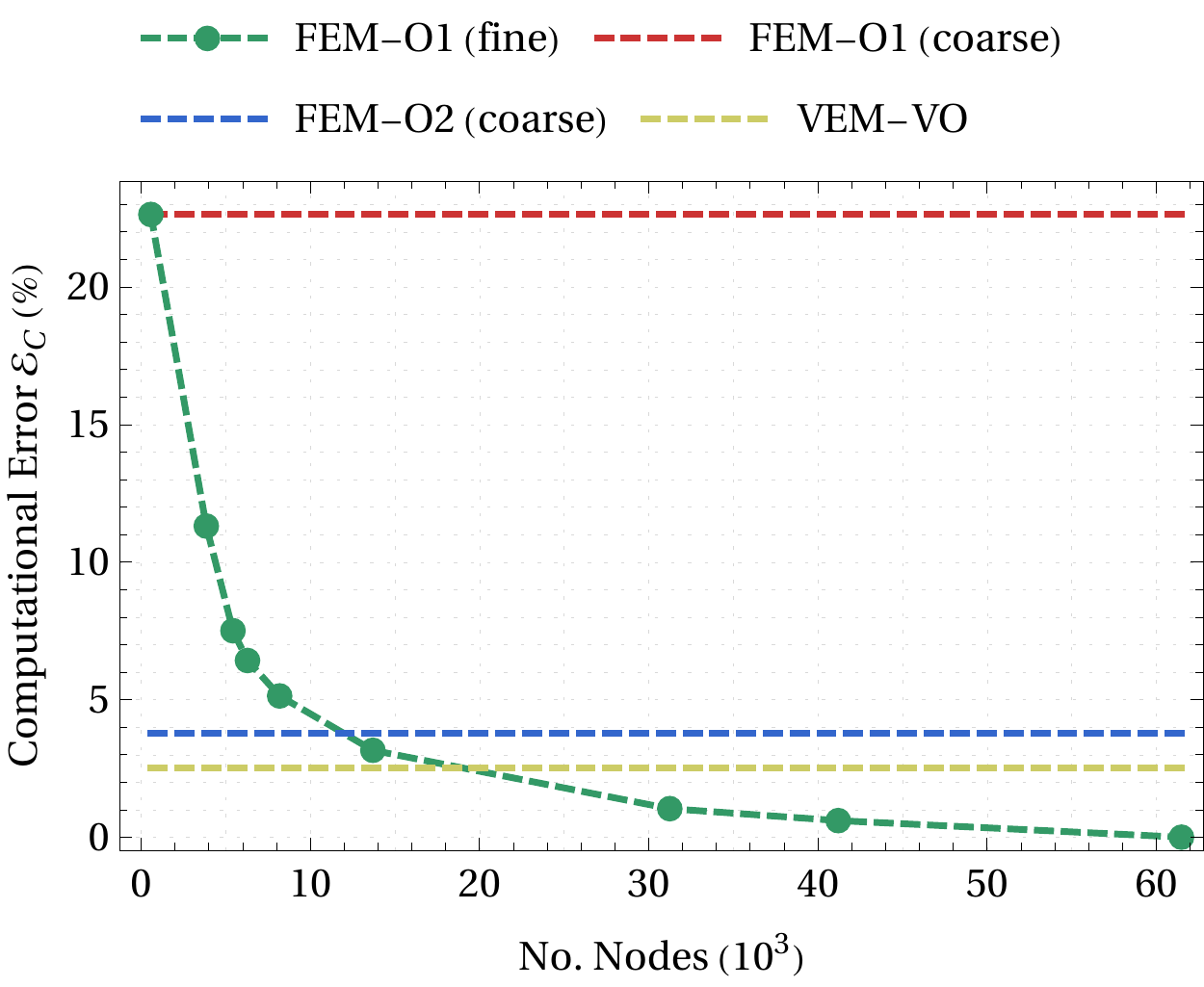}
        
        b)
    \end{minipage}
    \hfill
    \begin{minipage}{0.3\linewidth}
    \centering
        \includegraphics[width=\textwidth]{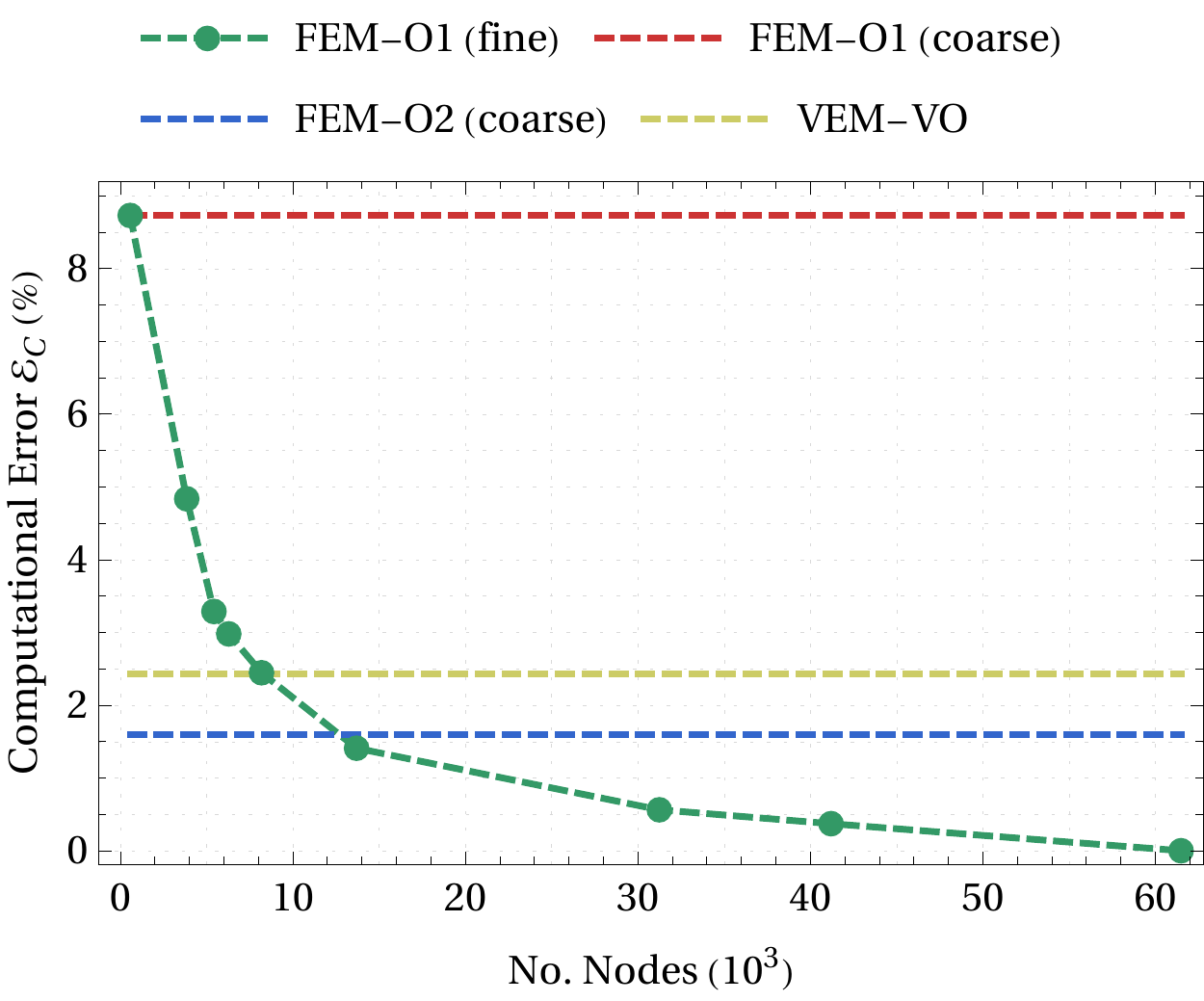}
        
        c)
    \end{minipage}
    \caption{Computational error $\mathcal{E}_{C}$ of $\mathrm{AlPO_{4}}$ with orthorhombic unit cell: 
    a) effective mechanical modulus $\overline{\mathbb{C}}$; b) effective electro-mechanical modulus $\overline{\mathbf{e}}$; c) effective dielectric modulus $\overline{\boldsymbol{\epsilon}}$}
    \label{fig:NormPlotsort2}
\end{figure}
\FloatBarrier
\begin{figure}[htp!]
\centering
    \begin{minipage}{0.3\linewidth}
    \centering
        \includegraphics[width=\textwidth]{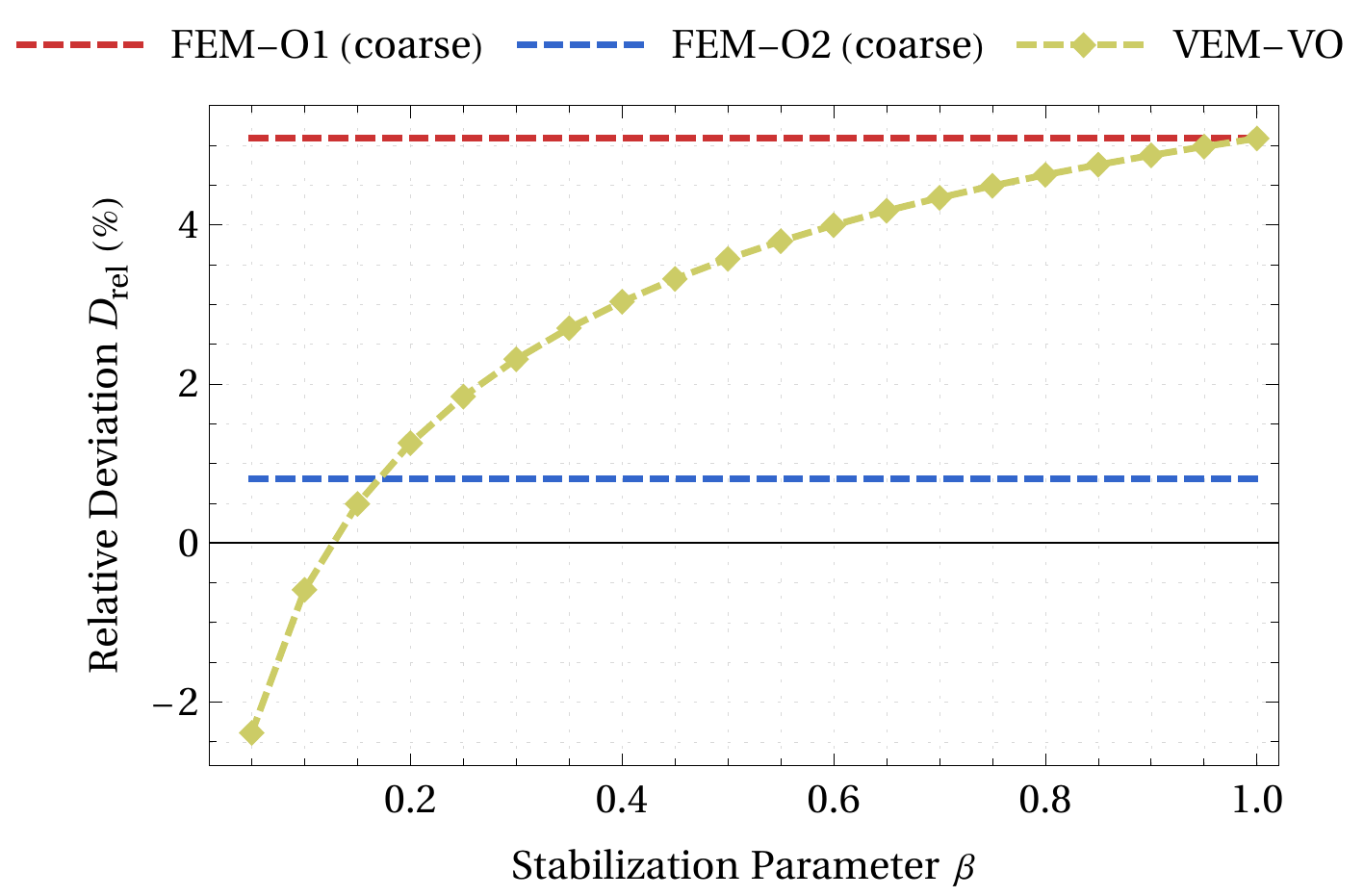}
        
        a)
    \end{minipage}
    \hfill
    \begin{minipage}{0.3\linewidth}
    \centering
        \includegraphics[width=\textwidth]{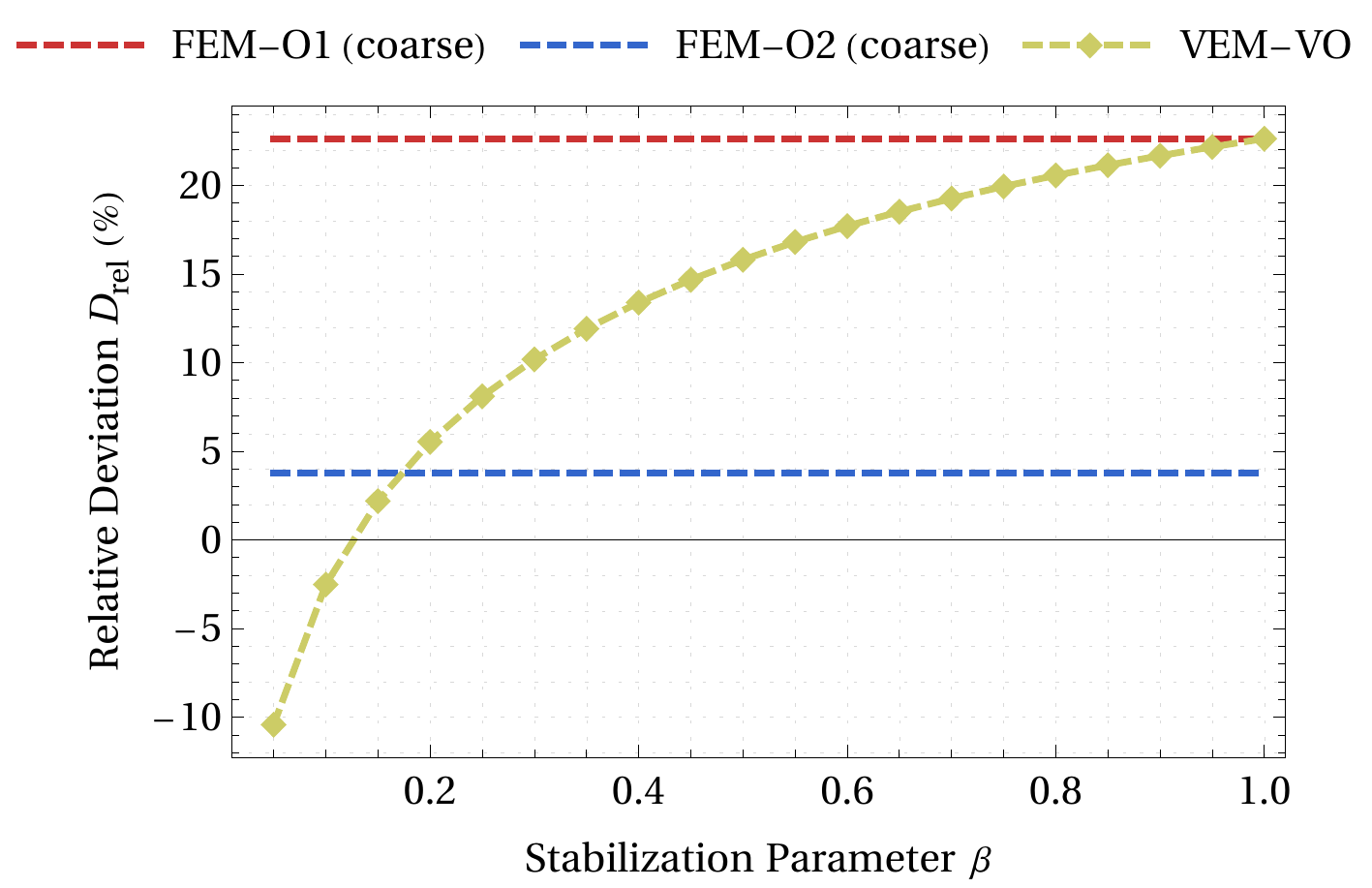}
        
        b)
    \end{minipage}
    \hfill
    \begin{minipage}{0.3\linewidth}
    \centering
        \includegraphics[width=\textwidth]{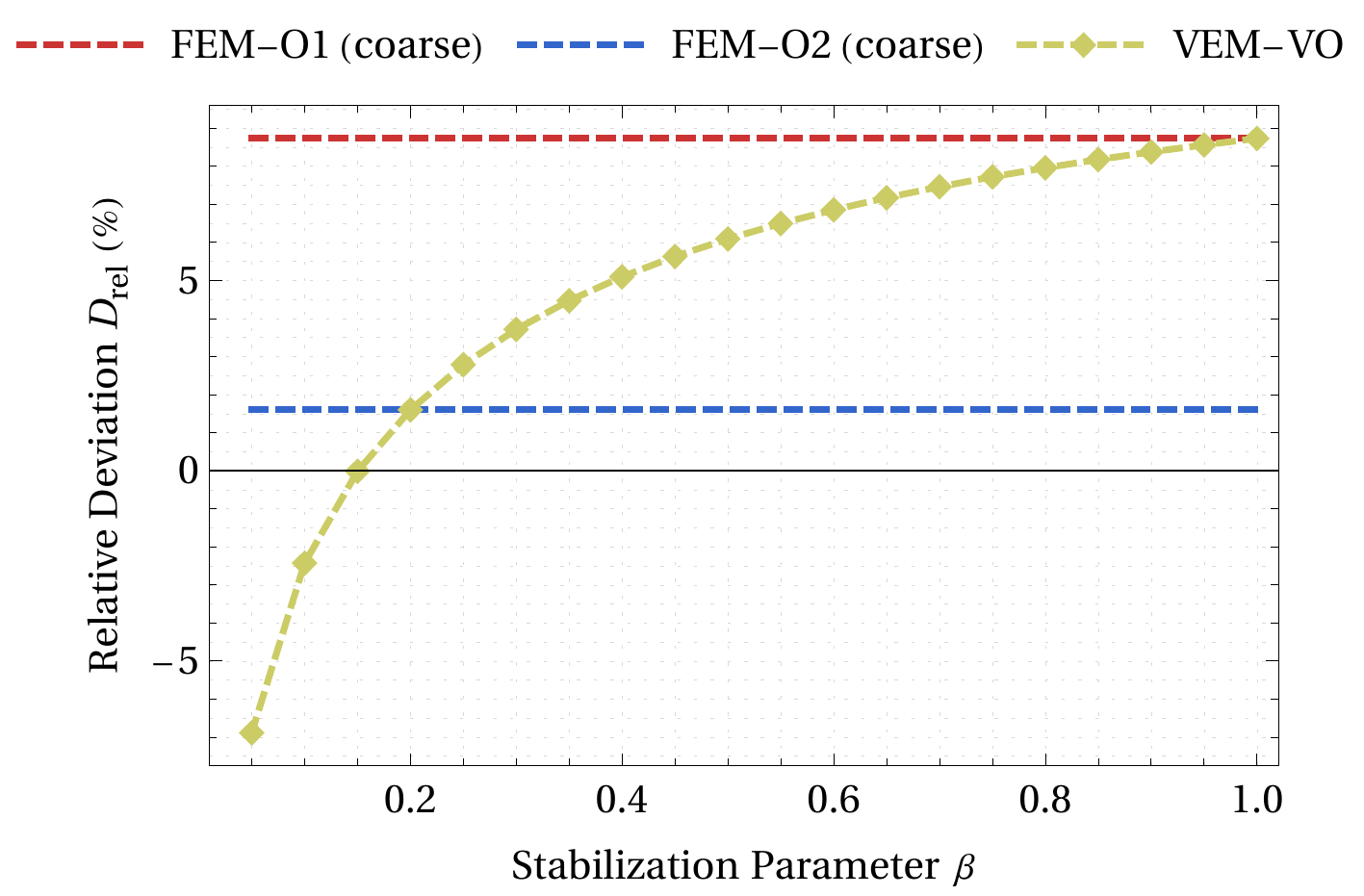}
        
        c)
    \end{minipage}
    \caption{Relative deviation $D_{\mathrm{rel}}$ of effective macroscopic moduli, material $\mathrm{AlPO_{4}}$, orthorhombic unit cell: a) mechanical modulus $\overline{\mathbb{C}}$; b) electro-mechanical modulus $\overline{\mathbf{e}}$; c) dielectric modulus $\overline{\boldsymbol{\epsilon}}$}
    \label{fig:NormpartbetaPlotsort2}
\end{figure}
\FloatBarrier
\begin{figure}[htp!]
\centering
    \begin{minipage}{0.3\linewidth}
    \centering
        \includegraphics[width=\textwidth]{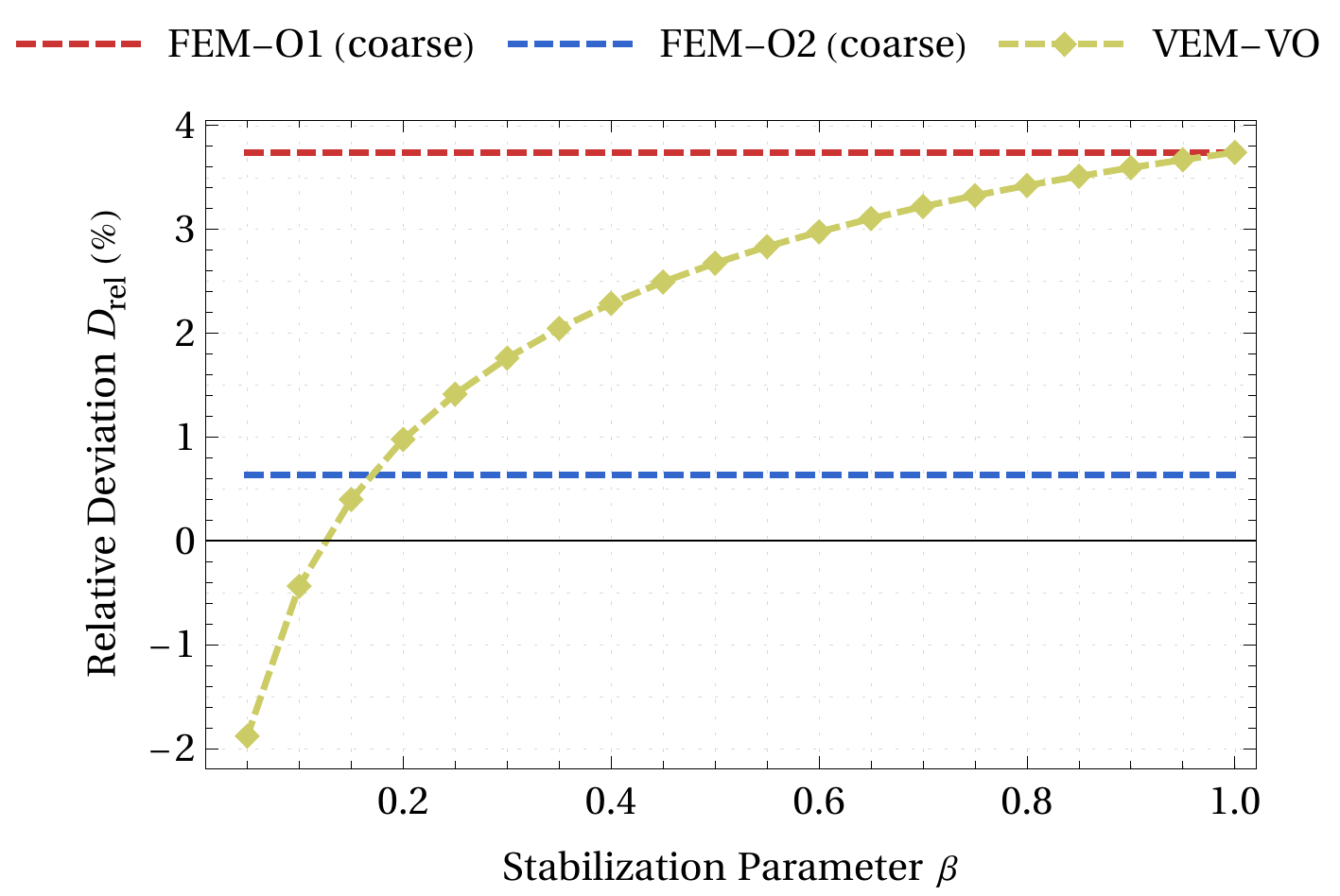}
        
        a)
    \end{minipage}
    \hfill
    \begin{minipage}{0.3\linewidth}
    \centering
        \includegraphics[width=\textwidth]{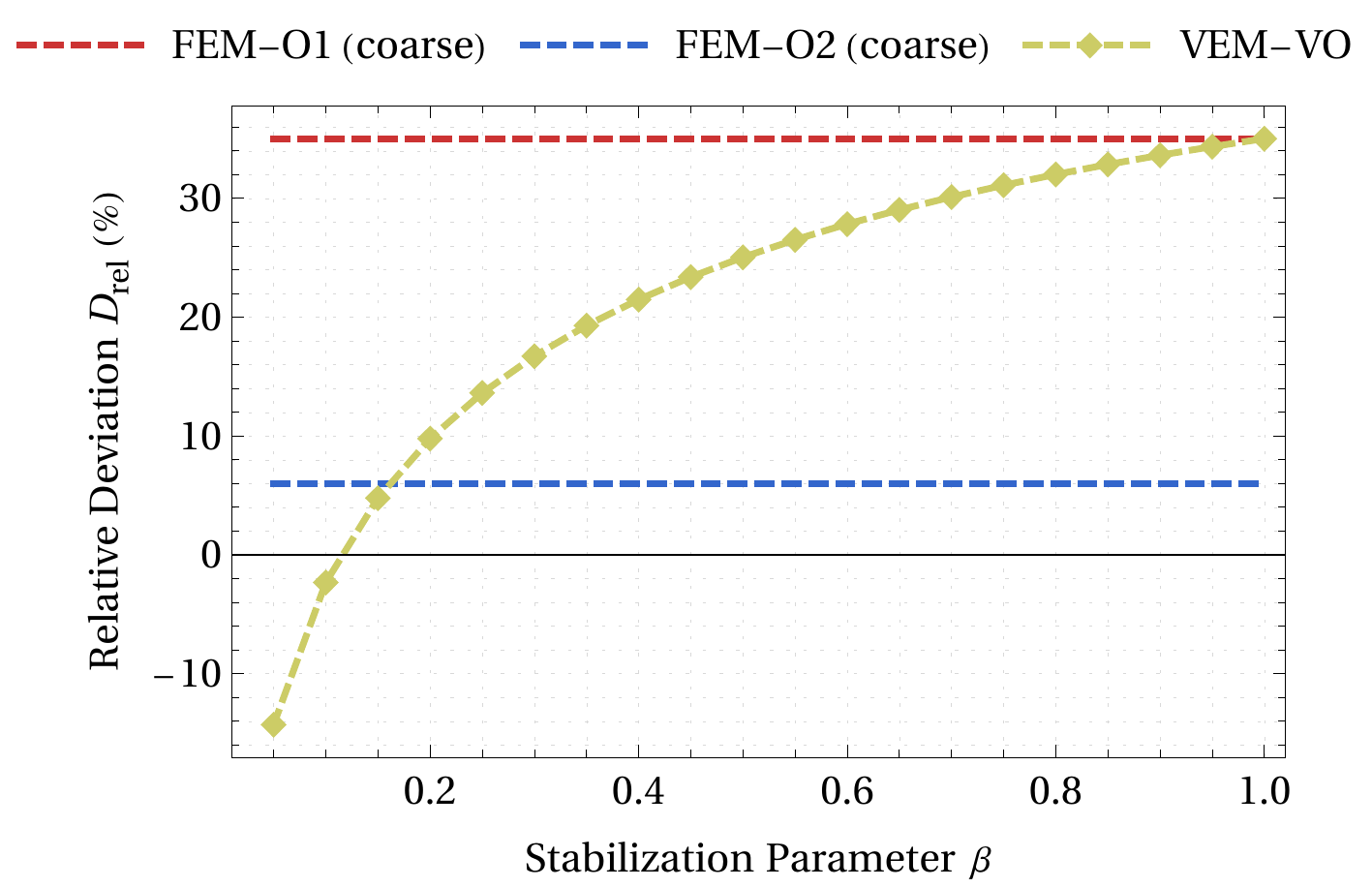}
        
        b)
    \end{minipage}
    \hfill
    \begin{minipage}{0.3\linewidth}
    \centering
        \includegraphics[width=\textwidth]{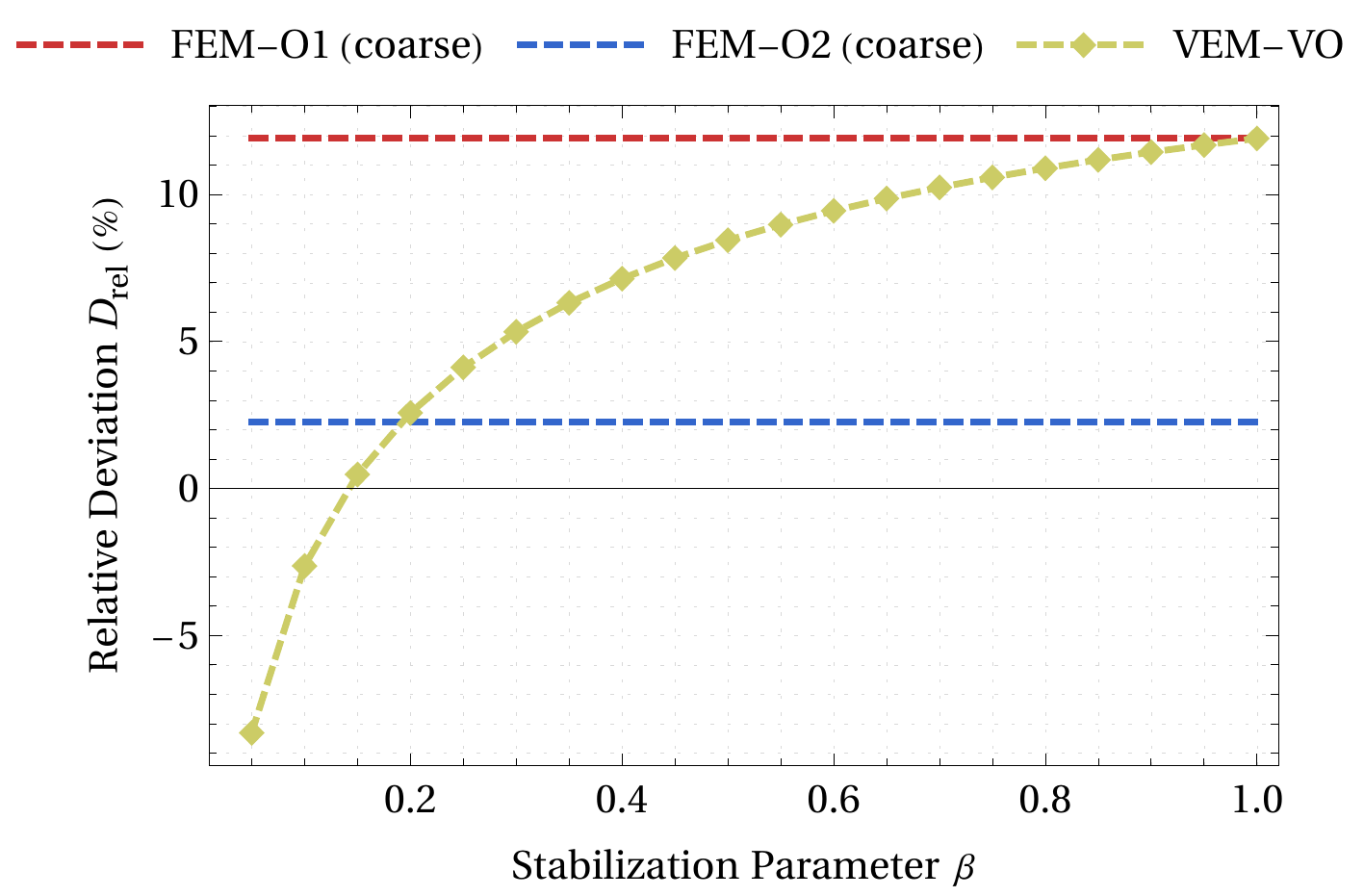}
        
        c)
    \end{minipage}
    \caption{Relative deviation $D_{\mathrm{rel}}$ of effective macroscopic moduli, material $\mathrm{GaPO_{4}}$, orthorhombic unit cell: a) mechanical modulus $\overline{\mathbb{C}}$; b) electro-mechanical modulus $\overline{\mathbf{e}}$; c) dielectric modulus $\overline{\boldsymbol{\epsilon}}$}
    \label{fig:NormpartbetaPlotsort3}
\end{figure}
\FloatBarrier
\begin{figure}[htp!]
\centering
    \begin{minipage}{0.3\linewidth}
    \centering
        \includegraphics[width=\textwidth]{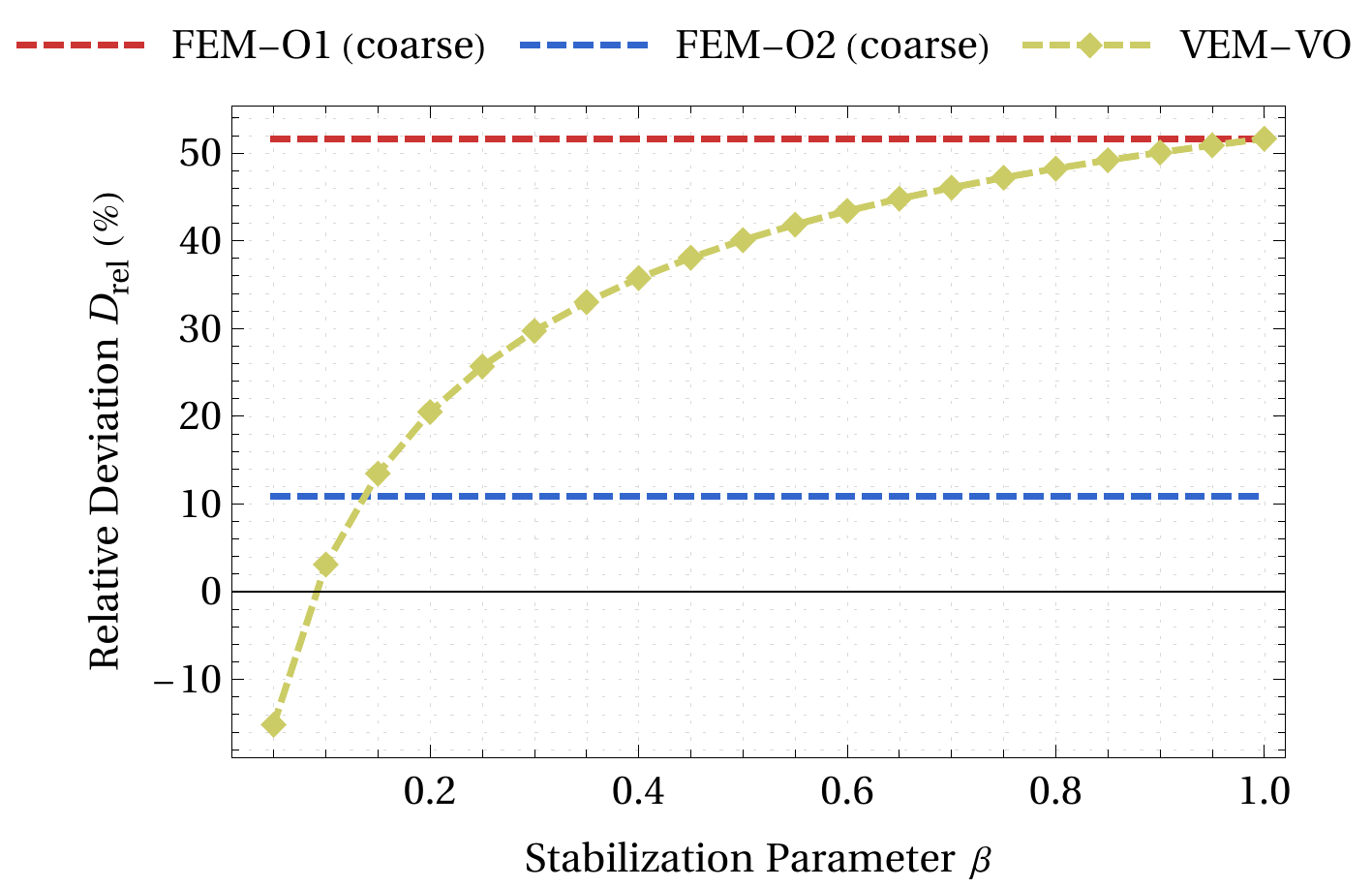}
        
        a)
    \end{minipage}
    \hfill
    \begin{minipage}{0.3\linewidth}
    \centering
        \includegraphics[width=\textwidth]{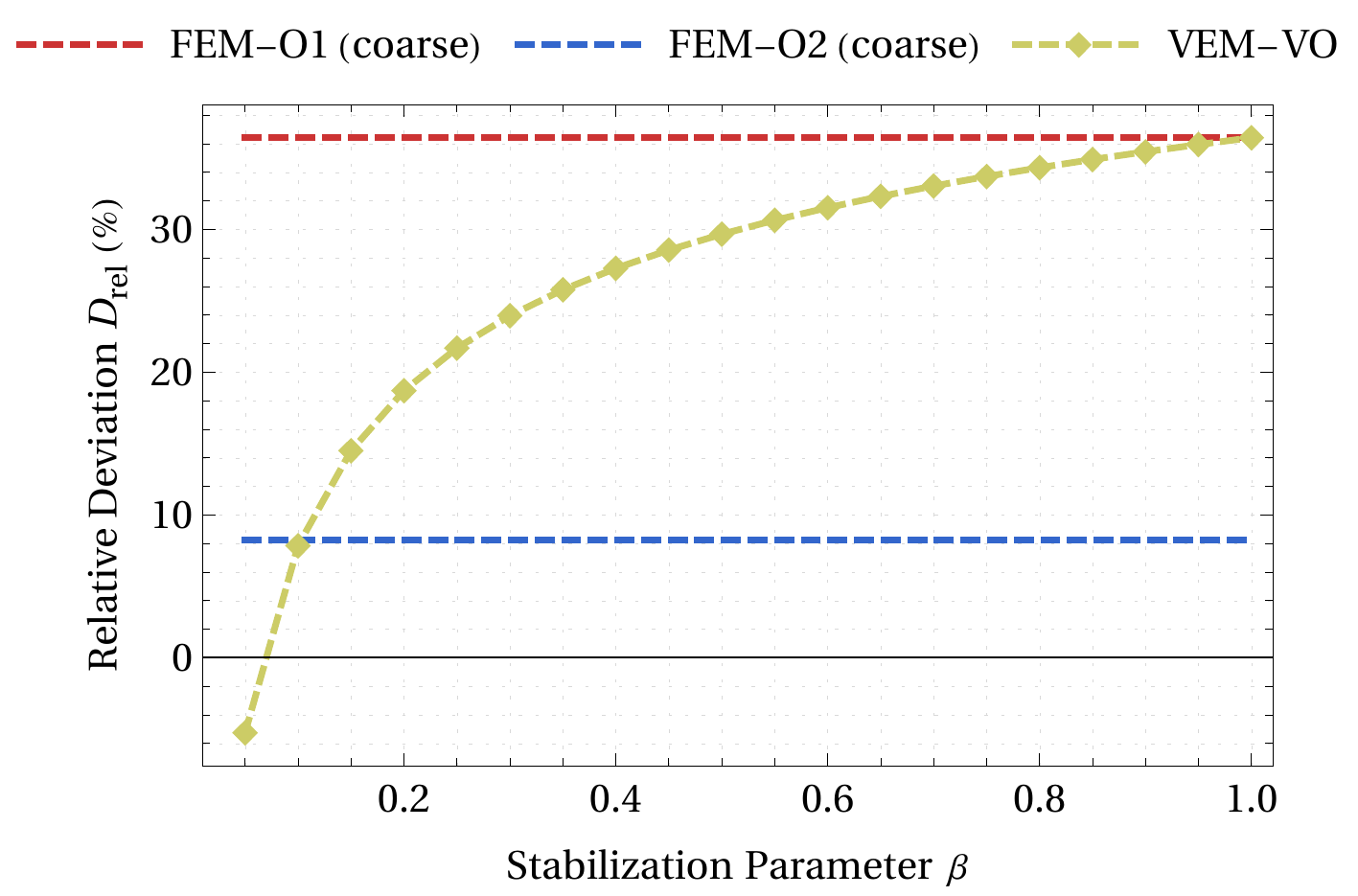}
        
        b)
    \end{minipage}
    \hfill
    \begin{minipage}{0.3\linewidth}
    \centering
        \includegraphics[width=\textwidth]{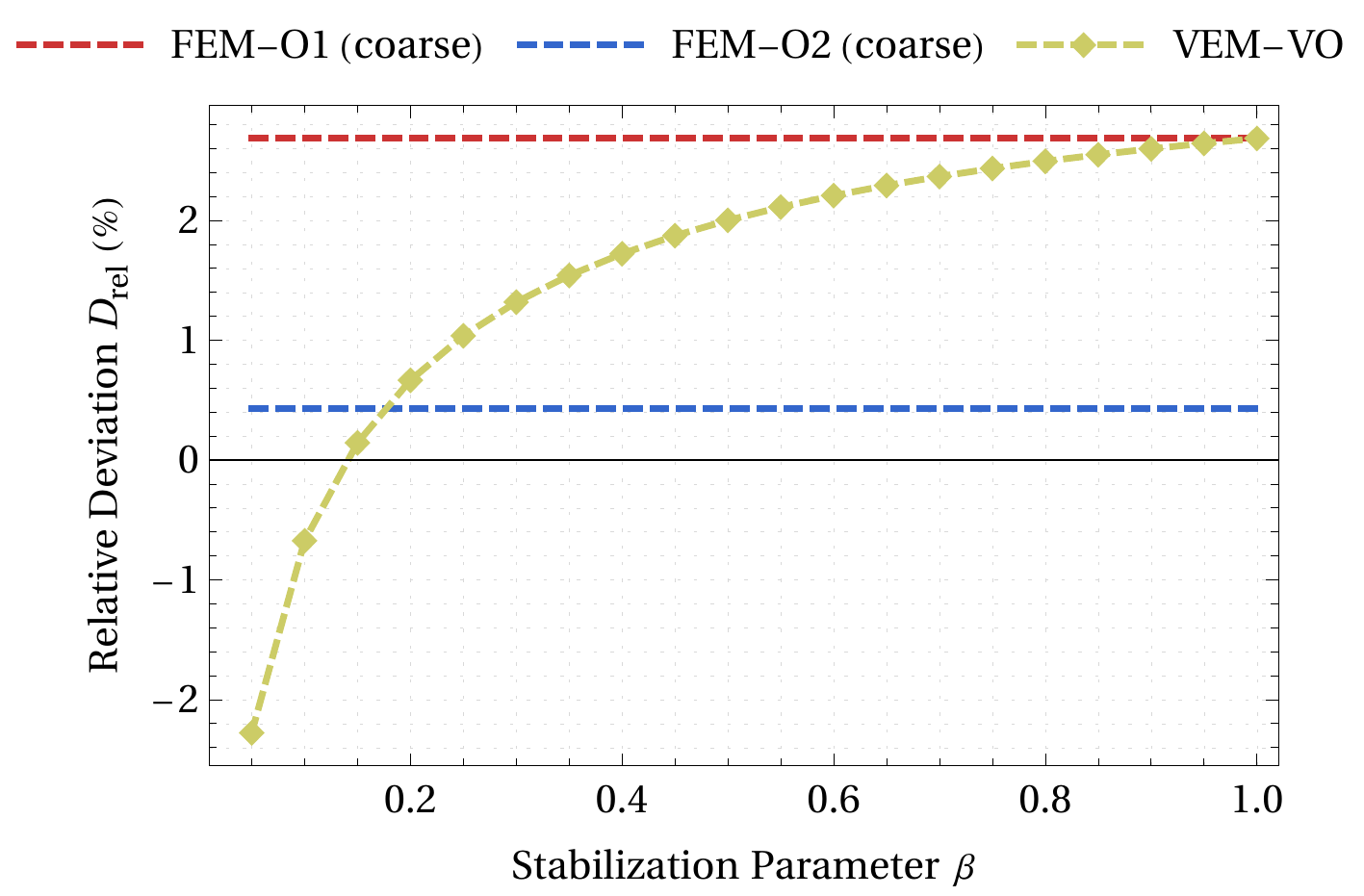}
        
        c)
    \end{minipage}
    \caption{Relative deviation $D_{\mathrm{rel}}$ of effective macroscopic moduli, material $\mathrm{MoS_{2}}$, trigonal unit cell: a) mechanical modulus $\overline{\mathbb{C}}$; b) electro-mechanical modulus $\overline{\mathbf{e}}$; c) dielectric modulus $\overline{\boldsymbol{\epsilon}}$}
    \label{fig:NormpartbetaPlotstri3}
\end{figure}
\FloatBarrier
\begin{figure}[htp!]
\centering
    \begin{minipage}{0.3\linewidth}
    \centering
        \includegraphics[width=\textwidth]{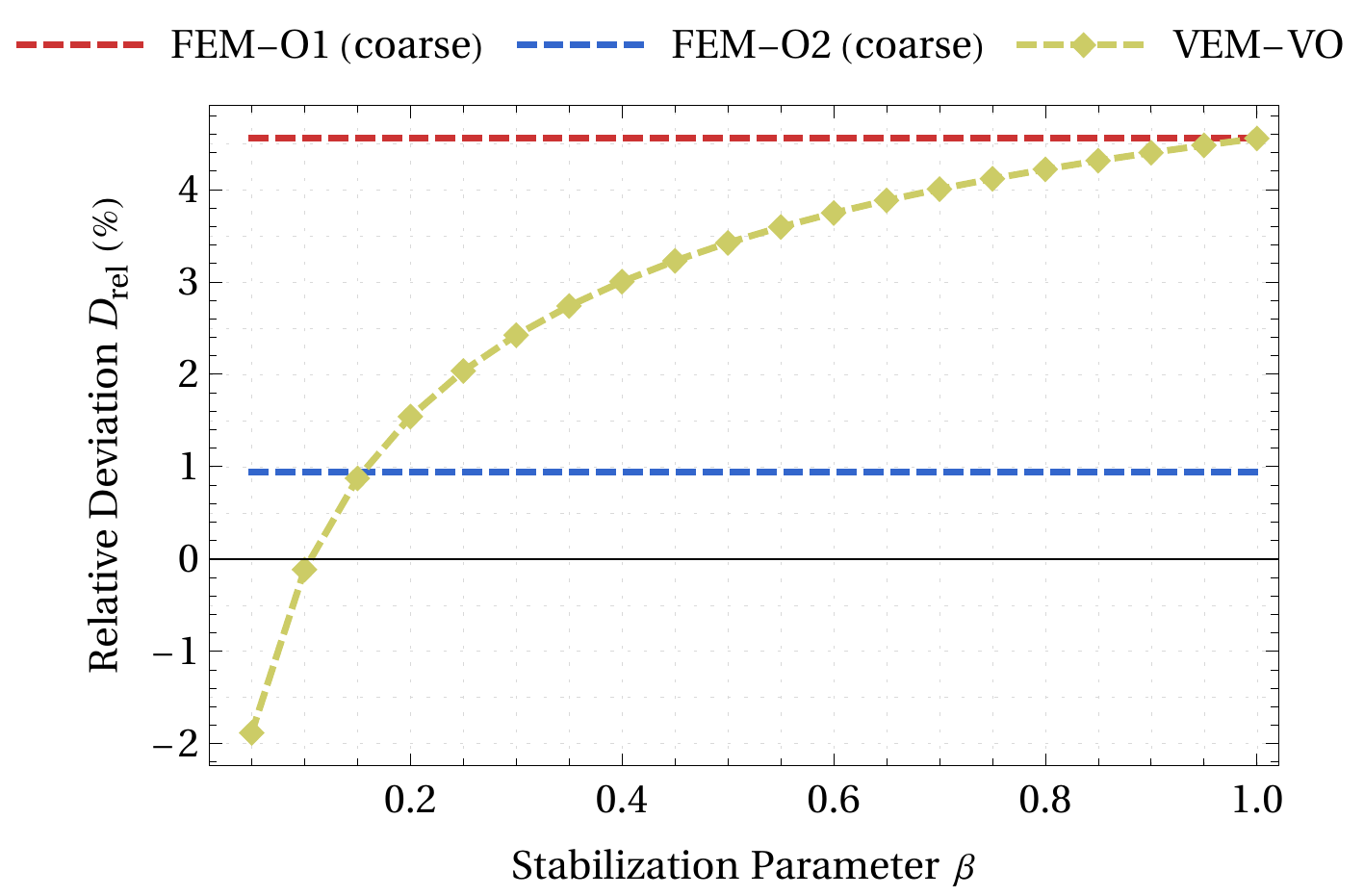}
        
        a)
    \end{minipage}
    \hfill
    \begin{minipage}{0.3\linewidth}
    \centering
        \includegraphics[width=\textwidth]{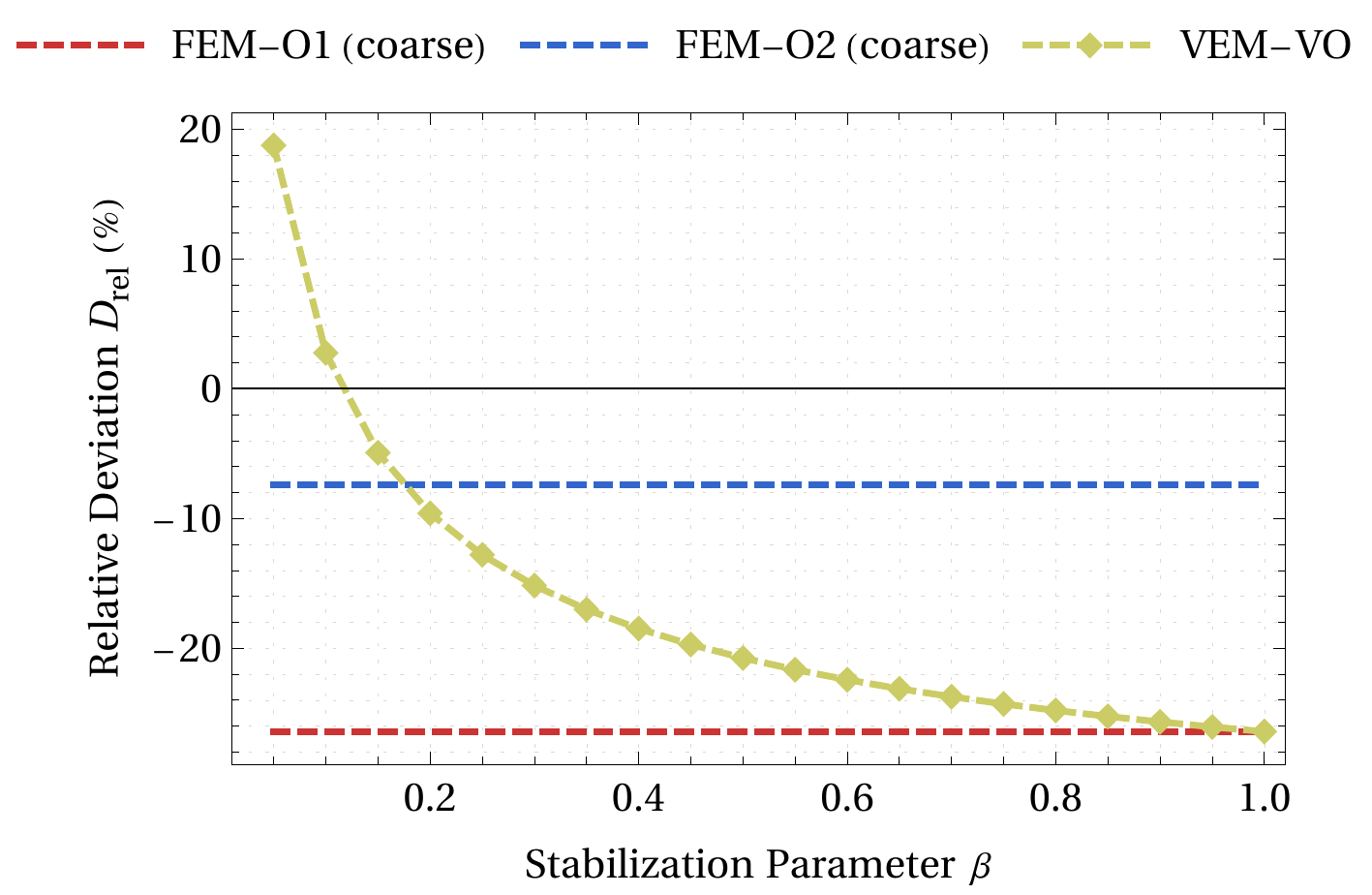}
        
        b)
    \end{minipage}
    \hfill
    \begin{minipage}{0.3\linewidth}
    \centering
        \includegraphics[width=\textwidth]{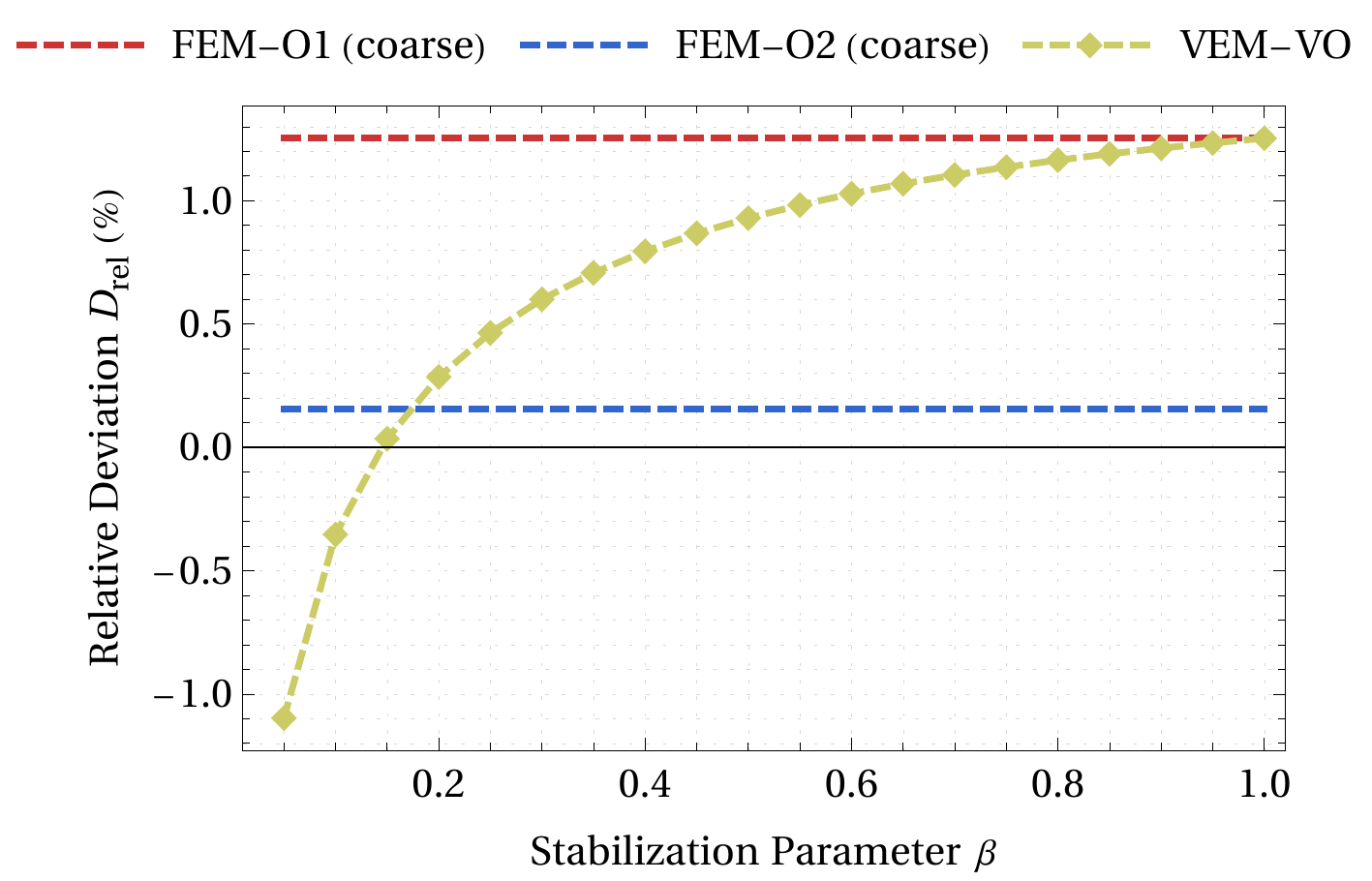}
        
        c)
    \end{minipage}
    \caption{Relative deviation $D_{\mathrm{rel}}$ of effective macroscopic moduli: a) mechanical modulus $\overline{\mathbb{C}}$, $\mathrm{BaNiO_{3}}$, hexagonal unit cell; b) electro-mechanical modulus $\overline{\mathbf{e}}$, $\mathrm{BN}$, hexagonal unit cell; c) dielectric modulus $\overline{\boldsymbol{\epsilon}}$, $\mathrm{BN}$, hexagonal unit cell}
    \label{fig:NormpartbetaPlotshexa23}
\end{figure}
\FloatBarrier
\begin{figure}[htp!]
\centering
    \begin{minipage}{0.3\linewidth}
    \centering
        \includegraphics[width=\textwidth]{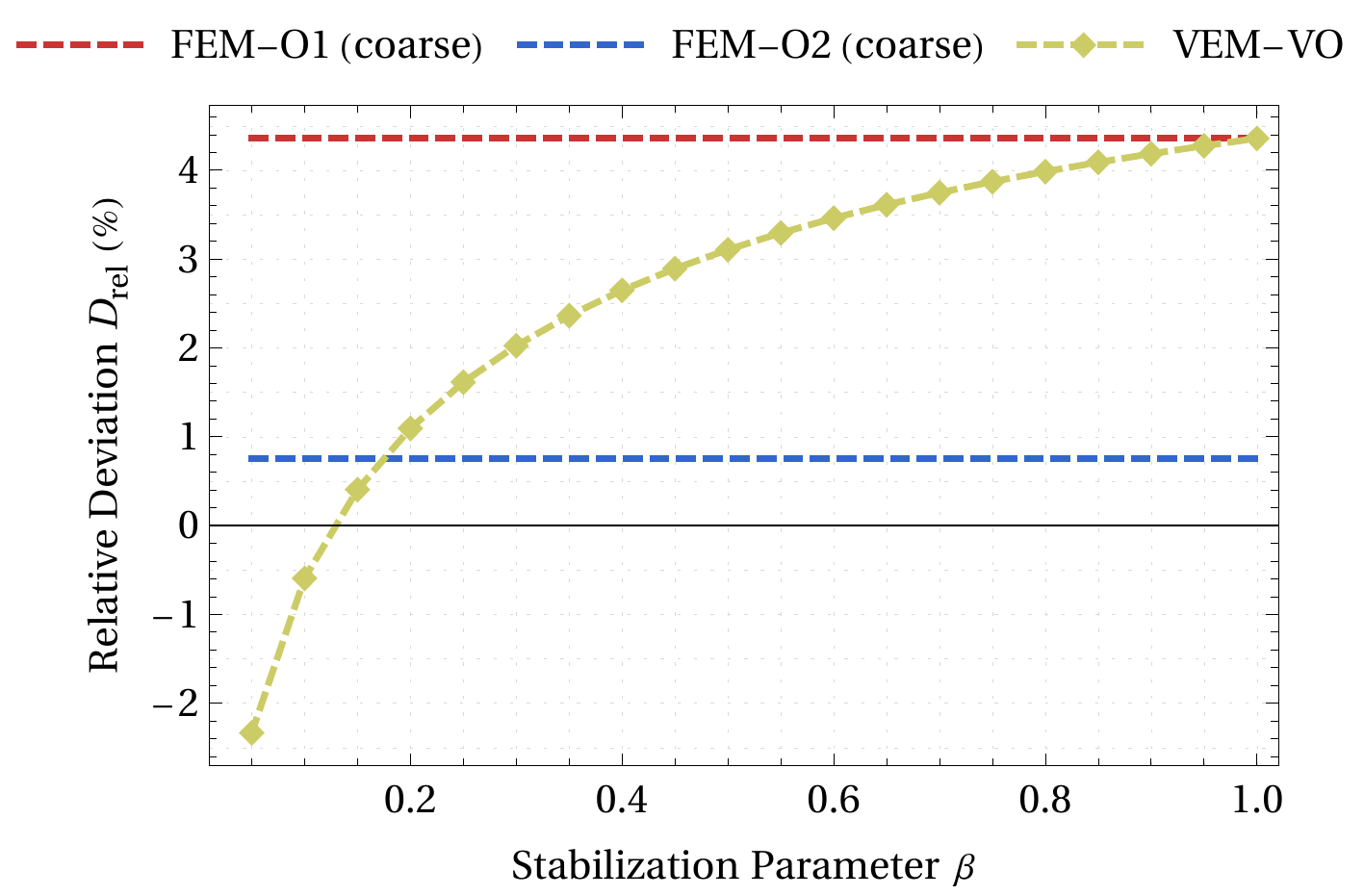}
        
        a)
    \end{minipage}
    \begin{minipage}{0.3\linewidth}
    \centering
        \includegraphics[width=\textwidth]{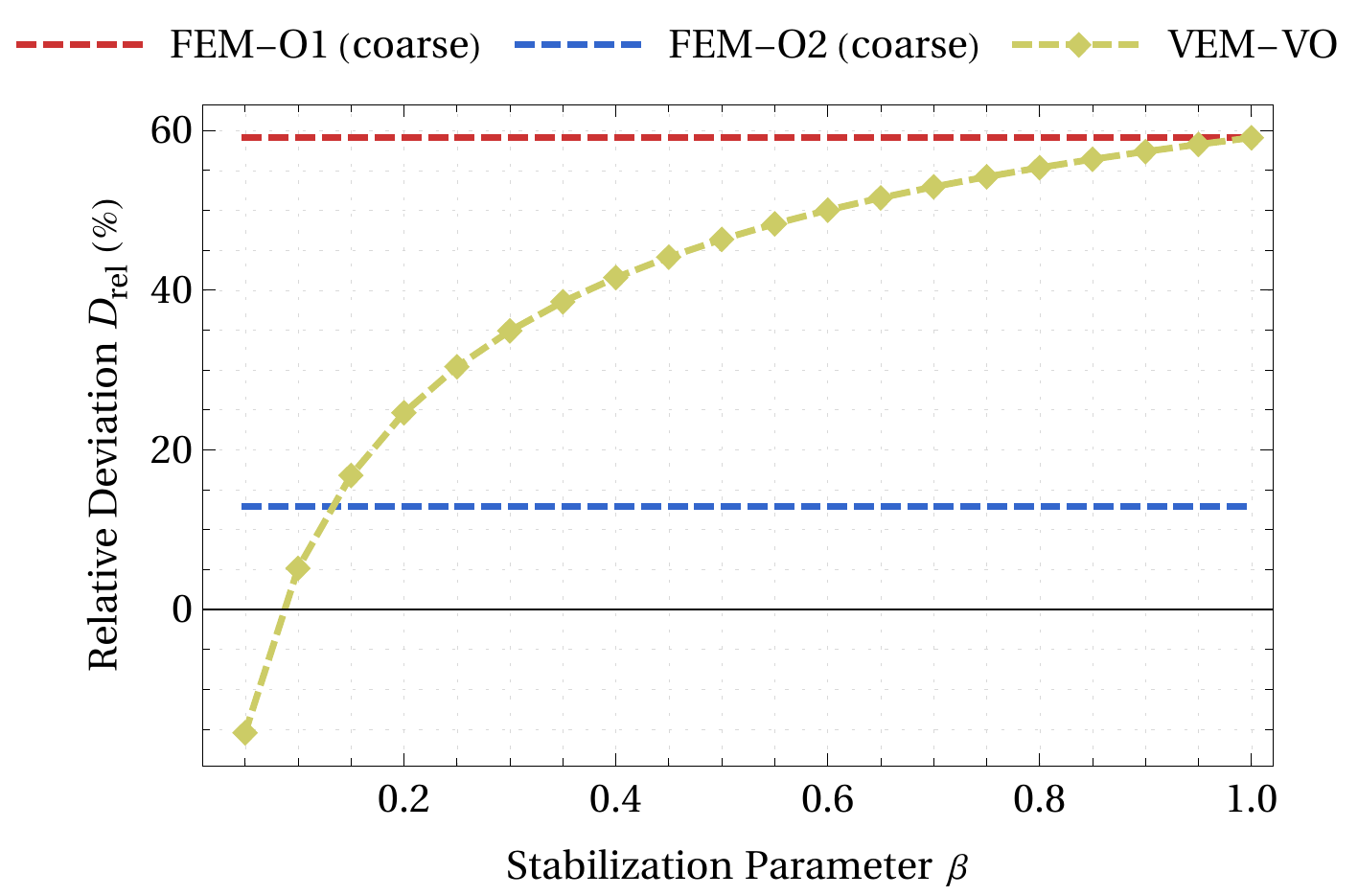}
        
        b)
    \end{minipage}
    \caption{Relative deviation $D_{\mathrm{rel}}$ of effective macroscopic modulus $\overline{\mathbb{G}}$: a) $\mathrm{GaPO_{4}}$, orthorhombic unit cell; b) $\mathrm{BN}$, hexagonal unit cell}
    \label{fig:NormGbetaPlotsort3hexa3}
\end{figure}
\FloatBarrier
\bibliography{NotesBIB}
\bibliographystyle{unsrt}
\end{document}